  \documentclass[a4paper,11pt]{article}
  \pdfoutput=1
  \usepackage{silence}
  \usepackage{amsmath}
  \usepackage{jheppub} 
  
\newcommand{\beq}{\begin{equation}}
\newcommand{\eeq}{\end{equation}}

\DeclareMathOperator*{\argmax}{arg\,max}

\usepackage[utf8]{inputenc}
\usepackage{amsthm}
\usepackage[T1]{fontenc}
\usepackage{mathtools}
\usepackage{bm}
\usepackage[usenames,dvipsnames]{xcolor}
\usepackage[normalem]{ulem}

\usepackage[ruled,vlined,linesnumbered]{algorithm2e}

\usepackage{refcount}
\newcommand{\footref}[1]{\hyperref[#1]{\footnotemark[\getrefnumber{#1}]}}

\allowdisplaybreaks 

  \usepackage{array}
  \usepackage{longtable}
  \usepackage{booktabs}
  
  \newcolumntype{P}[1]{>{\centering\arraybackslash}p{#1}} 
  \newcolumntype{M}[1]{>{\centering\arraybackslash}m{#1}} 
  \newcolumntype{B}[1]{>{\centering\arraybackslash}b{#1}} 
  
  \usepackage{dcolumn}
  \newcolumntype{.}{D{.}{.}{-1}} 

\makeatletter
\newcommand{\vast}{\bBigg@{3}}
\newcommand{\Vast}{\bBigg@{4}}
\makeatother

\begin{document}

\title{Finding Wombling Boundaries in LHC Data with Voronoi and Delaunay Tessellations}

\author{Konstantin T. Matchev,}
\author{Alexander Roman,}
\author{Prasanth Shyamsundar}
\affiliation{Institute for Fundamental Theory, Physics Department, University of Florida, Gainesville, FL 32611, USA}

\date{June 8 2020}

\emailAdd{matchev@ufl.edu}
\emailAdd{alexroman@ufl.edu}
\emailAdd{prasanths@ufl.edu}

\abstract{We address the problem of finding a wombling boundary in point data generated by a general Poisson point process, a specific example of which is an LHC event sample distributed in the phase space of a final state signature, with the wombling boundary created by some new physics. We discuss the use of Voronoi and Delaunay tessellations of the point data for estimating the local gradients and investigate methods for sharpening the boundaries by reducing the statistical noise. The outcome from traditional wombling algorithms is a set of boundary cell candidates with relatively large gradients, whose spatial properties must then be scrutinized in order to construct the boundary and evaluate its significance. Here we propose an alternative approach where we simultaneously form and evaluate the significance of all possible boundaries in terms of the total gradient flux. We illustrate our method with several toy examples of both straight and curved boundaries with varying amounts of signal present in the data. }

\dedicated{Dedicated to B.~P.~Delaunay.}

\maketitle
\flushbottom

\section{Introduction}

Collider data in high energy physics can be viewed, at least at the parton level, as a collection of points $\{\vec{\bm x}_1, \vec{\bm x}_2, \ldots, \vec{\bm x}_N\}$ in the relevant phase space ${\cal P}$ of the final state signature. The ultimate goal of a high-energy physics experiment is then to test whether the distribution of those $N$ points (commonly referred to as ``events'') in ${\cal P}$ follows the probability distribution predicted in some theory model by the fully differential cross-section
\begin{equation}
\frac{d\sigma}{d\vec{\bm x}} \equiv f(\vec{\bm x}, \left\{\alpha\right\}),
\label{fdef}
\end{equation}
where $\vec{\bm x}\in {\cal P}$ is a particular phase space point and $\left\{\alpha\right\}$ is a set of input model parameters such as particle masses, widths, couplings, etc.  More specifically, in searches for new physics (NP) beyond the standard model (SM), eq.~(\ref{fdef}) can be split into respective SM and NP contributions
\begin{equation}
f(\vec{\bm x}, \left\{\alpha\right\}) \equiv f_{SM} (\vec{\bm x}, \left\{\alpha_{SM}\right\})
+f_{NP}(\vec{\bm x}, \left\{\alpha_{SM}\right\}, \left\{\alpha_{NP}\right\}),
\label{f}
\end{equation}
where $\{\alpha_{SM}\}$ and $\{\alpha_{NP}\}$ label respectively the set of SM parameters and the set of additional NP parameters. Let the corresponding regions of phase space populated by SM and NP events be ${\cal P}_{SM}$ and ${\cal P}_{NP}$, then the respective total cross-sections are given by
\begin{subequations}
\begin{eqnarray}
\sigma_{SM} &=& \int_{{\cal P}_{SM}} f_{SM}(\vec{\bm  x}, \left\{\alpha_{SM}\right\}) \, d\vec{\bm x},
\label{sigma_SM}\\ [2mm]
\sigma_{NP} &=& \int_{{\cal P}_{NP}} f_{NP}(\vec{\bm  x}, \left\{\alpha_{SM}\right\}, \left\{\alpha_{NP}\right\}) \, d\vec{\bm x}.
\label{sigma_NP}
\end{eqnarray}
\end{subequations}

Since the standard model is well known, its distribution $f_{SM}$, a.k.a.~``the background'', is calculable theoretically (up to some fixed order in perturbation theory). However, once we account for the experimental realities, the result may suffer from non-negligible systematic uncertainties, particularly in the case of challenging signatures involving QCD and/or reducible backgrounds. This is the main roadblock in NP searches via counting experiments, where one focuses on a suitably chosen ``signal region'' ${\cal P}_{SR}\subset {\cal P}_{NP}$ and looks for an excess over the SM expectation $\int_{{\cal P}_{SR}} f_{SM} (\vec{\bm x}, \left\{\alpha_{SM}\right\})\, d {\vec{\bm x}}$.\footnote{In this paper, we have in mind the typical NP scenarios where ${\cal P}_{NP}$ is a subset of ${\cal P}_{SM}$, ${\cal P}_{NP}\subseteq {\cal P}_{SM}$. This is certainly the case for signatures where ${\cal P}_{SM}$ consists of the full phase space, ${\cal P}_{SM}={\cal P}$. If by any chance the signature happens to be such that ${\cal P}_{SM}\subset {\cal P}_{NP}$, the selection of the signal region is trivial: ${\cal P}_{SR}={\cal P}_{NP}-({\cal P}_{NP}\cap {\cal P}_{SM})$, and a counting experiment should already be good enough.}

Instead, in this paper we shall consider methods which could allow us to infer, at least in principle, the existence of the NP contribution $f_{NP}$ {\em without any prior knowledge} of the SM prediction $f_{SM}$. Recently there has been hightened interest in such ``blind'' or ``background-independent'' searches for NP, particularly using machine learning techniques \cite{Metodiev:2017vrx,Aguilar-Saavedra:2017rzt,Collins:2018epr,DeSimone:2018efk,Hajer:2018kqm,Heimel:2018mkt,Farina:2018fyg,Casa:2018avf,Cerri:2018anq,Collins:2019jip,Roy:2019jae,Dillon:2019cqt,Blance:2019ibf,Mullin:2019mmh,Nachman:2020lpy}. Here, instead of looking for an excess in the signal region ${\cal P}_{SR}$, we shall follow up on the idea of Refs.~\cite{Debnath:2015wra,Debnath:2015hva} to target directly the {\em boundary} $\partial {\cal P}_{NP}$ of the NP phase space region ${\cal P}_{NP}$, by using the fact that the combined distribution (\ref{f}) is non-differentiable anywhere on $\partial {\cal P}_{NP}$ where $f_{NP}$ is non-vanishing. As it turns out, the latter is a very safe assumption --- if anything, $f_{NP}$ is not only non-vanishing, but often enhanced and even {\em singular} on the boundary $\partial {\cal P}_{NP}$ \cite{Kim:2009si,Rujula:2011qn,DeRujula:2012ns,Agrawal:2013uka,Debnath:2016mwb,Debnath:2016gwz,Altunkaynak:2016bqe,Debnath:2018azt,Kim:2019prx,Matchev:2019bon}. Since the background distribution $f_{SM}$ is a smooth function across $\partial {\cal P}_{NP}$, the presence of $f_{NP}$ creates a discontinuous ``jump'' in the combined event density (\ref{f}), precisely at the location of the boundary $\partial {\cal P}_{NP}$ \cite{Debnath:2015wra,Debnath:2015hva}. We can thus reformulate the original problem of finding evidence of NP in the collider data as follows:
\begin{quote}
    Given a collection of $N$ points $\{\vec{\bm x}_1, \vec{\bm x}_2, \ldots, \vec{\bm x}_N\}$ in the phase space ${\cal P}$, identify (the locations of) any candidate wombling boundaries and estimate their statistical significance.
\end{quote}
The detection of such difference boundaries (or wombling boundaries, named after a pioneer in the field, W.~H.~Womble \cite{Womble}) is a well-known problem in spatial statistics, see, e.g., \cite{DaleFortin}. Broadly speaking, wombling is any of a number of techniques used for identifying zones of rapid change, typically in some quantity as it varies across some geographical or Euclidean space. Wombling techniques are being applied in a wide variety of disciplines, including computational ecology, anthropology, linguistics, geography and many others.\footnote{For example, wombling has been used to identify genetic boundaries in Eurasian human populations \cite{Barbujani1989, Barbujani1990}, language boundaries in Europe \cite{Oden1993}, transition zones in genetic, morphometric and physiological characteristics \cite{Bocquet-Appel1994}, boundaries of different types of vegetation \cite{Fortin1994,Fortin1997}, hospital admission rates for respiratory conditions \cite{Jacquez1995}, cancer rates \cite{JacquezGreiling2003}, metal concentrations in the Swiss Jura \cite{Gleyze2001} and other environmental data \cite{FortinDrapeau1995}. } In our case here, we shall be interested in identifying the phase space region in the vicinity of $\partial {\cal P}_{NP}$ where the {\em density} of points is changing significantly. 

Before going through the typical steps of a wombling analysis, several comments are in order. First, the discontinuous jump of the combined distribution (\ref{f}) across $\partial  {\cal P}_{NP}$ in practice will be smoothed out to some extent by the detector resolution and finite width effects, leading to a well-defined, finite, density gradient everywhere in ${\cal P}$. This precludes us from using methods specifically designed to detect image discontinuities such as ridges and cliffs but do not admit gradients \cite{Banerjee2006}. Second, a good wombling method should also be able to pick up any boundaries created within ${\cal P}_{SM}$ by interesting SM subprocesses, e.g., top, Higgs or heavy gauge boson production. This would guarantee an opportunity for the LHC experiments to start testing and validating the method with existing real data, before any NP discovery. Third, while most applications of wombling in the literature have been limited to two-dimensional data, the method shoud be readily generalizable to higher dimensions, if it is to be of any interest to the high-energy physics community where the dimensionality of the relevant phase space $\cal P$ is typically much higher (although in some special cases it can be reduced to 2 or even 1 through suitable projections preserving the boundary). Along those lines, it is also important to choose a good parametrization of ${\cal P}$, so that the dimensionality can be reduced by projecting out uninteresting degrees of freedom without washing out the wombling boundary. 

The main steps of a typical wombling analysis are the following \cite{Jacquez2000}.
\begin{itemize}
\item {\it Data preparation and preprocessing.} The starting point in wombling is a spatially referenced dataset
\begin{equation}
(\vec{x}_i, f_i), \quad i=1,2,\ldots,N,
\label{dataset}
\end{equation}
where a set of values $\{f_i\}$ for the function of interest $f(\vec{x})$ are obtained at some finite number $N$ of point locations $\{\vec{x}_i\}$. In some applications, e.g., for aerial and remotely sensed images, the locations $\{\vec{x}_i\}$ can be chosen by the experimenter --- then it may be convenient to arrange them in some kind of a regular lattice, as required by some wombling algorithms, including the original proposal in \cite{Womble}. Alternatively, when the data are gathered by an irregular or random design, eq.~(\ref{dataset}) is known as {\em point-referenced} or {\em geostatistical} data. While in many other fields of science the functional values $\{f_i\}$ for geostatistical data are obtained directly from field observations, for the Monte Carlo simulations used in high-energy physics the situation is more subtle --- each $f_i$ is supposed to be a measure of the local point density at $\vec{x}_i$ and needs to be evaluated as a preprocessing step. For this purpose, Refs.~\cite{Debnath:2015wra,Debnath:2015hva} proposed to consider the Voronoi tessellation in ${\cal P}$ of the dataset (\ref{dataset}), since the geometric volume $v_i$ of the Voronoi cell containing $\vec{x}_i$ provides a natural local estimator of the point density at $\vec{x}_i$ \cite{Okabe1992}:
\begin{equation}
    f_i \sim \frac{1}{v_i}.
\label{feq1overv}
\end{equation}
While Voronoi tessellations have been widely used in many other fields of science, they seem to be underutilized in high energy physics where their application has been limited to jet clustering \cite{Cacciari:2011ma} and the partitioning of the signature phase space into search regions, as implemented in the {\tt SLEUTH} algorithm \cite{Knuteson:2003dm,Knuteson:2004nj,Knuteson:2005ev} which was used to perform model-independent new physics searches at D0 \cite{Abbott:2001ke,Abbott:2000fb,Abbott:2000gx}, HERA \cite{Aktas:2004pz} and CDF \cite{Aaltonen:2007dg,Aaltonen:2007ab,Aaltonen:2008vt}. Yet, the Voronoi approach\footnote{In the context of treating high energy collider data as a point dataset (\ref{dataset}), it is worth mentioning the recent idea of Ref.~\cite{Mullin:2019mmh} to consider $\{\vec{x}_i\}$ as a graph network of weighted nodes, which is somewhat orthogonal, but similar in spirit to the Voronoi approach.} is ideally suited for finding interesting (e.g., singular) features in $f(\vec{x})$, since it preserves the maximum spatial resolution in the data~\cite{Cappellari:2009sc}. For example, the standard approach of binning the data in order to obtain a local density estimate necessarily throws away a certain amount of useful information, and is associated with some arbitrariness in the exact choice of binning  \cite{Debnath:2015wra}. 
\item {\it Gradient estimation.} Once we have the point-referenced dataset (\ref{dataset}), the next step is usually to estimate the magnitude of the local gradient $\vec{\nabla} f$ of the function $f(\vec{x})$, since any zone of rapid change is necessarily associated with large values for $|\vec{\nabla} f|$. In calculating the gradient, one has to overcome the fact that the data is a) discrete and b) irregularly sampled. One possible approach is to obtain a continuous approximation for $f(\vec{x})$ via some spatial interpolation method \cite{Steed1984,Okabe1992}. Unfortunately, interpolation techniques tend to smooth out not only the noise but also small local discontinuities, which can result in masking some true boundaries \cite{Fortin1994}. For this reason, Refs.~\cite{Debnath:2015wra,Debnath:2015hva} explored several boundary detection techniques (further developed and illustrated in \cite{Debnath:2016mwb,Debnath:2016gwz,DebnathPhD}) which continued to use the Voronoi tessellation of the data and the fundamental relation (\ref{feq1overv}). Among the different options studied in \cite{Debnath:2015wra,DebnathPhD},  the normalized standard deviation (sometimes also called the coefficient of variance) of the volumes of the neighboring cells emerged as a viable measure of the magnitude of the local gradient within a given Voronoi cell. In this paper, we shall pursue a somewhat orthogonal and more traditional approach, known in the literature as ``triangulation wombling'' \cite{Fortin1994,FortinDrapeau1995}, which makes use of the dual Delaunay tessellation of the data (\ref{dataset}). Even though the two types of tessellations are dual to each other\footnote{In graph theory, the dual graph of a plane graph $G$ is a graph that has a vertex for each face of $G$. Correspondingly, each edge $e$ of $G$ has a corresponding dual edge, whose endpoints are the dual vertices corresponding to the faces on either side of $e$. Some care must be exercised for the Delaunay edges along the convex hull of the point set --- their Voronoi duals are infinite rays which can be turned into finite line segments by adding an artificial point at infinity which serves as a common endpoint for all the rays. In our analysis, this complication will not arise since we will perform our analysis in the interior of ${\cal P}$.}, the Delaunay version seems more natural for the specific problem at hand of calculating gradients --- this will be further discussed and illustrated in  Section~\ref{sec:VoronoiDelaunay} below.
\item {\em Tagging selection.} Having constructed the Delaunay or Voronoi tessellation and obtained estimates for the local gradients, the next task is to identify elements of the tessellation (edges or vertices) which are likely to be located on or near a wombling boundary. This is typically accomplished with a cut on the ranked values of $|\vec\nabla f|$ for all elements in the tessellation, e.g., selecting elements whose calculated gradients are in a certain upper percentile\footnote{In NP scenarios where the signal density is additionally {\em enhanced} on the boundary $\partial {\cal P}_{NP}$, Ref.~\cite{Debnath:2016mwb} proposed a two-dimensional cut, simultaneously targeting both large gradients and large values of the function.}. However, using such a simple threshold for tagging boundary elements has been viewed as somewhat arbitrary and rather subjective \cite{Jacquez2000} --- in the absence of any robust guidelines, typical values used in the literature range in the $5^{th}$ to $10^{th}$ percentile \cite{Barbujani1989,Barbujani1990,FortinDrapeau1995,Jacquez1995}. Furthermore, for any given value of the threshold, there will always be a certain number of elements passing the cut, including elements ``in the bulk'', i.e., away from any wombling boundaries, and one has to design a prescription on how to deal with such false positives. Obviously, a value for the cut which is too stringent will miss many true boundary elements, while a value which is too generous will bring about a lot of false positives. Finally, given that the dimensionalities of ${\cal P}_{NP}$ and $\partial {\cal P}_{NP}$ necessarily differ by one, the concept of a ``boundary element'' can be open to interpretation --- how close to the boundary does an element have to be in order to be considered a ``boundary element''?
\item {\em Agglomeration.} The previous step results in a collection of tagged boundary elements scattered throughout ${\cal P}$, so now the question is how to use that information to reconstruct the complete boundary. As a first step, one can start forming sub-boundaries by linking adjacent tagged boundary elements, possibly subject to some additional criteria, e.g., that the directions\footnote{The two requirements --- that the gradients are large {\em and} that their directions are correlated --- can be conveniently encoded in the scalar (dot) product of the gradient vectors of neighboring elements \cite{Debnath:2015wra}.} of their gradients are within $30^\circ$ of each other \cite{Barbujani1990}. This agglomeration procedure will result in a graph whose nodes are the tagged boundary elements \cite{DaleFortin2010}. The properties of this graph can then be studied to determine its statistical significance \cite{Oden1993,Fortin1994} (see the next item) and to get some idea about the shape of the boundary. 

At this point it is worth mentioning that in collider physics applications there can be situations where the shape of the boundary is parametrically known. This is precisely the case with ``simplified model'' NP searches at the LHC \cite{Alves:2011wf} --- once the event topology is assumed, the geometry of the NP final state phase space ${\cal P}_{NP}$ is also fixed. As an example, consider a sequence of three two-body decays, which is the classic squark signature in supersymmetry (SUSY) \cite{Martin:1997ns}. The relevant phase space ${\cal P}_{NP}$ is three-dimensional and can be parametrized by the invariant masses of the three pairs of visible decay products. The equation for the boundary $\partial {\cal P}_{NP}$ is known analytically \cite{Costanzo:2009mq,Agrawal:2013uka,Kim:2015bnd} in terms of just four parameters --- the masses of the SUSY particles participating in the decay chain. In that situation, Ref.~\cite{Debnath:2016gwz} proposed to bypass {\em both} of the last two steps (the tagging and the agglomeration) altogether and instead fit the equation for the surface $\partial {\cal P}_{NP}$ to the full tessellation. Operationally this was done by computing a quantity inspired by Bayesian wombling (see next bullet), namely, a two-dimensional surface integral of the gradient magnitude over the boundary surface, normalized to the total area of the surface:
\begin{equation}
\frac{\int_{\partial {\cal P}_{NP}} da\ |\vec\nabla f|}{\int_{\partial {\cal P}_{NP}} da}.
\label{eq:surfacewombling}
\end{equation}
In Ref.~\cite{Debnath:2016gwz}, it was demonstrated that this quantity is maximized for the true values of the SUSY masses, resulting in a novel method for SUSY mass measurements.

\item {\it Significance estimation.} Any wombling algorithm as described so far will produce numerical results regardless of whether a true pattern exists or not. The crucial question now is to assess the likelihood that the observed pattern could have been produced from random fluctuations instead of a true boundary. One possible approach, known as sub-boundary statistics \cite{Oden1993,Fortin1994}, is to analyze the properties of the graph mentioned in the previous step, formed out of the tagged boundary elements. Strictly speaking, sub-boundary statistics tests whether the different components of the graph are sufficiently contiguous (and not whether the rates of change are sufficiently large) \cite{Jacquez2000,DaleFortin}. To this end, one looks at (distributions of) quantities which would characterize coherent boundaries formed from connected boundary elements, such as: the total number of subgraphs, the number of single-node subgraphs, the maximum and the mean of the length and/or the diameter of the subgraphs, the superfluity, etc. An alternative approach, named Bayesian wombling, starts with an ansatz for the shape of the wombling line (in two spatial dimensions) and then computes the average flux of the two-dimensional gradient field through all possible such lines \cite{Banerjee2006,Banerjee2010,Gelfand2015}. The idea is that the average flux will be maximized when the ansatz matches the true wombling boundary. As already mentioned in relation to eq.~(\ref{eq:surfacewombling}), the advantage of this approach is that it avoids the subjectiveness associated with the steps of tagging and agglomeration. With either approach, one has to specify a null hypothesis, in order to quantify the confidence level. Unlike other fields of science, where the null hypothesis may not be immediately obvious and one typically has to rely on a randomization scheme \cite{DaleFortin}, in high-energy physics the null hypothesis is well defined --- it is the SM. 
\end{itemize}

In this paper we further develop and refine the Voronoi boundary detection methods from Refs.~\cite{Debnath:2015wra,Debnath:2016mwb,Debnath:2016gwz,DebnathPhD}. As before, the main goal will be to outline a method for discovering new physics in LHC collider data by identifying wombling boundaries in phase space. The paper is structured around the five typical steps of algorithmic wombling described above. The novel elements in the analysis presented here are the following:
\begin{itemize}
\item In addition to the Voronoi tessellations utilized in \cite{Debnath:2015wra,Debnath:2016mwb,Debnath:2016gwz,DebnathPhD}, here we also consider Delaunay tessellations. In Section~\ref{sec:VoronoiDelaunay} we shall briefly review the two types of tessellations and outline the range of new possibilities offered by the use of a Delaunay tessellation for our purposes. In particular, in Section~\ref{sec:boundaryobjects} we shall illustrate the four possible types of boundary elements, discuss their relations to each other, and how each can be potentially targeted in the tagging step of the wombling algorithm.
\item Unlike previous work in high-energy physics, here our main tool for calculating the local gradients will be the Delaunay tessellation (often referred to as ``triangulation'' since in two dimensions the Delaunay polygons are triangles). Section~\ref{sec:gradientestimation} is devoted to the topic of gradient estimation --- after a brief review  in Section~\ref{sec:voronoigradients} of previous work on estimating gradients from a Voronoi tessellation, in Section~\ref{sec:triangulation} we shall describe the gradient calculation from the Delaunay triangulation. In the process, we shall pay special attention to techniques for reducing the random fluctuations in the obtained gradient values --- three such procedures and their interplay and optimization are discussed in Section~\ref{sec:denoising}. 
\item Different techniques for tagging boundary elements will be discussed in Section~\ref{sec:tagging}. In addition to tagging Voronoi cells \cite{Debnath:2015wra,Debnath:2016mwb}, here we shall also be interested in tagging Voronoi edges, Delaunay cells and Delaunay edges as well. Although this is not our main focus here, in Section~\ref{sec:agglomeration} we shall illustrate how these tagging methods can be used for agglomeration.
\item The main results of the paper are presented in Sections~\ref{sec:womblingcontinuous}-\ref{sec:significance}. We follow the approach of Bayesian wombling \cite{Banerjee2006,Banerjee2010,Gelfand2015}, which lets us avoid the intricacies and uncertainties of the tagging and agglomeration steps. To gain some intuition, in Section~\ref{sec:womblingcontinuous} we first go over a toy example where the function $f(\vec{x})$ can be sampled continuously from a distribution which resembles real data including finite width effects and detector resolution. Then in Section~\ref{sec:womblingdiscrete} we study point-referenced data of the type (\ref{dataset}) with a straight (Section~\ref{sec:discretestraight}) or circular (Section~\ref{sec:discretecircular}) boundary. The corresponding estimates of the statistical significance of the obtained wombling boundaries are performed in Section~\ref{sec:significance}. For simplicity, the illustrative examples in the main body of the paper use data generated from uniform background distributions. For completeness, in Appendix \ref{sec:nonuniform} we also present results for two additional examples in which the background distribution is not uniform, but varies according to a power law (Section~\ref{sec:ramp}) or an exponential (Section~\ref{sec:exp}).
\end{itemize}

\section{Voronoi and Delaunay tessellations of point data}
\label{sec:VoronoiDelaunay}

\subsection{Simulation Details}
\label{sec:simdetails}

Virtually all wombling studies in the literature have been concerned with {\em two-dimensional} point data generated, e.g., from field samples taken within a certain geographical area, from remotely sensed images, etc. In this paper, we shall continue to work in two-dimensions, but this will be done only for clarity of the presentation, since it is difficult to visualize Voronoi and Delaunay tessellations in more than two dimensions; the methods which we shall describe will be applicable to higher dimensional data as well. To be specific, we shall consider the Cartesian plane where the data points are specified by their coordinates $(x_i,y_i)$, so that the dataset (\ref{dataset}) reduces to
\begin{equation}
(x_i, y_i, f_i), \quad i=1,2,\ldots,N.
\label{dataset2d}
\end{equation}
For concreteness, we shall choose our field of view to be the unit square, $0\le x\le 1$, $0\le y\le 1$, although data will be generated beyond the boundaries of the unit square --- this will eliminate any spurious boundary effects like clipping which would modify the statistical properties of the Voronoi cells near the boundaries \cite{Koufos2019}. Following Refs.~\cite{Debnath:2015wra,Debnath:2015hva,Debnath:2016mwb}, the datasets will be generated according to (\ref{f}) with the following assumptions for $f_{SM}$ and $f_{NP}$:
\begin{itemize}
\item {\em Background.} As in previous work \cite{Debnath:2015wra,Debnath:2015hva,Debnath:2016mwb}, our proxy for the SM background distribution  $f_{SM}$ will be the uniform distribution\footnote{Other choices for the background distribution will be considered in Appendix~\ref{sec:nonuniform}.}
\beq
f_{SM} = constant.
\label{fSMeqC}
\eeq
The exact value of the constant will depend on the normalization: for pure-background samples within the unit square the constant is 1, while for background plus signal samples, it will depend on the relative strength of the signal. Strictly speaking, the assumption (\ref{fSMeqC}) is unrealistic from the point of view of a high energy physicist, since $f_{SM}$ is in general a function of the kinematic variables parametrizing the phase space ${\cal P}$. Nevertheless, it is good enough for our purposes here --- the important point is that any realistic SM distribution is {\em very weakly varying} across the boundary $\partial {\cal P}_{NP}$, which justifies the use of (\ref{fSMeqC}) for our model-independent toy examples below. A typical background distribution of $N=500$ points within the unit square is shown in the left panels of Figs.~\ref{fig:datasetsFlat} and \ref{fig:datasetsCircle}. It is evident that such a distribution does not have any obvious features and any wombling boundary would have to be created purely by chance.
\item {\em Signal with a flat boundary.} As our first example of a hypothesized NP signal we shall consider a distribution $f_{NP}$ populating a region ${\cal P}_{NP}$ with a flat boundary. Again following Refs.~\cite{Debnath:2015wra,Debnath:2015hva,Debnath:2016mwb}, we shall take the boundary to be the vertical line at $x=0.5$ and the corresponding signal distribution $f_{line}$ to be flat and non-zero only to the left of the boundary:
\beq
f_{line}=2H(0.5-x),
\label{fLine}
\eeq
where $H$ is the Heavyside step function. When adding this signal to the background distribution, it is important to specify the mixing ratio. For this purpose, Refs.~\cite{Debnath:2015wra,Debnath:2015hva,Debnath:2016mwb} introduced a parameter $\rho$ which measures the ratio of the event densities $\frac{dn}{da}$ on the two sides of the boundary\footnote{Since we are interested in detecting a wombling boundary, the parameter $\rho$ as defined here is more suitable than the more familiar ratio $S/B$ of signal to background inside the signal region ${\cal P}_{SR}$. Note that with our setup, the two are related as $S/B=\rho-1$.}:
\beq
\rho \equiv \frac{~~\lim\limits_{x\to 0.5^-} \left( \displaystyle\frac{dn}{da} \right)~~}{\lim\limits_{x\to 0.5^+} \left(\displaystyle\frac{dn}{da}\right) } .
\label{rhodef}
\eeq
\begin{figure}[t]
 \centering
 \includegraphics[width=.3\textwidth]{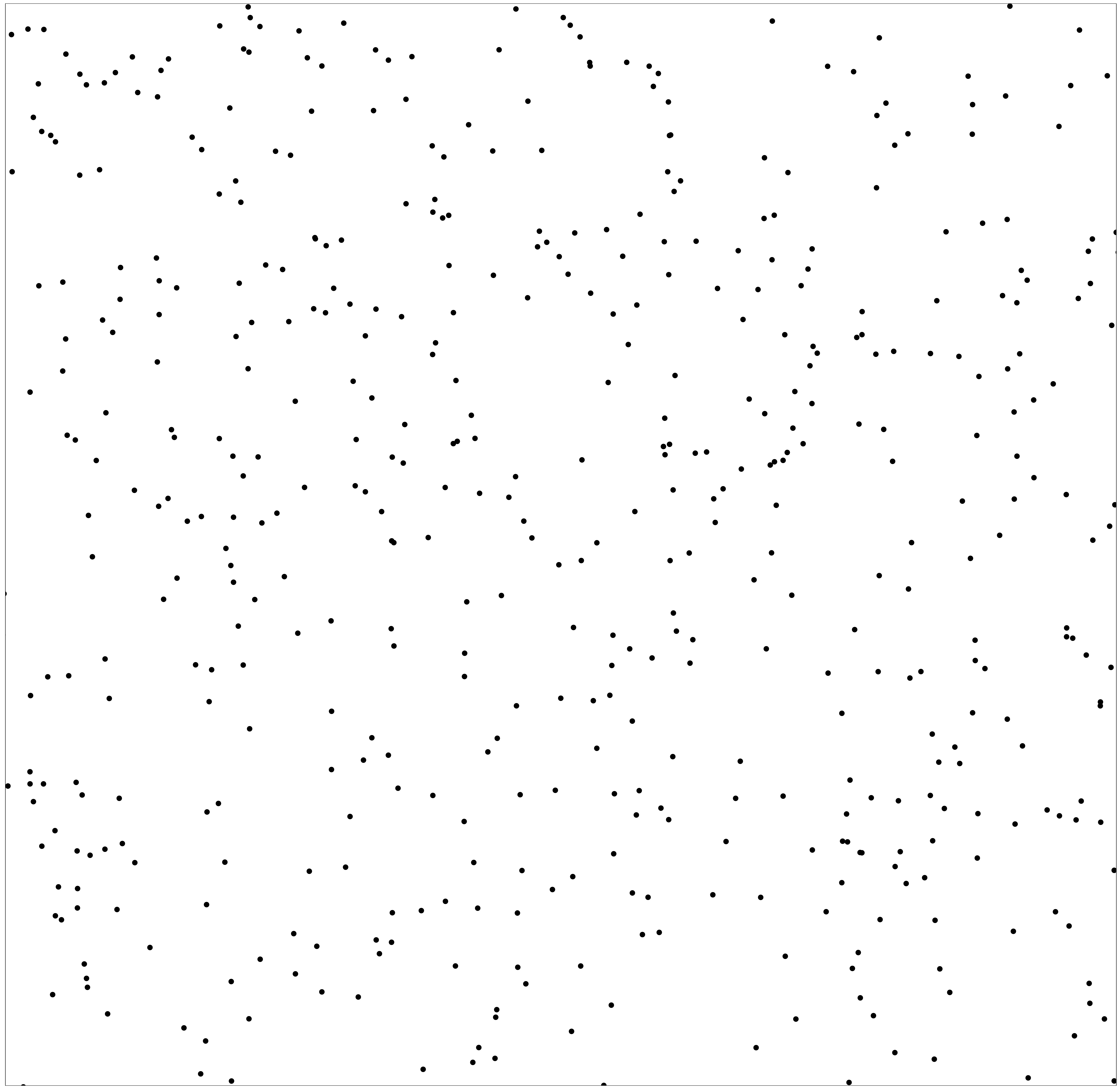}~~
 \includegraphics[width=.3\textwidth]{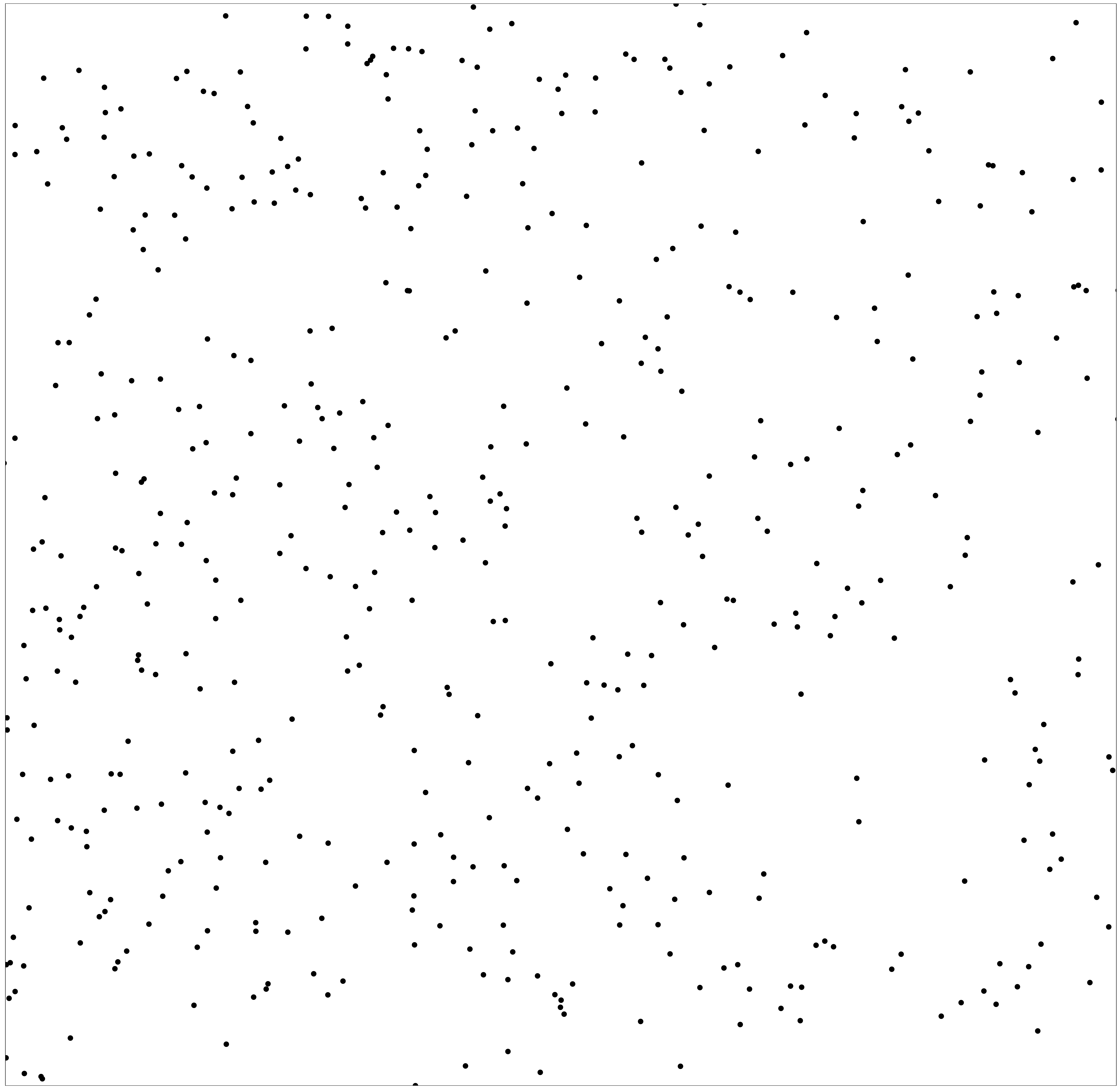}~~
 \includegraphics[width=.3\textwidth]{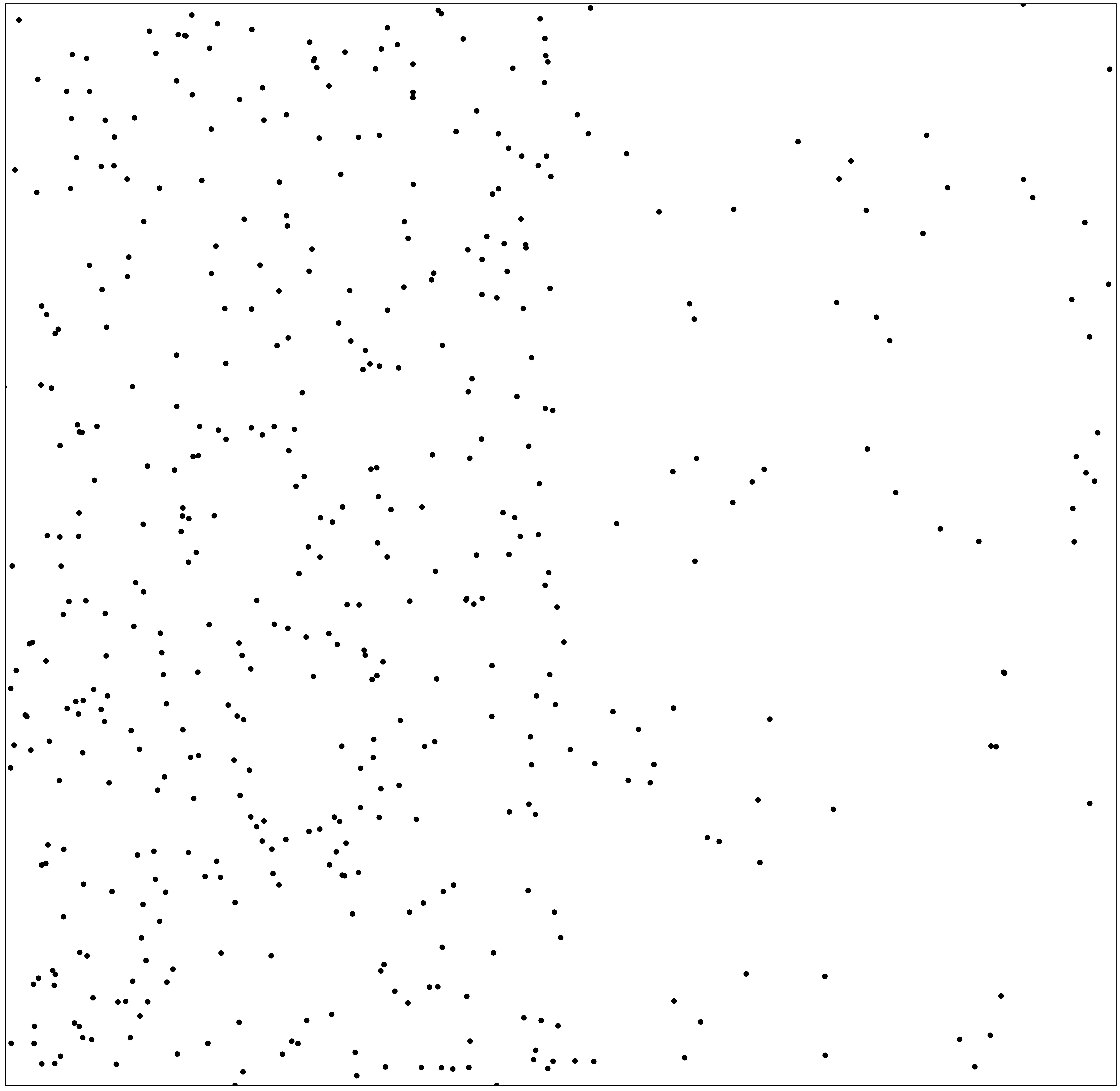}
\caption{\label{fig:datasetsFlat} Typical simulated point data sets used in our studies of straight-line boundaries. Here we show $N=500$ points within the unit square, distributed according to: background only (left panel); background with an additionally injected signal (\ref{fLine}) in the left half-plane with $\rho=1.5$ (middle panel) or $\rho=5$ (right panel).}
\end{figure}
Combining (\ref{fSMeqC}) and (\ref{fLine}), the unit-normalized total distribution (\ref{f}) on the unit square\footnote{In order to declutter the notation, in what follows we shall omit the prefactor of $H(x)H(1-x)H(y)H(1-y)$ which confines us to the unit square.} for a signal with a flat boundary reads \cite{Debnath:2015wra}
\beq
f = \frac{2}{\rho+1}\Bigl[\rho\, H(0.5-x)+H(x-0.5)\Bigr].
\label{fSMfLine}
\eeq

In the middle and right panel of Fig.~\ref{fig:datasetsFlat} we show distributions of $N=500$ points according to (\ref{fSMfLine}) with $\rho=1.5$ and $\rho=5$, respectively. In the latter case, the value of $\rho$ is sufficiently large that the boundary is clearly visible with the naked eye. As in Refs.~\cite{Debnath:2015wra,Debnath:2015hva,Debnath:2016mwb} (which considered an even more extreme value of $\rho=6$), the case of relatively large $\rho$ is meant mostly for illustration --- it makes it easier to visualize the benefits from the various wombling and denoising techniques introduced below. Our real target will be the case of relatively low values of $\rho$ as shown in the middle panel of Fig.~\ref{fig:datasetsFlat}, where it is rather difficult to discern any apparent wombling boundary. 
\begin{figure}[t]
 \centering
 \includegraphics[width=.3\textwidth]{points_background_0lloyds}~~
 \includegraphics[width=.3\textwidth]{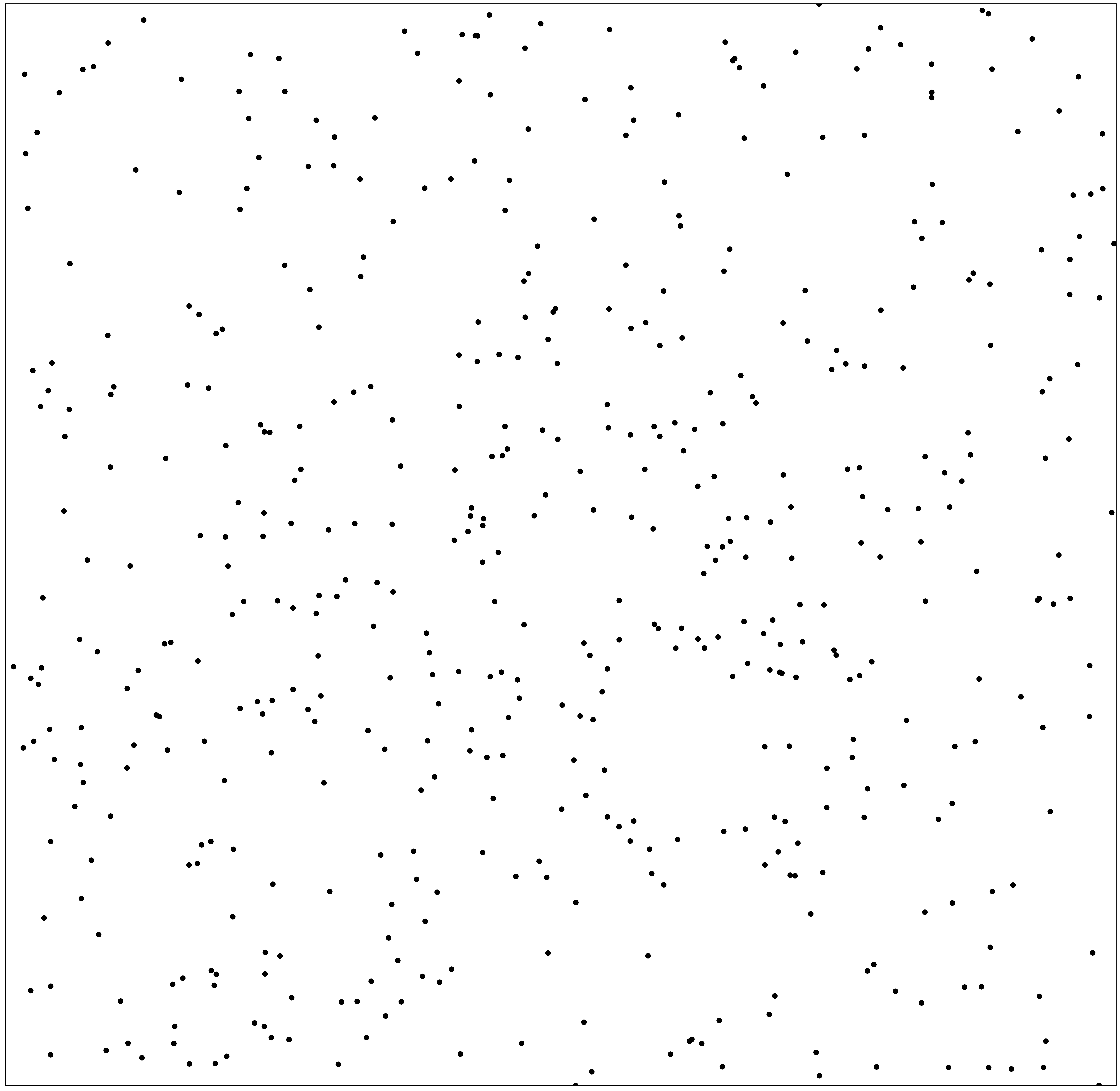}~~
 \includegraphics[width=.3\textwidth]{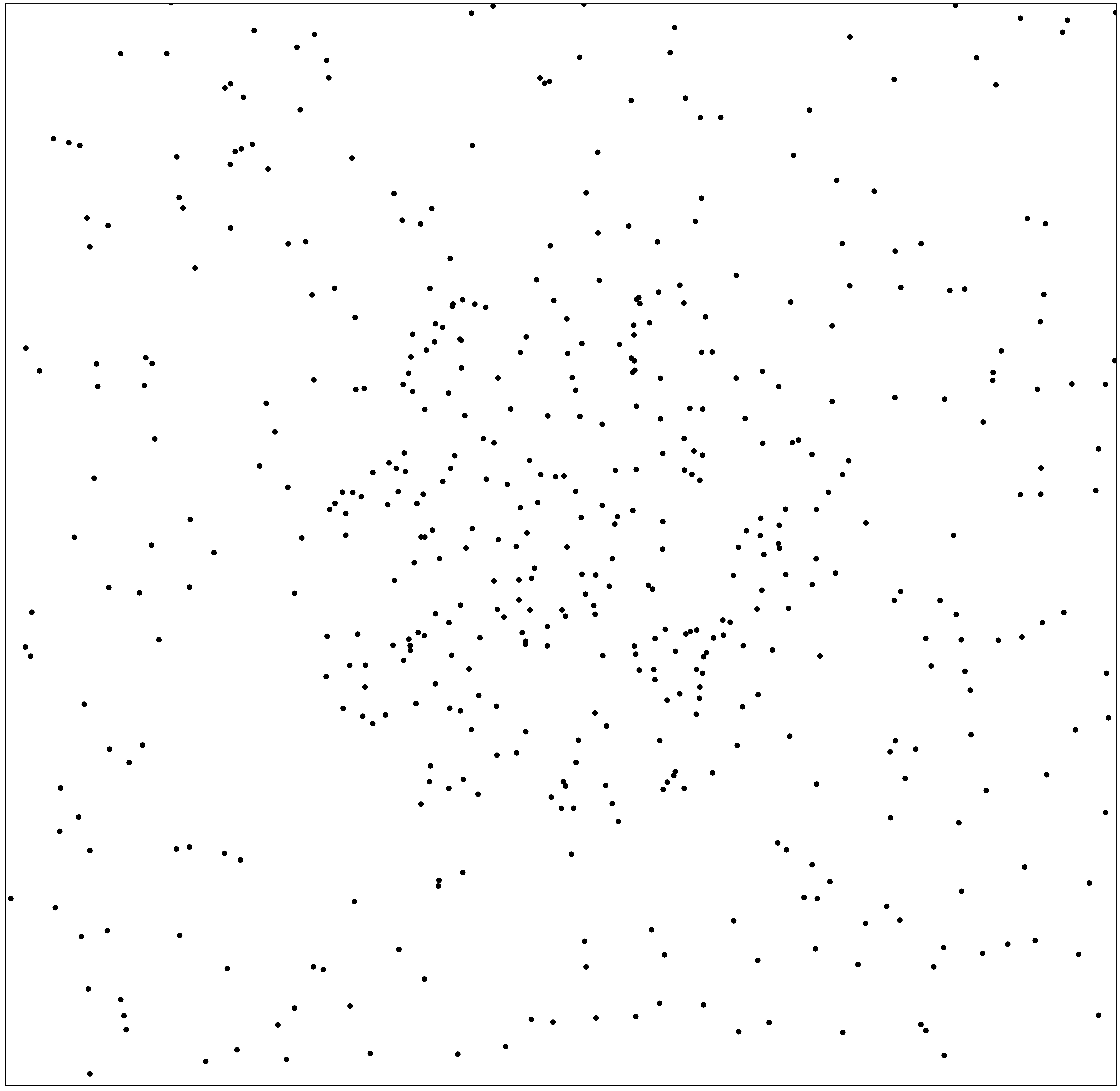}
\caption{\label{fig:datasetsCircle} The same as Fig.~\ref{fig:datasetsFlat}, but for the circular signal (\ref{fCircle}) with $R=0.25$.}
\end{figure}
\item {\em Signal with a circular boundary.} For completeness, we shall also consider an example of a signal in a domain bounded by a curvilinear boundary. Following \cite{Debnath:2016mwb}, we shall take the signal distribution $f_{circle}$ to be confined to a circular region of radius $R<0.5$ centered at $(x,y)=(0.5,0.5)$:
\beq
f_{circle}=\frac{1}{\pi R^2}\, H\Bigl(R-\sqrt{(x-0.5)^2+(y-0.5)^2}\Bigl),
\label{fCircle}
\eeq
so that the combined total distribution (\ref{f}) becomes
\begin{eqnarray}
f=\frac{1}{(\rho-1)\pi R^2+1}&\Bigl[\,\rho\, H\Bigl(R-\sqrt{(x-0.5)^2+(y-0.5)^2}\Bigl) \nonumber \\ 
&~~+ H\Bigl(\sqrt{(x-0.5)^2+(y-0.5)^2}-R\Bigl)\Bigr],
\label{fSMfCircle}
\end{eqnarray}
with the $\rho$ parameter suitably defined in analogy to (\ref{rhodef}) as the ratio of point densities across the circular boundary. The middle and right plots in Fig.~\ref{fig:datasetsCircle} depict typical distributions of $N=500$ points according to (\ref{fSMfLine}) with $\rho=1.5$ and $\rho=5$, respectively. As before, the boundary for $\rho=5$ is clearly visible, but the case of $\rho=1.5$ appears much more challenging.
\end{itemize}

\subsection{Voronoi and Delaunay tessellations}

\begin{figure}[t]
 \centering
 \includegraphics[width=.4\textwidth]{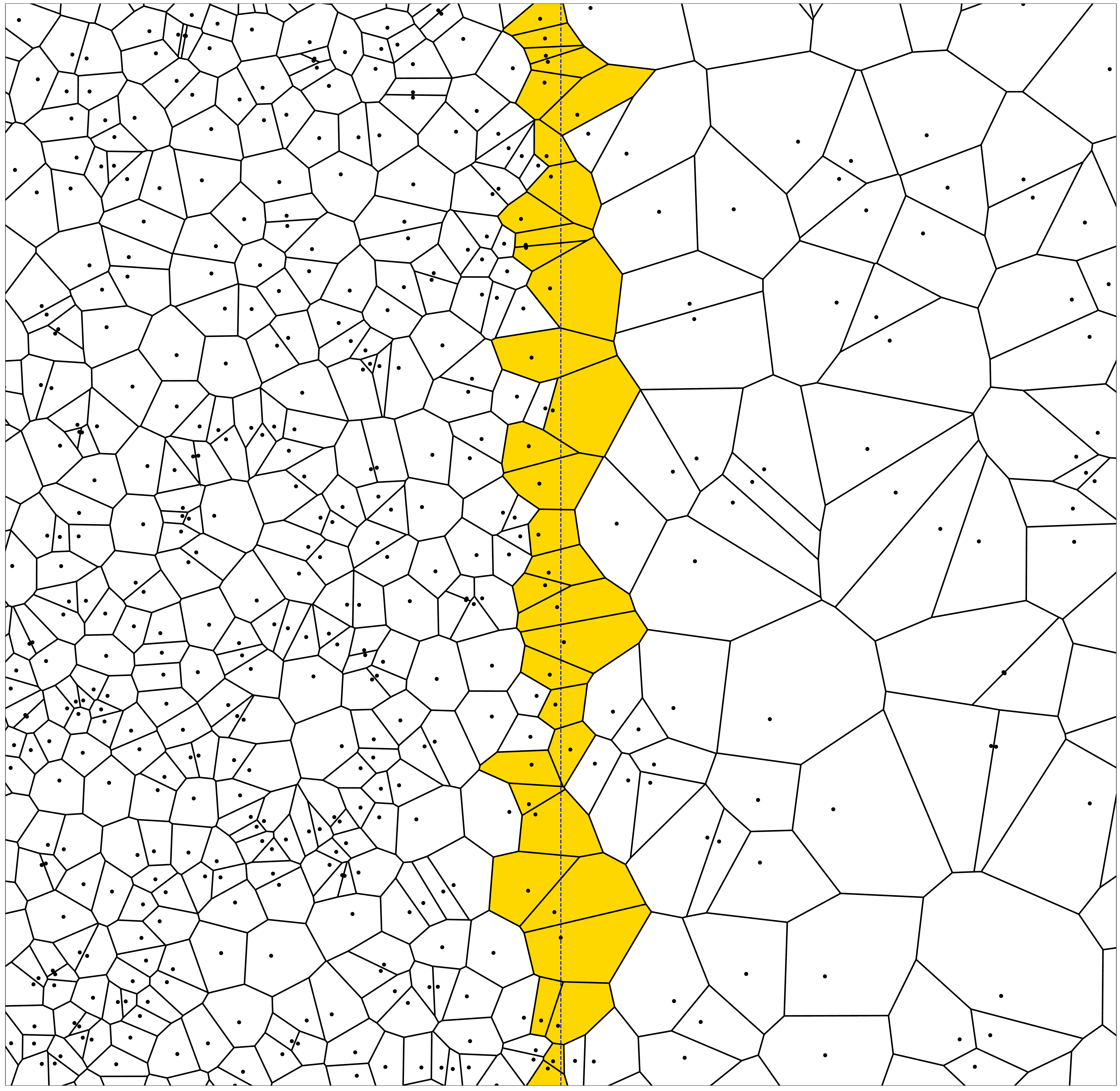}~~~
 \includegraphics[width=.4\textwidth]{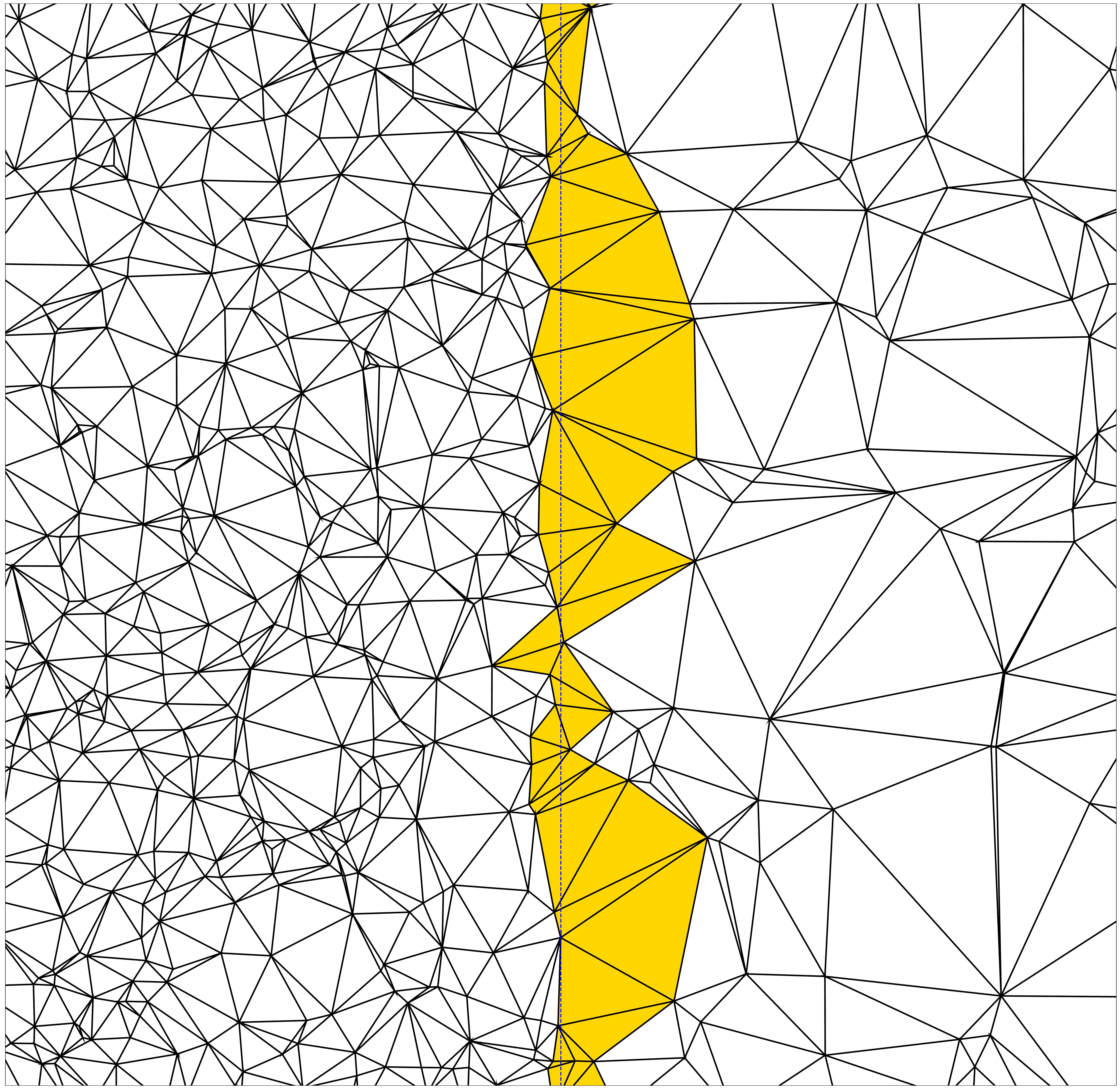}\\ [10pt]
 \includegraphics[width=.4\textwidth]{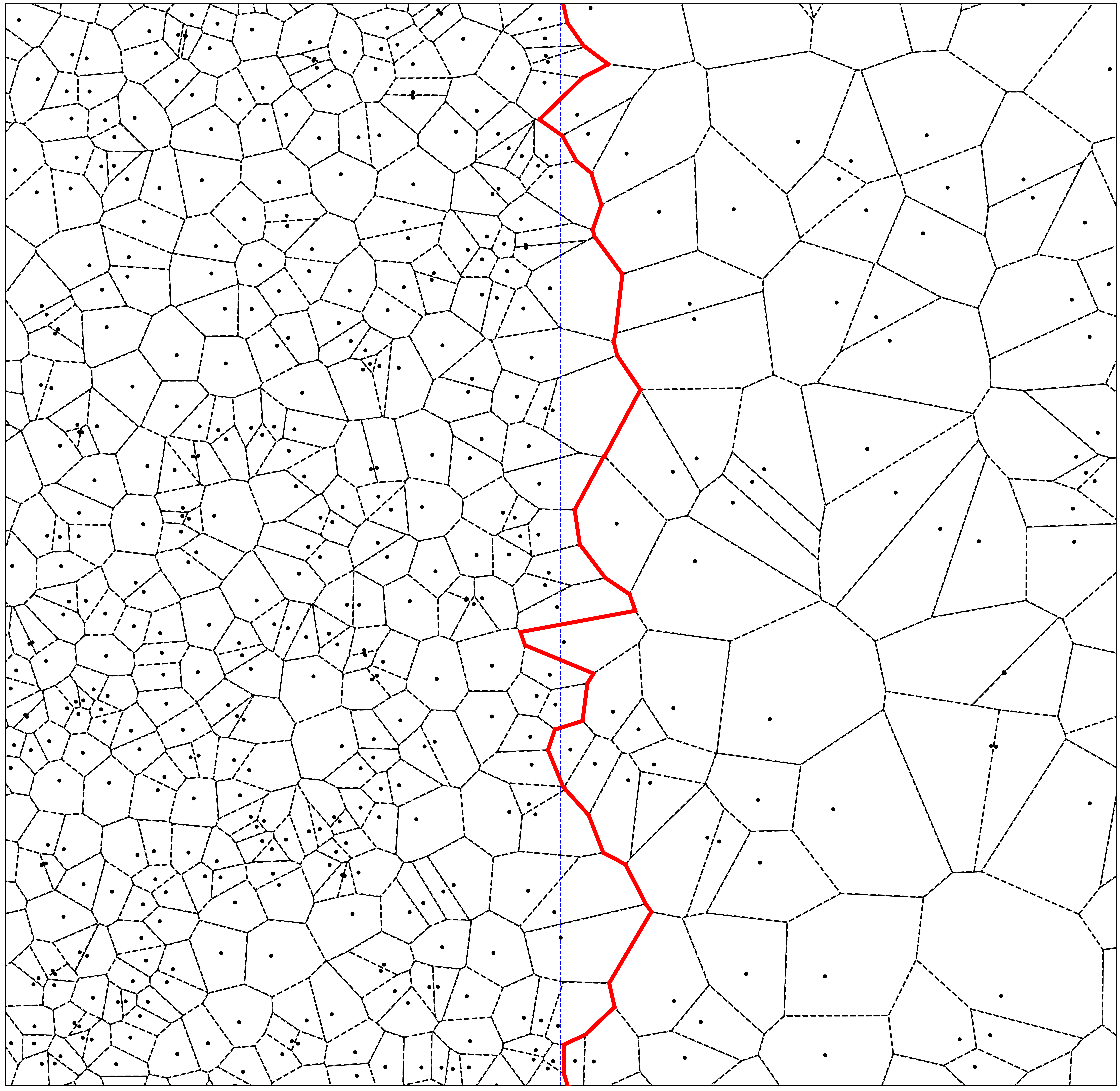}~~~
 \includegraphics[width=.4\textwidth]{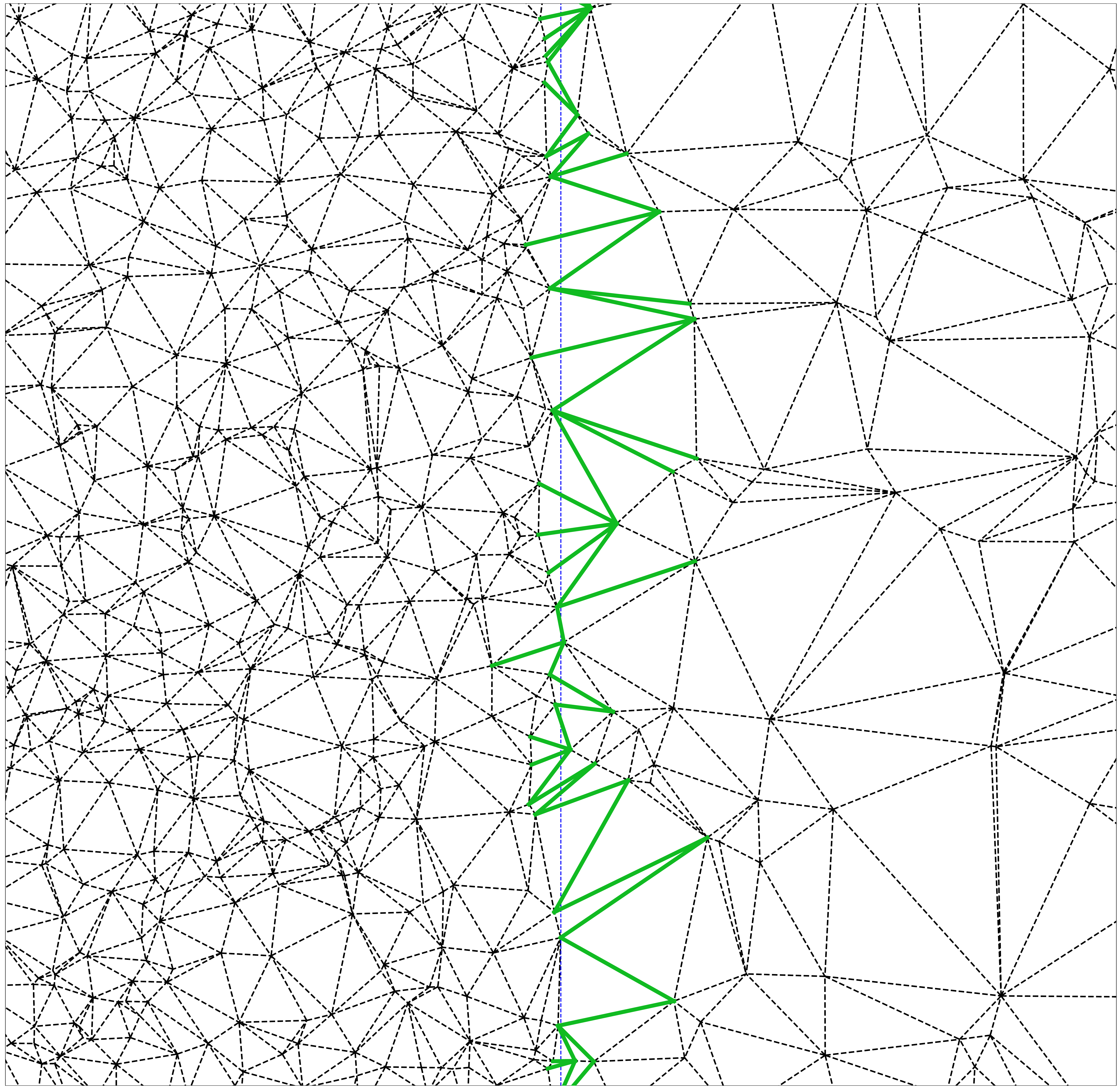}
\caption{\label{fig:boundaryobjects} The Voronoi and Delaunay tessellations of a typical point data distributed according to eq.~(\ref{fSMfLine}) with $N=500$ and $\rho=4$. The left (right) panels show the Voronoi (Delaunay) tessellation. The vertical blue dashed line marks the theoretical boundary line at $x=0.5$. The yellow-shaded cells in the top panels illustrate the boundary cells defined in Sec.~\ref{sec:boundaryobjects} for the Voronoi and Delaunay case, respectively. In the bottom panels, the tessellations are outlined with dotted lines while the solid lines represent the boundary Voronoi edges (red lines) and their dual Delaunay edges (green lines) as defined in Sec.~\ref{sec:boundaryobjects}. 
}
\end{figure}

The Voronoi and Delaunay tessellations of the point data sample in the right panel of Fig.~\ref{fig:datasetsFlat} are illustrated in Fig.~\ref{fig:boundaryobjects}. In the left panels, which show the Voronoi tessellation, the data points appear as dots, while in the right panels, which illustrate the Delaunay tessellation, the data points are located at the vertices of the Delaunay triangles and are not explicitly shown. 

As illustrated in the left panels of Fig.~\ref{fig:boundaryobjects}, a Voronoi tessellation of $N$ points (often referred to as {\em generators}) in the plane is constructed as follows (see, e.g., \cite{Okabe1992}). Every location in the plane is assigned to the closest member of the point set. If a location happens to be equally close to two (or more) generator points, it is assigned to all of those points; all such eqidistant locations form the {\em edges} of the Voronoi graph.  The set of locations assigned to a given member of the point set  forms the Voronoi {\em cell} corresponding to that generator point; as seen in Fig.~\ref{fig:boundaryobjects}, the Voronoi cells in the plane are polygons, with the corresponding generator point located somewhere in the polygon's interior, but not necessarily at its geometric center. Note that Voronoi polygons have different shapes and sizes; in particular, the number of edges of a polygon varies greatly, the average number being no more than six \cite{Okabe1992}. An endpoint of a Voronoi edge is called a Voronoi {\em vertex}; alternatively, a vertex may be defined as a point shared by three (or more) Voronoi edges. When each vertex belongs to three and only three edges, the Voronoi tessellation is non-degenerate; as seen in Fig.~\ref{fig:boundaryobjects} this will be our case as well, since the probability of generating a degenerate vertex in Monte Carlo sampled data is vanishingly small.

The right panels in Fig.~\ref{fig:boundaryobjects} depict the corresponding Delaunay tessellation of the same data. Since the Voronoi and Delaunay tessellations are dual to each other, one way to construct the Delaunay tessellation is to start from the Voronoi diagram and join all pairs of generator points whose Voronoi polygons share a common Voronoi edge. If the Voronoi tessellation is non-degenerate, each Voronoi vertex belongs to exactly three Voronoi edges, which in turn define a triangular polygon in the Delaunay tessellation (see Fig.~\ref{fig:boundaryobjects}); for this reason the Delaunay tessellation is often referred to as a {\em triangulation}. The described procedure\footnote{There are alternative methods to construct the Delaunay triangulation, e.g., using the property that the interiors of the circumcircles of Delaunay triangles are empty circles, i.e., contain no points from the dataset.} also manifestly pairs up all Voronoi edges with their corresponding dual edges in the Delaunay tessellation; if {\em each} such pair of dual edges from the two tessellations has a common point, i.e., the dual edges cross each other, the Delaunay triangulation is known as Pitteway triangulation \cite{Pitteway,McLain1976}. As we shall see explicitly below in Fig.~\ref{fig:correlations}, our datasets will generally {\em not} lead to Pitteway triangulations; so it will be important to keep in mind that some dual pairs of edges may be slightly offset and not intersect each other.

In what follows, our discussion will often switch back and forth between the two types of tessellations, so at this point it may be useful to build some intuition by recapping some of the relationships between the constituent objects of the Voronoi and Delaunay tessellations for a dataset consisting of $N$ points:
\begin{itemize}
\item Each data point defines both a Voronoi polygon {\em and} a corresponding Delaunay vertex; the total number of Voronoi polygons or Delaunay vertices is thus $N$. The number of sides of a Voronoi polygon is equal to the number of Delaunay edges joining at the corresponding Delaunay vertex.
\item Each Voronoi edge has a dual Delaunay edge; the total number of Voronoi edges is therefore equal to the total number of Delaunay edges and is on the order of, but slightly less than, $3N$ \cite{Okabe1992}.
\item Each Voronoi vertex defines a corresponding Delaunay triangle; the total number of such objects is on the order of, but slightly less than, $2N$ \cite{Okabe1992}.
\end{itemize}

In summary, we have the following duality relations between the elements of the Voronoi and Delaunay tessellations:
\begin{subequations}
\begin{eqnarray}
{\rm Voronoi\ cell} &\longleftrightarrow& {\rm Delaunay\ vertex}, \label{VCDV}\\[2pt]
{\rm Voronoi\ edge} &\longleftrightarrow& {\rm Delaunay\ edge}, \label{VEDE}\\[2pt]
{\rm Voronoi\ vertex} &\longleftrightarrow& {\rm Delaunay\ triangle}. \label{VVDT}
\end{eqnarray}
\label{eq:dualities}
\end{subequations}
We can formalize these relations by introducing some notation. Let us use Latin indices to label elements from the Voronoi tessellation and Greek indices to label elements from the Delaunay triangulation. Then let $\{P_i\}$ be the set of generator points, $\{V_i\}$ be the set of Voronoi cells, $\{V_{ij}\}$ be the set of Voronoi edges and $\{V_{ijk}\}$ be the set of Voronoi vertices. Note that each Voronoi edge can be uniquely identified by the labels $i$ and $j$ of the pair of Voronoi cells which it separates, and similarly, for non-degenerate tessellations, a Voronoi vertex $V_{ijk}$ is labelled by exactly three indices since it is the meeting point of the edges $V_{ij}$, $V_{jk}$ and $V_{ki}$. Also let $N_i$ be the number of edges (or equivalently, neighboring polygons) of the $i^{\rm th}$ Voronoi cell $V_i$. With regards to the Delaunay triangulation, let $\{D_\alpha\}$ be the set of Delaunay triangles, $\{D_{\alpha\beta}\}$ be the set of Delaunay edges and $\{D_{\alpha_1\alpha_2\alpha_3...\alpha_{N_i}}\}$ be the set of Delaunay vertices. As before, each edge $D_{\alpha\beta}$ can be identified by the labels $\alpha$ and $\beta$ of the Delaunay triangles which it separates, and each vertex can be identified by the labels $\{\alpha_1, \alpha_2, ...,\alpha_{N_i}\}$ of the $N_i$ Delaunay triangles which are sharing it. Then the duality relations (\ref{eq:dualities}) can be written as
\begin{subequations}
\begin{eqnarray}
P_i \longleftrightarrow V_i &\longleftrightarrow& D_{\alpha_1\alpha_2\alpha_3...\alpha_{N_i}}, \label{dual1}\\[2pt]
V_{ij} &\longleftrightarrow& D_{\alpha\beta}, \label{dual2}\\[2pt]
V_{ijk} &\longleftrightarrow& D_\alpha, \label{dual3} \\[2pt]
N_{i} &\longleftrightarrow& N_\alpha = 3. \label{dual4}
\end{eqnarray}
\label{eq:dualitiesformalized}
\end{subequations}
Note the trade-off in complexity --- in the Voronoi case, vertices are labelled with exactly 3 indices, but the number of polygon edges $N_i$ varies, while in the Delaunay case, the number of polygon edges is always 3, but vertices are labelled with a varying number of indices.

\subsection{Candidate boundary objects}
\label{sec:boundaryobjects}

Having described the Voronoi and Delaunay tessellations of the data, before proceeding to the next two stages of gradient computation and tagging, it is worth pausing for a moment to discuss which elements of the tessellation are best suited for describing a wombling boundary. In Fig.~\ref{fig:boundaryobjects} the theoretical boundary at $x=0.5$ was marked with a vertical blue dashed line, but this was done only to guide the eye, since in reality both the existence and location of the boundary will be a priori unknown. In practice, we need to identify individual elements of the tessellations located at (or close to) the boundary which could be targeted by the wombling analysis.

\begin{figure}[t]
 \centering
 \includegraphics[width=.4\textwidth]{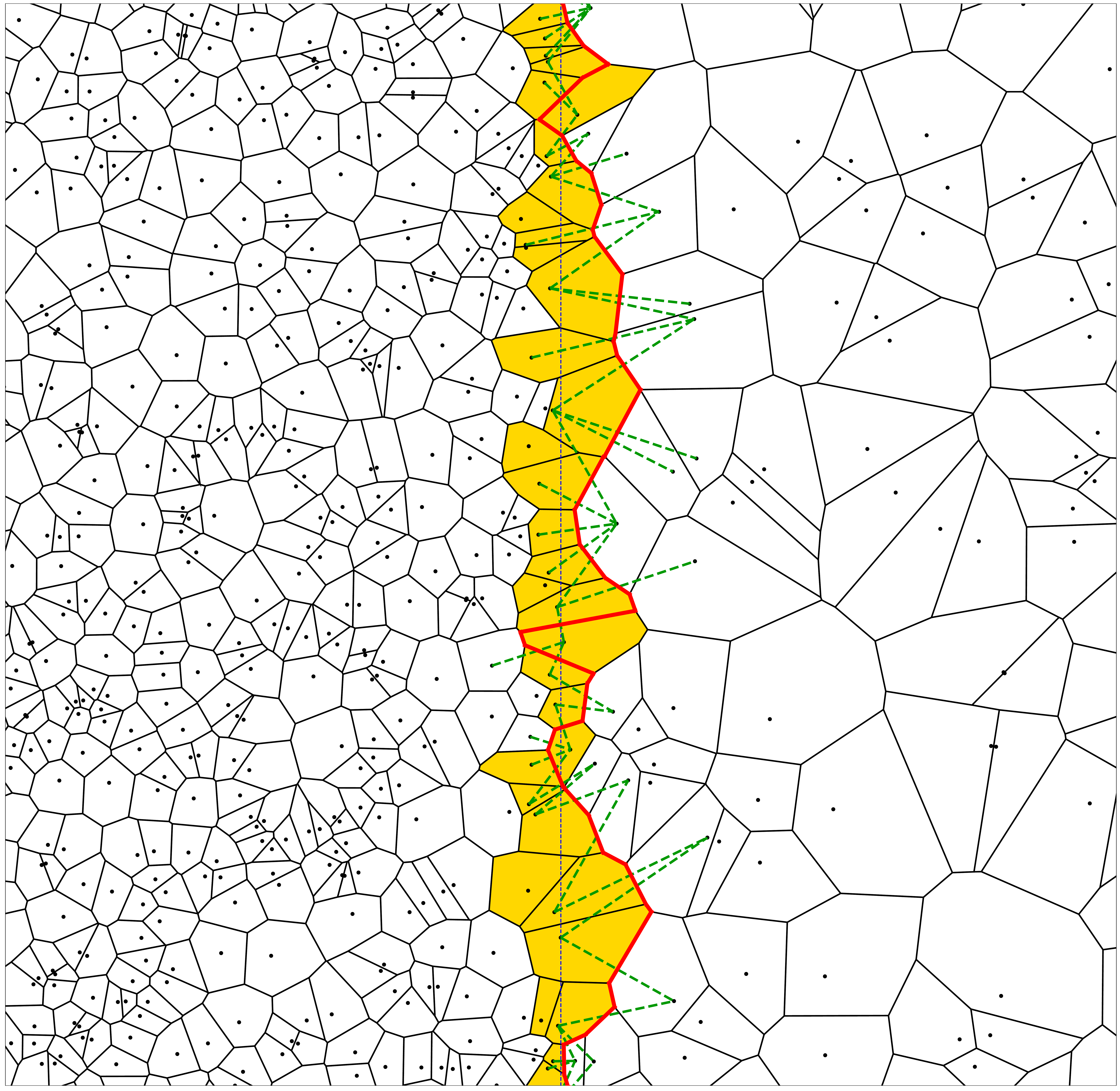}~~~
 \includegraphics[width=.4\textwidth]{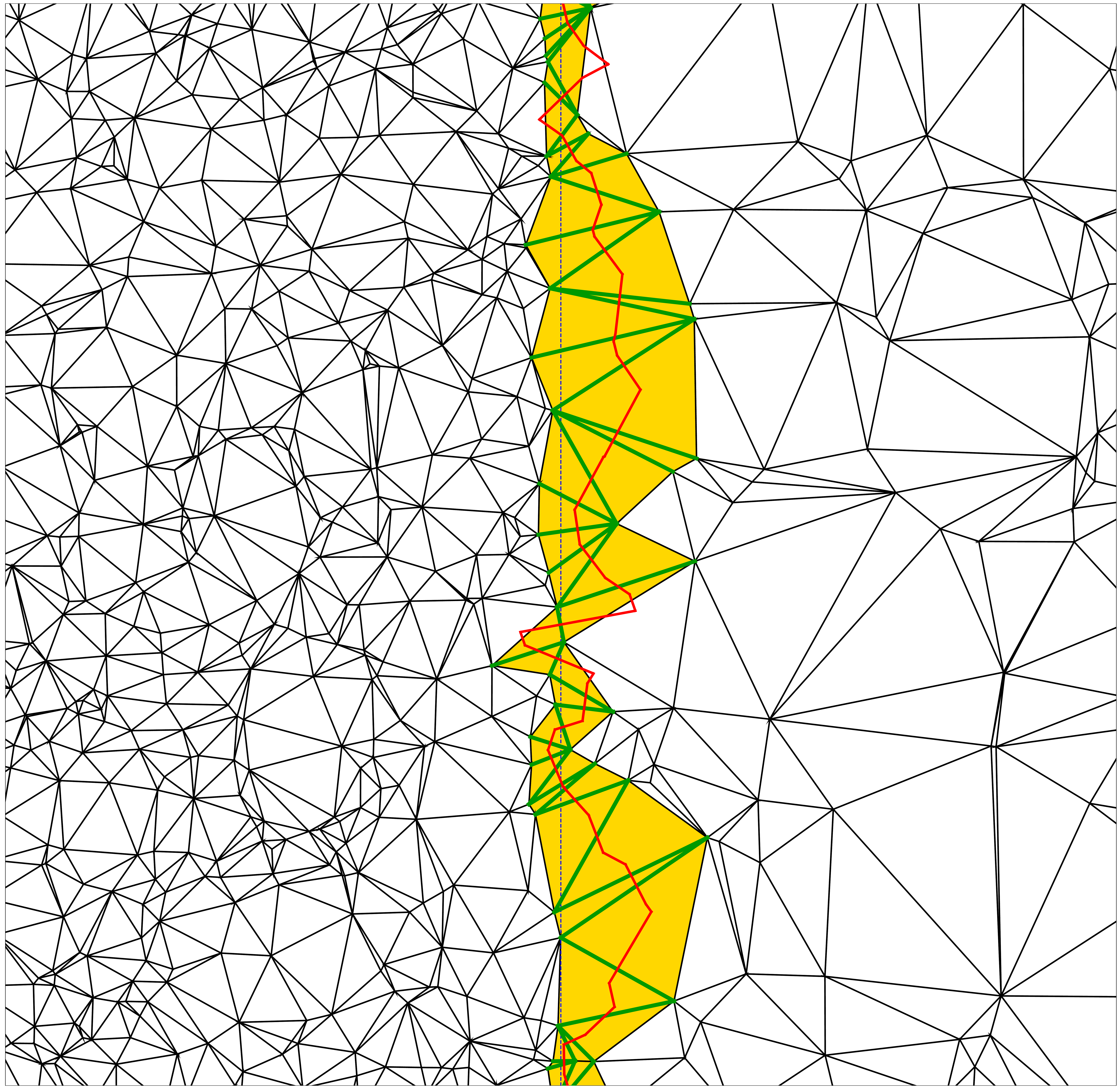}
\caption{\label{fig:correlations} Left: The Voronoi tessellation of the data shown in Fig.~\ref{fig:boundaryobjects} illustrating the relationships between Boundary Voronoi Cells (yellow-shaded), Boundary Voronoi Edges (red solid lines) and Boundary Delaunay Edges (green dashed lines). Right: Delaunay tessellation of the data shown in Fig.~\ref{fig:boundaryobjects} illustrating the relationships between Boundary Delaunay Triangles (yellow-shaded), Boundary Voronoi Edges (red solid lines) and Boundary Delaunay Edges (green solid lines). 
}
\end{figure}

By including the Delaunay tessellation into our discussion, we obtain three new possibilities in addition to the approach of Refs.~\cite{Debnath:2015wra,Debnath:2015hva}. The four panels in Fig.~\ref{fig:boundaryobjects} illustrate these four options:
\begin{itemize}
    \item {\em Boundary Voronoi Cells (BVCs)}, shown in the upper left panel in Fig.~\ref{fig:boundaryobjects}. When working with the Voronoi tessellation, this is the most natural and perhaps only option. A Boundary Voronoi Cell was defined as any Voronoi cell which is crossed  by the theoretical boundary \cite{Debnath:2015wra,Debnath:2015hva}. In the upper left panel of Fig.~\ref{fig:boundaryobjects} and in the left panel of Fig.~\ref{fig:correlations}, the BVCs are shaded in yellow.
    \item {\em Boundary Delaunay Triangles (BDTs)}, shown in the upper right panel in Fig.~\ref{fig:boundaryobjects}. When working with the Delaunay triangulation, in analogy we can now define a Boundary Delaunay Triangle to be any Delaunay triangle which is crossed by the theoretical boundary. In the upper right panel of Fig.~\ref{fig:boundaryobjects} and in the right panel of Fig.~\ref{fig:correlations}, the BDTs are shaded in yellow.
    \item {\em Boundary Delaunay Edges (BDEs)}, shown in the lower right panel in Fig.~\ref{fig:boundaryobjects}. Similarly, we can define a Boundary Delaunay Edge to be any Delaunay edge which is crossed by the theoretical boundary. In the lower right panel of Fig.~\ref{fig:boundaryobjects} and in the two panels of Fig.~\ref{fig:correlations}, the BDEs are indicated with green lines.
    \item {\em Boundary Voronoi Edges (BVEs)}, shown in the lower left panel in Fig.~\ref{fig:boundaryobjects}. Finally, using the duality relation (\ref{VEDE}) we can define a Boundary Voronoi Edge to be any Voronoi edge which is dual to a Boundary Delaunay Edge. In the lower left panel of Fig.~\ref{fig:boundaryobjects} and in the two panels of Fig.~\ref{fig:correlations}, the BVEs are indicated with red lines.
\end{itemize}
Note that the last three options all rely on the Delaunay triangulation and would not have been possible if we were only considering the Voronoi tessellation of the data.

Given the duality relations (\ref{eq:dualities}) between the Voronoi and Delaunay tessellations, the different categories of boundary objects defined above are related to each other. These relationships are exhibited in Fig.~\ref{fig:correlations}, where we simply superimpose some of the results from Fig.~\ref{fig:boundaryobjects} in order to better see the existing correlations. Figs.~\ref{fig:boundaryobjects} and \ref{fig:correlations} demonstrate that all four definitions lead to a contiguous set of boundary objects strung along the theoretical boundary. Now the question becomes how to tag these boundary objects with a suitable algorithm using their geometric properties.

\section{Estimation of local gradients from the tessellation}
\label{sec:gradientestimation}

\subsection{Gradient estimation from a Voronoi tessellation}
\label{sec:voronoigradients}

\begin{figure}[t]
 \centering
 \includegraphics[width=.3\textwidth]{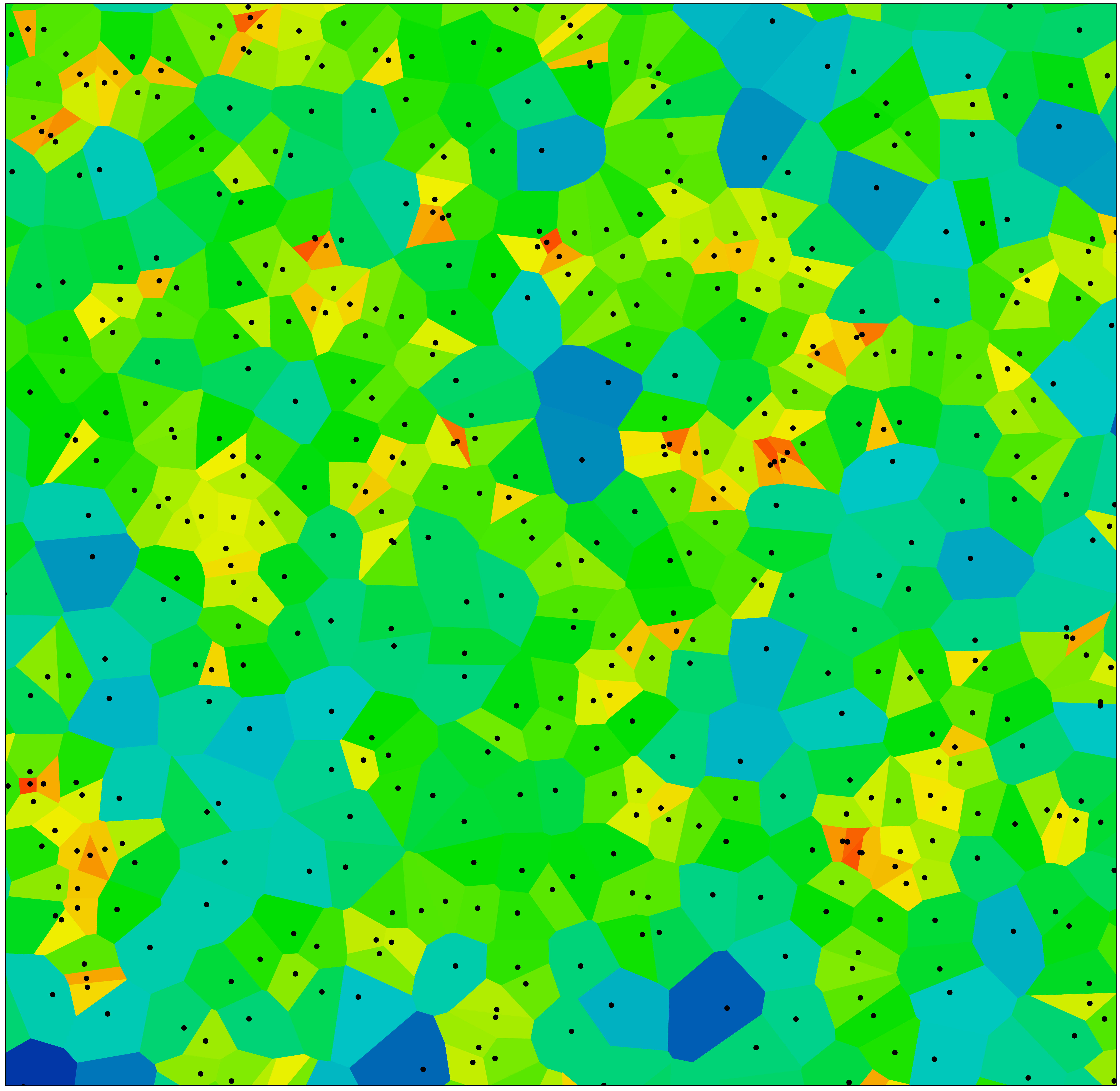}~~
 \includegraphics[width=.3\textwidth]{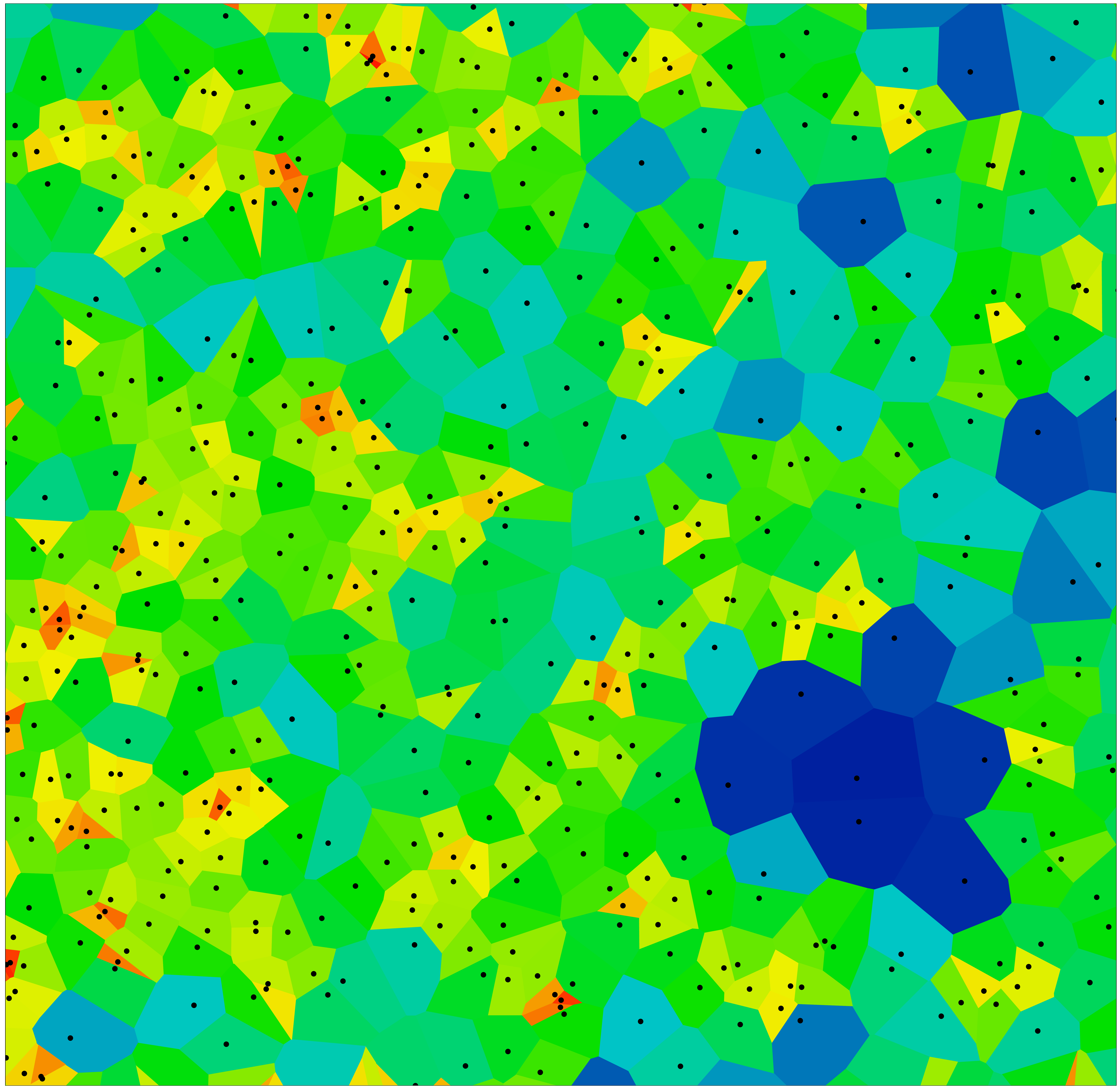}~~
 \includegraphics[width=.3\textwidth]{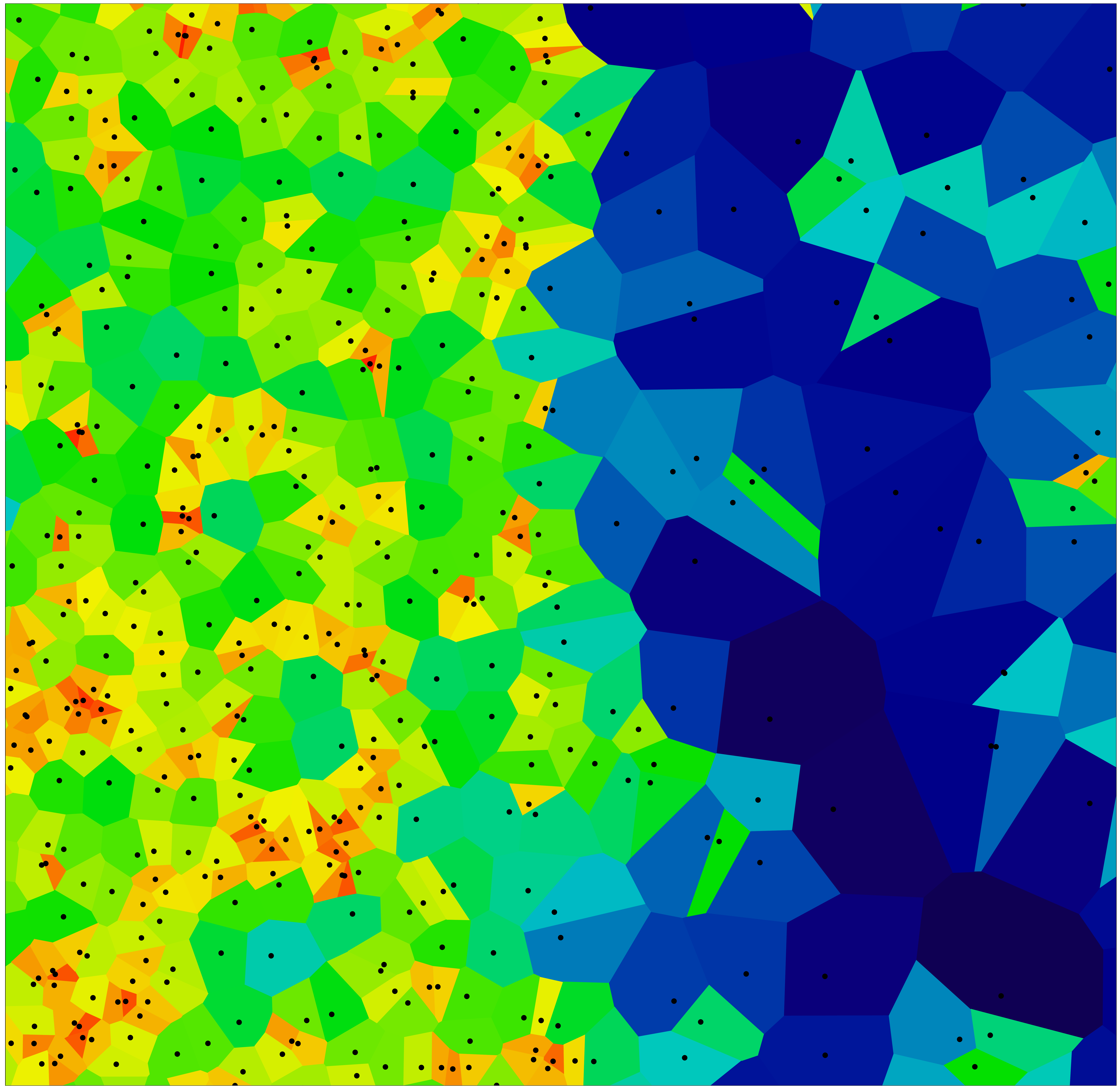}
\caption{\label{fig:voronoiareas} Pictorial illustration of the identification (\ref{feq1overa}) for the case of the three point datasets shown in Fig.~\ref{fig:datasetsFlat}. The Voronoi cells are color-coded by their inverse areas with warm (cold) colors indicating large (small) values of $f$.}
\end{figure}

As discussed in the Introduction, the Voronoi tessellation provides a natural estimate (\ref{feq1overv}) for the {\em values} of the function $f$ at the location of each generator point. In our case, since we are dealing with a two-dimensional dataset (\ref{dataset2d}), eq.~(\ref{feq1overv}) reduces to 
\begin{equation}
    f_i \sim \frac{1}{a_i},
\label{feq1overa}
\end{equation}
where $a_i$ is the area of the $i^{\rm th}$ Voronoi cell $V_i$. The identification (\ref{feq1overa}) is pictorially illustrated in Fig.~\ref{fig:voronoiareas} for the point data examples from Fig.~\ref{fig:datasetsFlat}. Unfortunately, the area $a_i$ by itself does not tell us anything about the {\em gradient} of the function $f(x,y)$ --- for this purpose, we need to compare $a_i$ to the areas of the surrounding Voronoi cells.  Let ${\cal N}_i=\{j_1, j_2, \ldots, j_{N_i}\}$ be the set of indices labelling the neighboring Voronoi cells, i.e., the Voronoi cells sharing an edge with $V_i$. By taking the neighboring cells one at a time, $j\in {\cal N}_i$, one can compute directional derivatives $(\nabla_{\hat n_{ij}} f)_i$ in the direction of the $j^{\rm th}$ neighboring cell, i.e., along the unit vector
$$
\hat n_{ij} = \frac{1}{\sqrt{(x_j-x_i)^2+(y_j-y_i)^2}} \, (x_j-x_i,y_j-y_i),
$$
as \cite{Debnath:2015wra}
\beq
(\nabla_{\hat n_{ij}} f)_i = (a_ia_j)^{\frac{3}{4}}\frac{f_j-f_i}{\sqrt{(x_j-x_i)^2+(y_j-y_i)^2}},
\label{directionalderivative}
\eeq
where the prefactor of $(a_ia_j)^{\frac{3}{4}}$ was included to make the directional derivative dimensionless. Since each Voronoi cell $V_i$ has a varying number of edges $N_i\equiv |{\cal N}_i|$, there will be a different number of directional derivatives available at each point $i$, but they can all be fitted to the expected distribution from the true gradient, thus producing an estimate $\vec{G}_i$ of the gradient vector at the $i^{\rm th}$ generator point $P_i$ \cite{Debnath:2015wra}. Another variable explored in Ref.~\cite{Debnath:2015wra} was the relative standard deviation $\bar{\sigma}_i$ of the areas of the neighboring cells,
\begin{equation}
\label{defvar}
\bar{\sigma}_i \equiv \frac{\sigma_i}{\bar{a}_i} \equiv 
\frac{1}{\bar{a}_i}\, \sqrt{\sum_{j\in {\cal N}_i} \frac{\left(a_j-\bar{a}_i\right)^2}{N_i-1}},
\end{equation}
where 
$$
\bar{a}_i\equiv \frac{1}{N_i}\, \sum_{j\in {\cal N}_i} a_j
$$ 
is the mean area of the neighbors of the $i^{\rm th}$ Voronoi cell. As demonstrated in Refs.~\cite{Debnath:2015wra,Debnath:2016mwb,DebnathPhD}, among the different possibilities, the relative standard deviation (\ref{defvar}) performed rather well in tagging the BVCs. Of course, those studies were utilizing only the Voronoi tessellation, which is not ideal for computing gradients. The Delaunay triangulation, on the other hand, provides a more natural framework for the gradient estimation\footnote{A purist might say that, since the two tessellations are dual to each other, strictly speaking there is nothing more to be gained from the Delaunay tessellation that could not have already be obtained from the Voronoi tessellation. While this may be technically correct, we found the Delaunay tessellation useful in hinting at some new techniques and ideas as discussed below.}, as will be discussed in the following subsection.

\subsection{Triangulation wombling from a Delaunay tessellation}
\label{sec:triangulation}

The Delaunay triangulation leads to a natural method for computing the gradient known as ``triangulation wombling'' \cite{Fortin1994,FortinDrapeau1995}. The starting point is the observation that for each Delaunay triangle $D_\alpha$, the functional values at its three vertices are known from (\ref{feq1overa}). Three points are enough to fit a plane, whose slope will provide an estimate of the gradient vector $\vec{G}_\alpha$ to be associated with the Delaunay triangle $D_\alpha$. Recall the duality relation (\ref{dual3}) which maps the triangle $D_\alpha$ to its three  vertices which carry indices $i$, $j$ and $k$ in the Voronoi tessellation. We can then parametrize the plane defined by $D_\alpha$ as 
\beq
f = G_{\alpha x}\, x + G_{\alpha y}\, y + C,
\eeq
where $C$ is some constant. Applying this relation at each vertex, we obtain three  independent equations
\begin{subequations}
\begin{eqnarray}
f_i &=& G_{\alpha x}\, x_i + G_{\alpha y}\, y_i + C, \\[2pt]
f_j &=& G_{\alpha x}\, x_j + G_{\alpha y}\, y_j + C, \\[2pt]
f_k &=& G_{\alpha x}\, x_k + G_{\alpha y}\, y_k + C, 
\end{eqnarray}
\end{subequations}
which can be solved for the gradient $\vec{G}_\alpha$ and the constant $C$ as \cite{Fortin1994,FortinDrapeau1995}
\beq
\left(
\begin{array}{c}
G_{\alpha x} \\ G_{\alpha y} \\ C 
\end{array}
\right)
=
\left(
\begin{array}{ccc}
x_i & y_i & 1 \\
x_j & y_j & 1 \\
x_k & y_k & 1 
\end{array}
\right)^{-1}
\left(
\begin{array}{c}
f_i \\ f_j \\ f_k
\end{array}
\right).
\label{Gvec}
\eeq
From here, a wombling analysis would typically focus on the magnitude of the gradient 
\beq
G_\alpha = \sqrt{G_{\alpha x}^2+G_{\alpha y}^2}
\label{Galpha}
\eeq
and proceed to select Delaunay triangles $D_\alpha$ with relatively large values of $G_\alpha$, typically in the top $10^{\rm th}$ percentile of all cells, as candidates for boundary elements.

\begin{figure}[t]
 \centering
 \includegraphics[height=.3\textwidth]{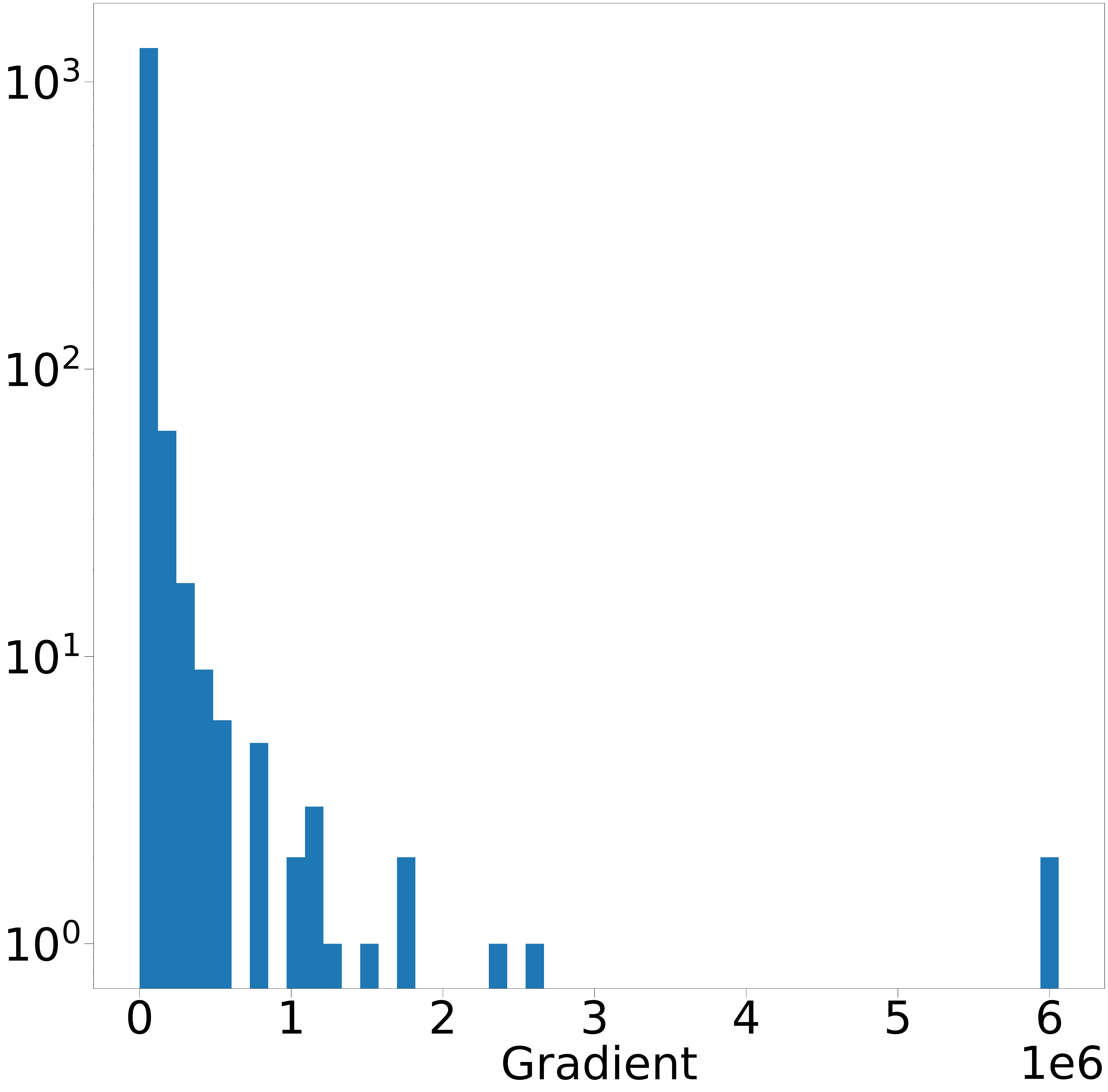}~~
 \includegraphics[height=.3\textwidth]{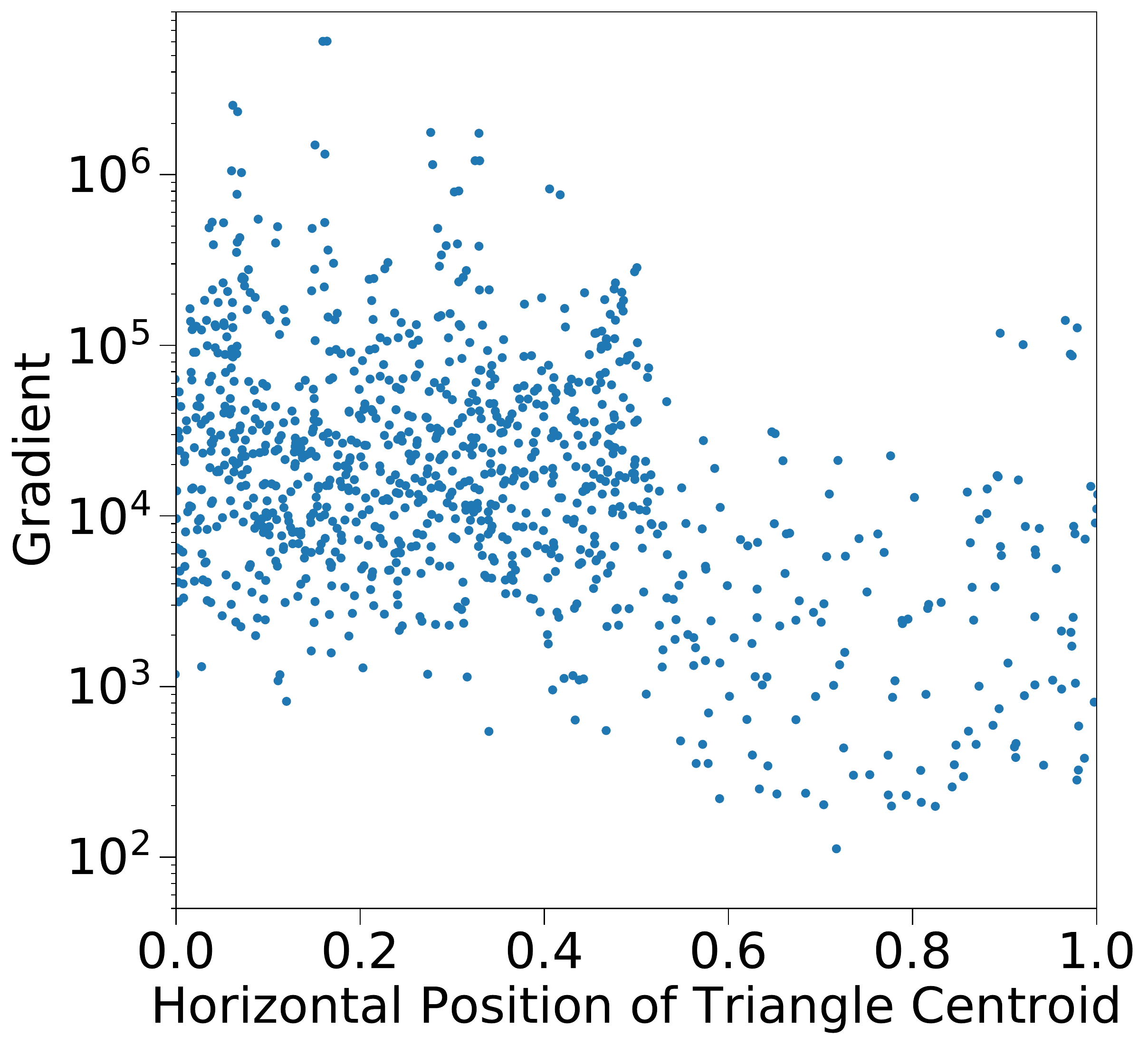}~~
 \includegraphics[height=.3\textwidth]{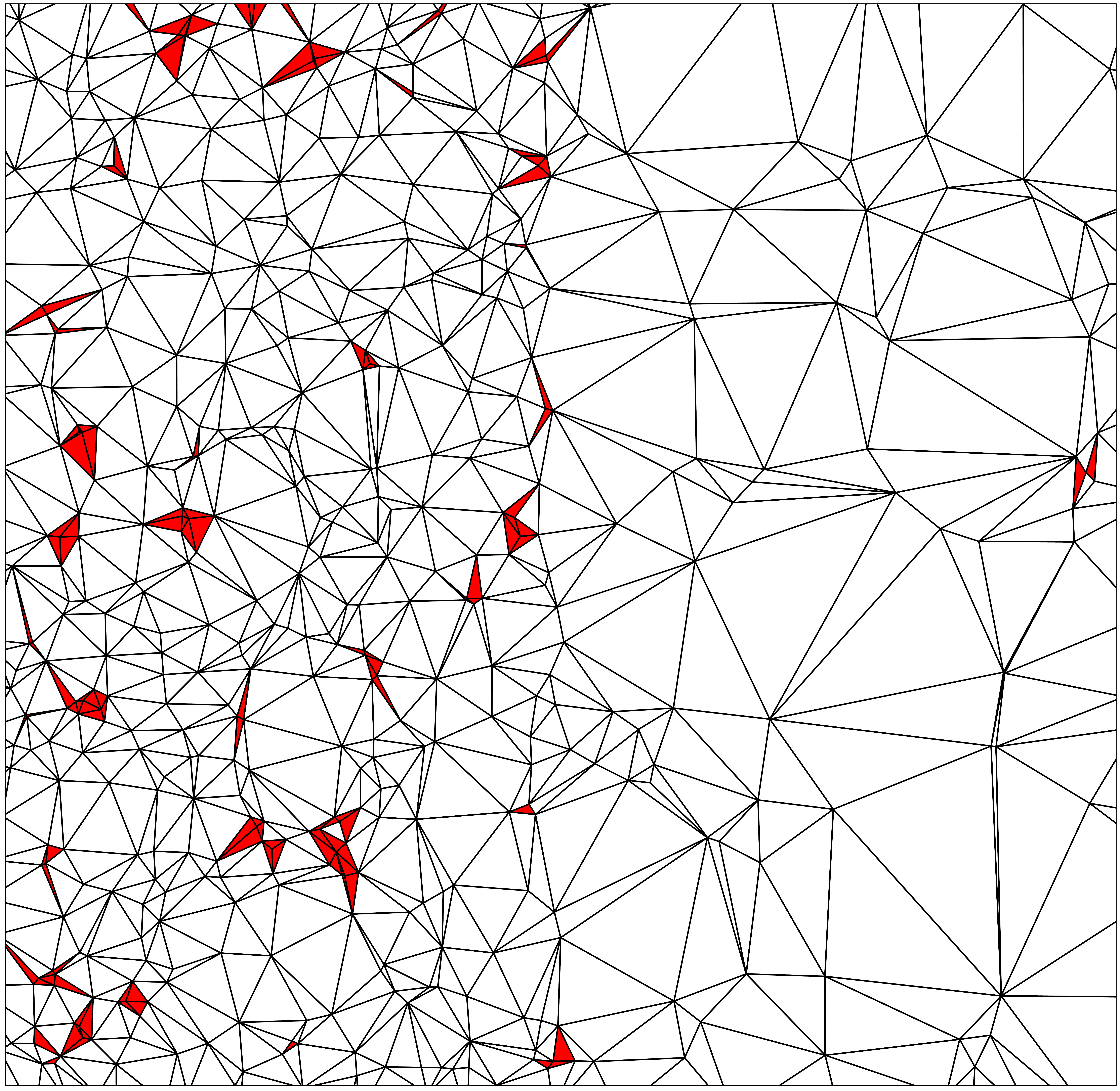}
\caption{\label{fig:unscaled_gradients} Left panel: Distribution of the magnitudes $G_\alpha$ of the gradient vectors $\vec{G}_\alpha$ computed in the process of triangulation wombling. Middle panel: A scatter plot of $G_\alpha$ versus the horizontal position of the centroid of the corresponding Delaunay triangle $D_\alpha$. Right panel: the Delaunay triangles (red-shaded) within the top $10^{\rm th}$ percentile as ranked by gradient magnitudes.
}
\end{figure}

However, the straightforward application of this procedure leads to a problem which is illustrated in Fig.~\ref{fig:unscaled_gradients} for the case of $\rho=5$. In the left panel we show the distributions of gradient magnitudes  $G_\alpha$ as calculated by the triangulation wombling method just described. We see that the distribution is a very steeply falling function with a long tail --- most Delaunay cells have relatively small gradients and only a small fraction populates the large $G_\alpha$ tail. Now, if the cells on the tail were predominantly BDTs, the method would have succeeded and there would be no problem, but unfortunately, that is not the case. In the middle panel of Fig.~\ref{fig:unscaled_gradients} we show a scatter plot of the calculated gradient magnitudes $G_\alpha$ versus the horizontal position of the corresponding Delaunay triangle $D_\alpha$ as given by its centroid. We notice that the gradients computed for cells in the dense region ($x<0.5$) tend to be much larger than the gradients in the sparse region ($x>0.5$) (this is to be expected, since statistical fluctuations scale as $\sqrt{f}$). As a result, the cells with large gradients will tend to be more or less uniformly distributed in the dense region, with no relation to the boundary at $x=0.5$. This is confirmed in the right panel of Fig.~\ref{fig:unscaled_gradients}, where we identify with red shading the Delaunay triangles whose gradients are in the top $10^{\rm th}$ percentile of gradient magnitudes. As anticipated from the result in the middle panel, the red-shaded Delaunay triangles are located almost entirely in the dense region and there is no apparent clustering near the boundary. This means that in order to properly tag the BDTs, we must first pre-process the computed gradients $\vec{G}_\alpha$ in order to mitigate the effect of the statistical noise. 

\subsection{Denoising}
\label{sec:denoising}

In this subsection we outline three different procedures for denoising the computed gradient vectors  $\vec{G}_\alpha$. As we shall demonstrate, each of them has the desired effect, and the optimal approach will be some combination of the three, although finding out the exact proportions is beyond the scope of this paper.

\subsubsection{Rescaling of the naive gradients}
\label{sec:gradrescaling}

\begin{figure}[t]
 \centering
 \includegraphics[height=.3\textwidth]{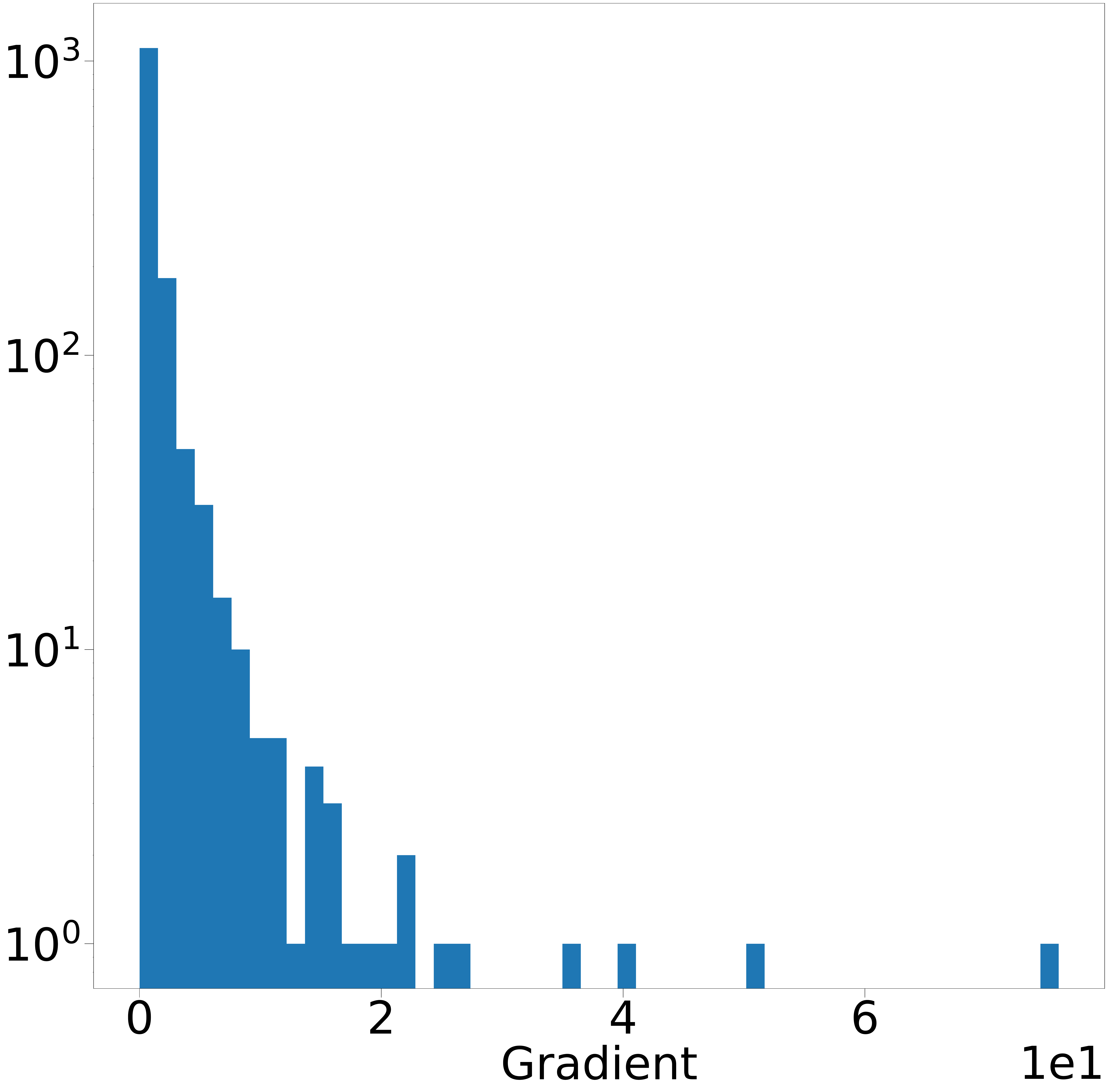}~~
 \includegraphics[height=.3\textwidth]{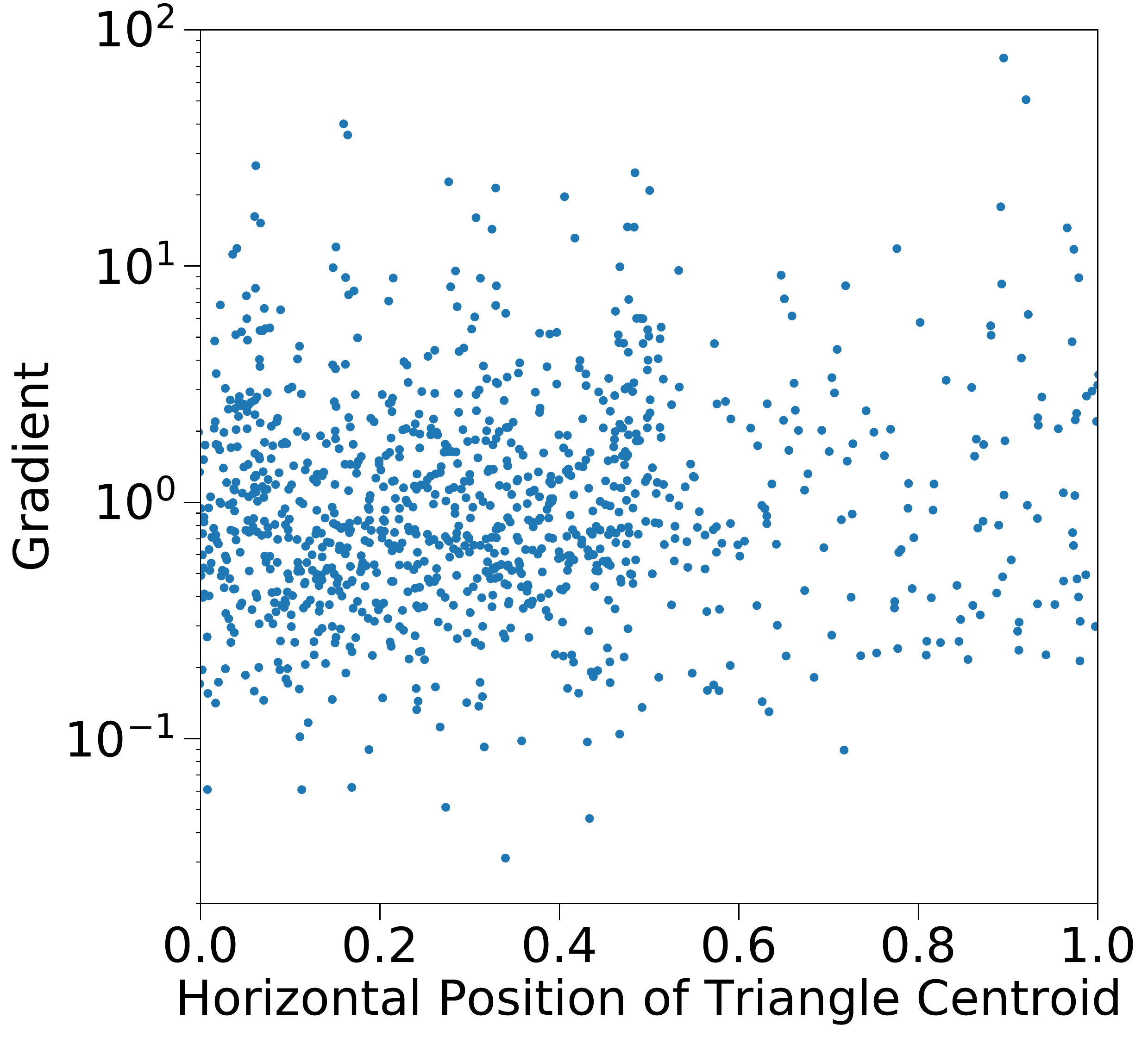}~~
 \includegraphics[height=.3\textwidth]{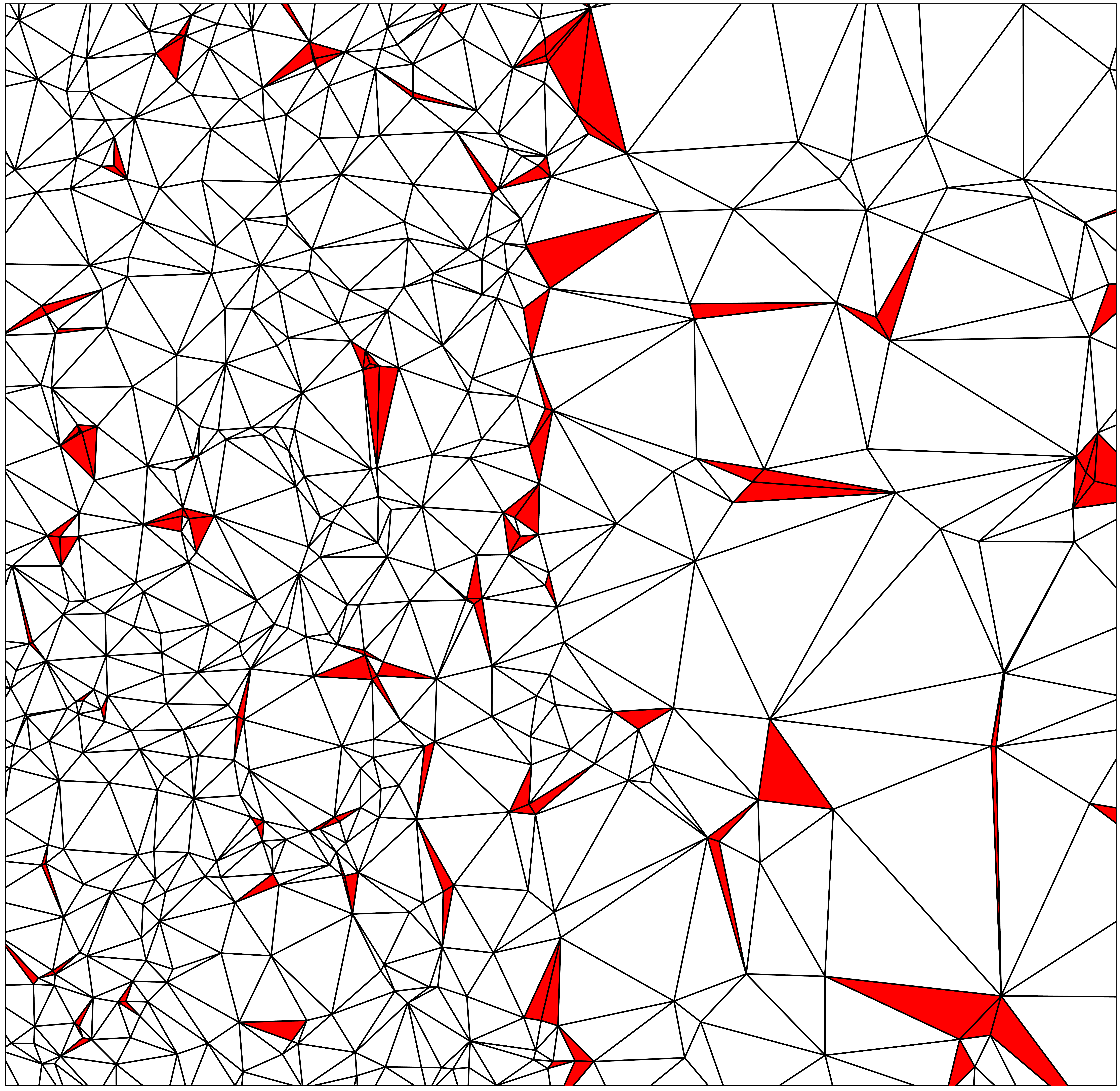}
\caption{\label{fig:scaled_gradients} The same as Fig.~\ref{fig:unscaled_gradients}, but after rescaling the calculated gradients as in eq.~(\ref{Grescaling}).
}
\end{figure}

The first approach is to rescale each calculated gradient magnitude $G_\alpha$ as 
\beq
G_\alpha \to \tilde G_\alpha \equiv G_\alpha\, \sqrt{a_i a_j a_k},
\label{Grescaling}
\eeq
where $a_i$, $a_j$ and $a_k$ are the areas of the Voronoi cells centered on the three vertices of the Delaunay triangle $D_\alpha$ (recall the duality relation (\ref{dual3})). The idea behind the rescaling (\ref{Grescaling}) is to render the rescaled gradient $\tilde G_\alpha$ ``dimensionless'' with respect to the cell areas $a_i$, and therefore insensitive to the overall density of points.
Fig.~\ref{fig:scaled_gradients} demonstrates that the rescaling (\ref{Grescaling}) does have the desired effect. 
In the left panel, the distribution of rescaled gradient magnitudes $\tilde G_\alpha$ still follows the same trend as in Fig.~\ref{fig:unscaled_gradients}, but the numerical values have been reduced by several orders of magnitude. More importantly, the middle panel of Fig.~\ref{fig:scaled_gradients} confirms that the rescaled gradient magnitudes are now uniformly consistent across the two regions, with a cluster of points with relatively large $\tilde G_\alpha$ beginning to emerge near the theoretical boundary. The right panel of Fig.~\ref{fig:scaled_gradients} shows the updated plot of the Delaunay triangles falling in the top $10^{\rm th}$ percentile of $\tilde G_\alpha$. We observe that, while the Delaunay cells tagged as BDTs are still scattered throughout the field of view, a significant fraction of them appears at the location of the boundary, which further validates the rescaling procedure (\ref{Grescaling}). That is why from now on, unless specified otherwise, we shall always work with gradient vectors which have been rescaled as in (\ref{Grescaling}).

\subsubsection{Lloyd steps uniformization (LSU)}
\label{sec:LSU}

Another method for removing unwanted noise fluctuations in the data was explored in \cite{Debnath:2015wra} and involved the so called Voronoi relaxation of the data by means of Lloyd's algorithm \cite{Lloyd1982}. A closer inspection of the Voronoi tessellations depicted in Figs.~\ref{fig:boundaryobjects} and \ref{fig:correlations} reveals that after the tessellation is constructed, the generator points can be found pretty much anywhere within the Voronoi polygon --- near the center of the cell, close to an edge, or somewhere in between. The idea of the Voronoi relaxation is to make the whole Voronoi structure more uniform by performing several steps (or iterations) of Lloyd's algorithm, where at each iteration, the generator point is moved to the centroid of the corresponding Voronoi cell and the tessellation is redone. 

\begin{figure}[t]
 \centering
 \includegraphics[height=.2\textwidth]{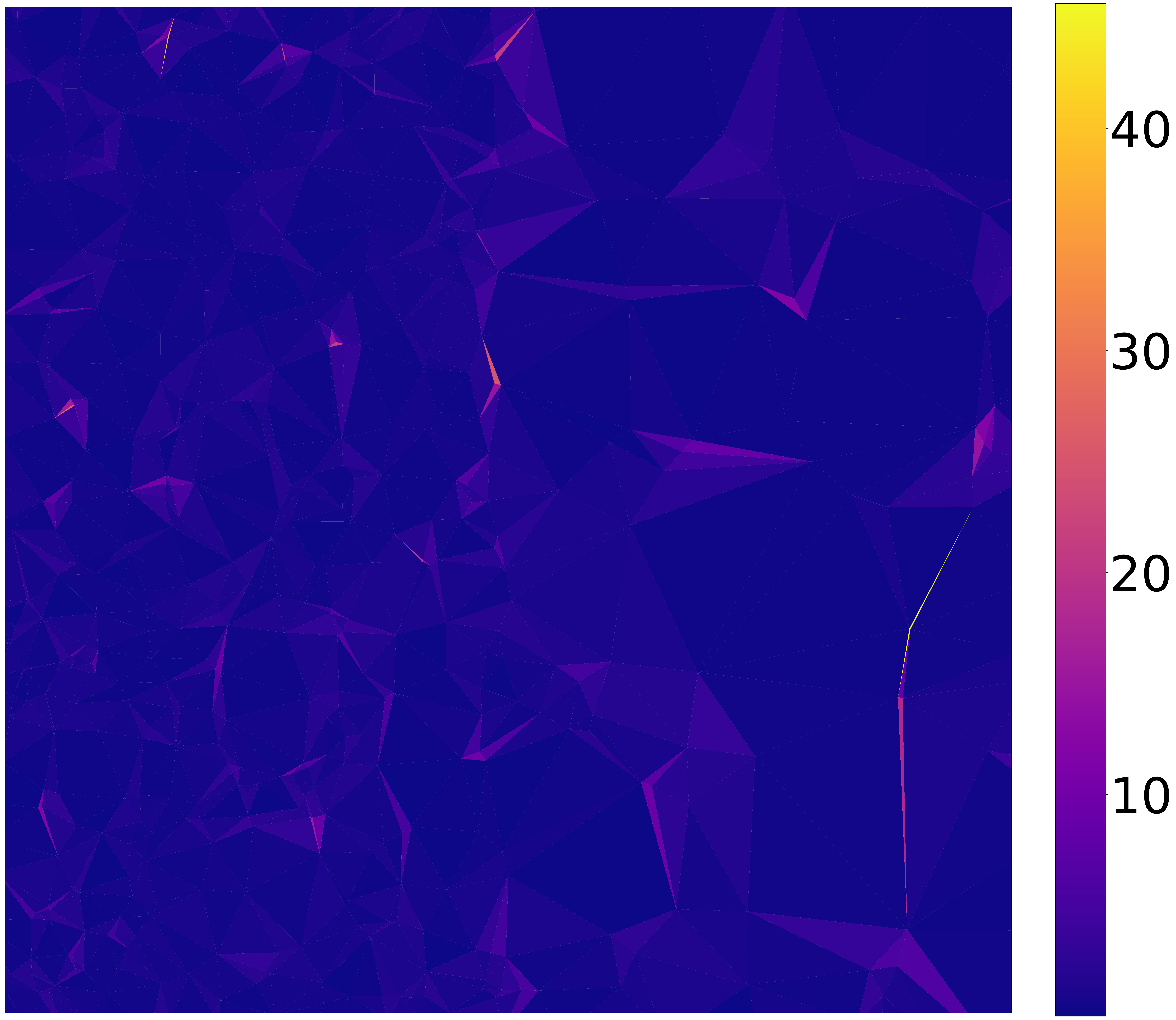}~~
 \includegraphics[height=.2\textwidth]{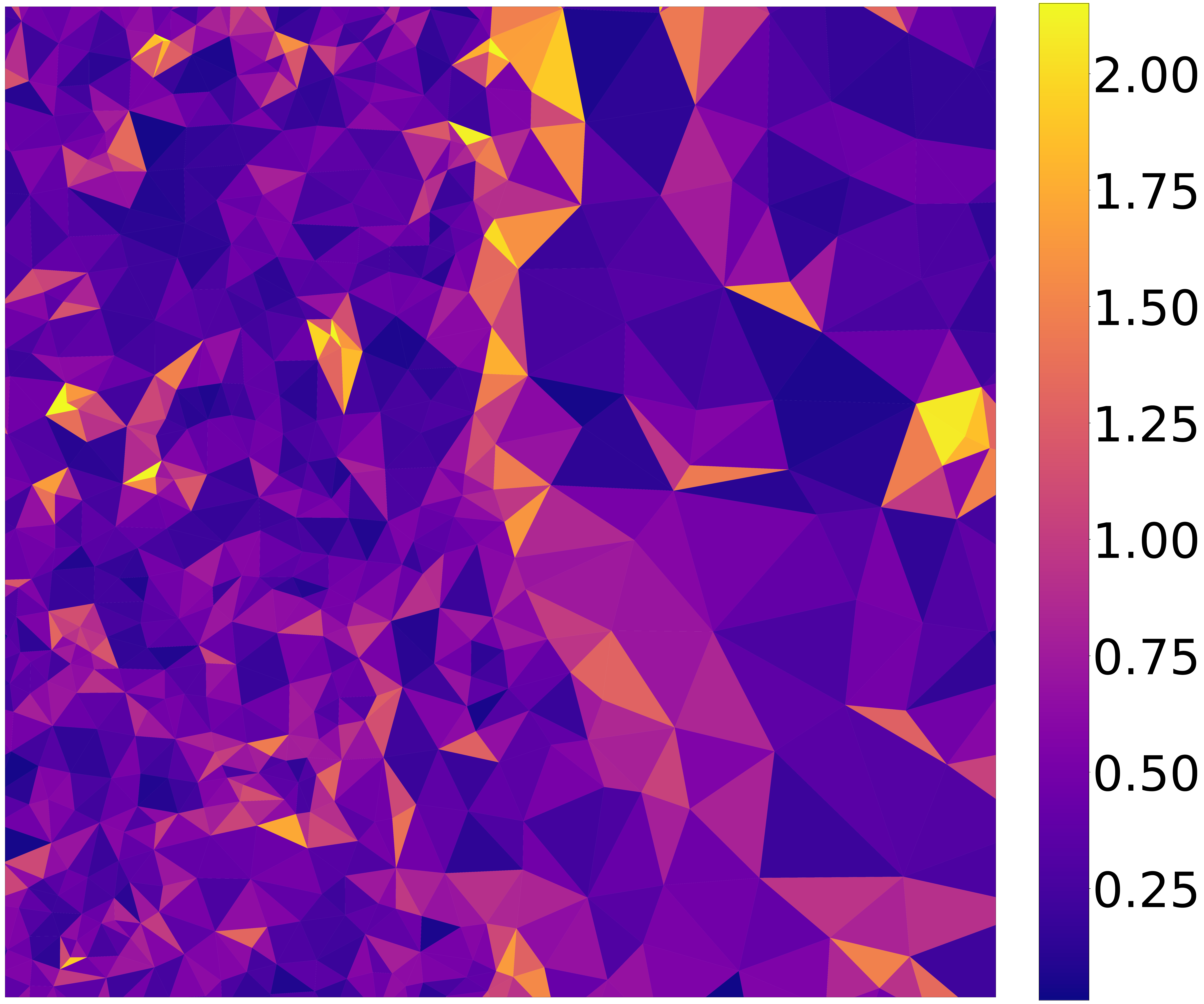}~~
 \includegraphics[height=.2\textwidth]{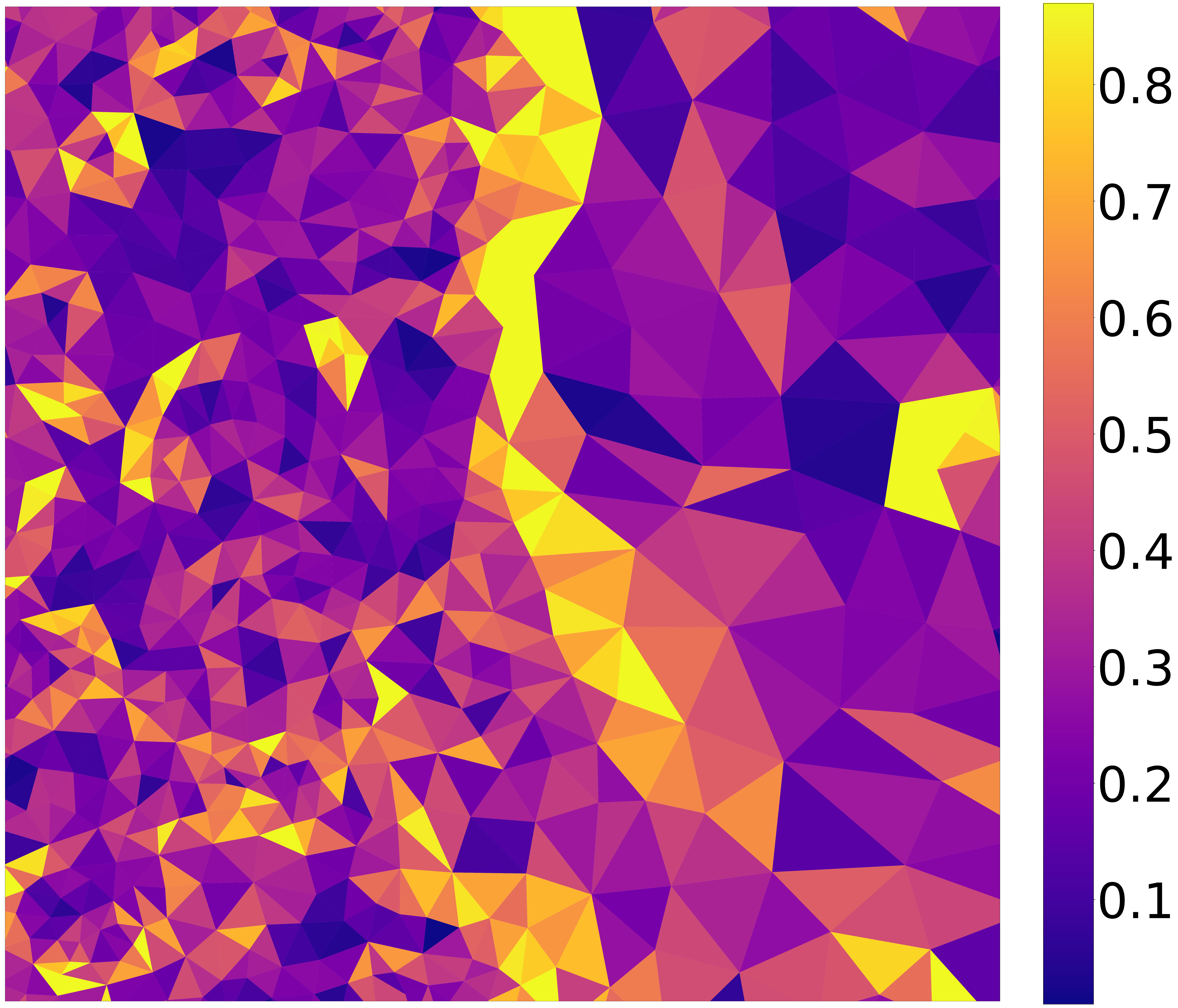}~~
\includegraphics[height=.2\textwidth]{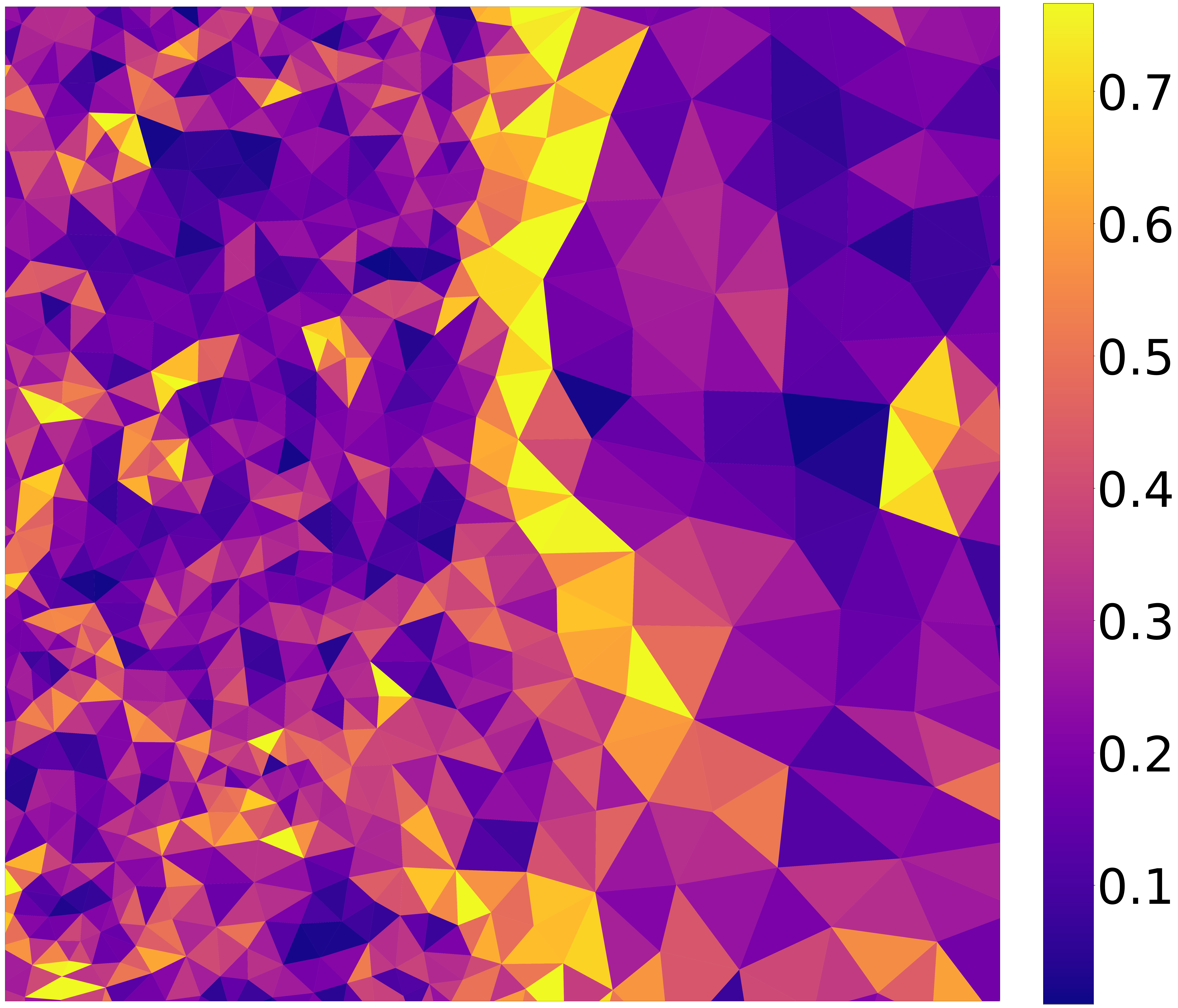}\\[5pt]
 \includegraphics[height=.2\textwidth]{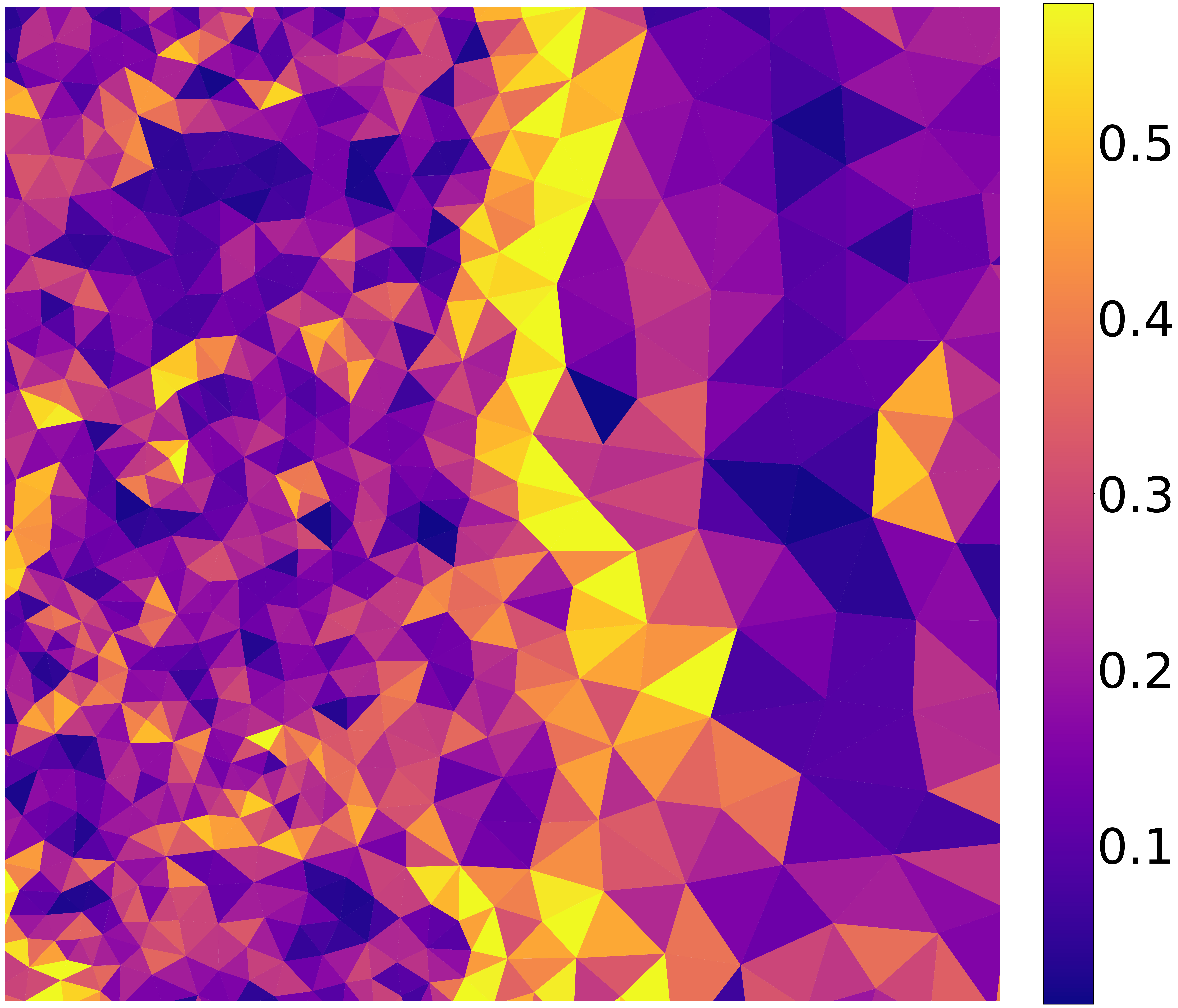}~~
 \includegraphics[height=.2\textwidth]{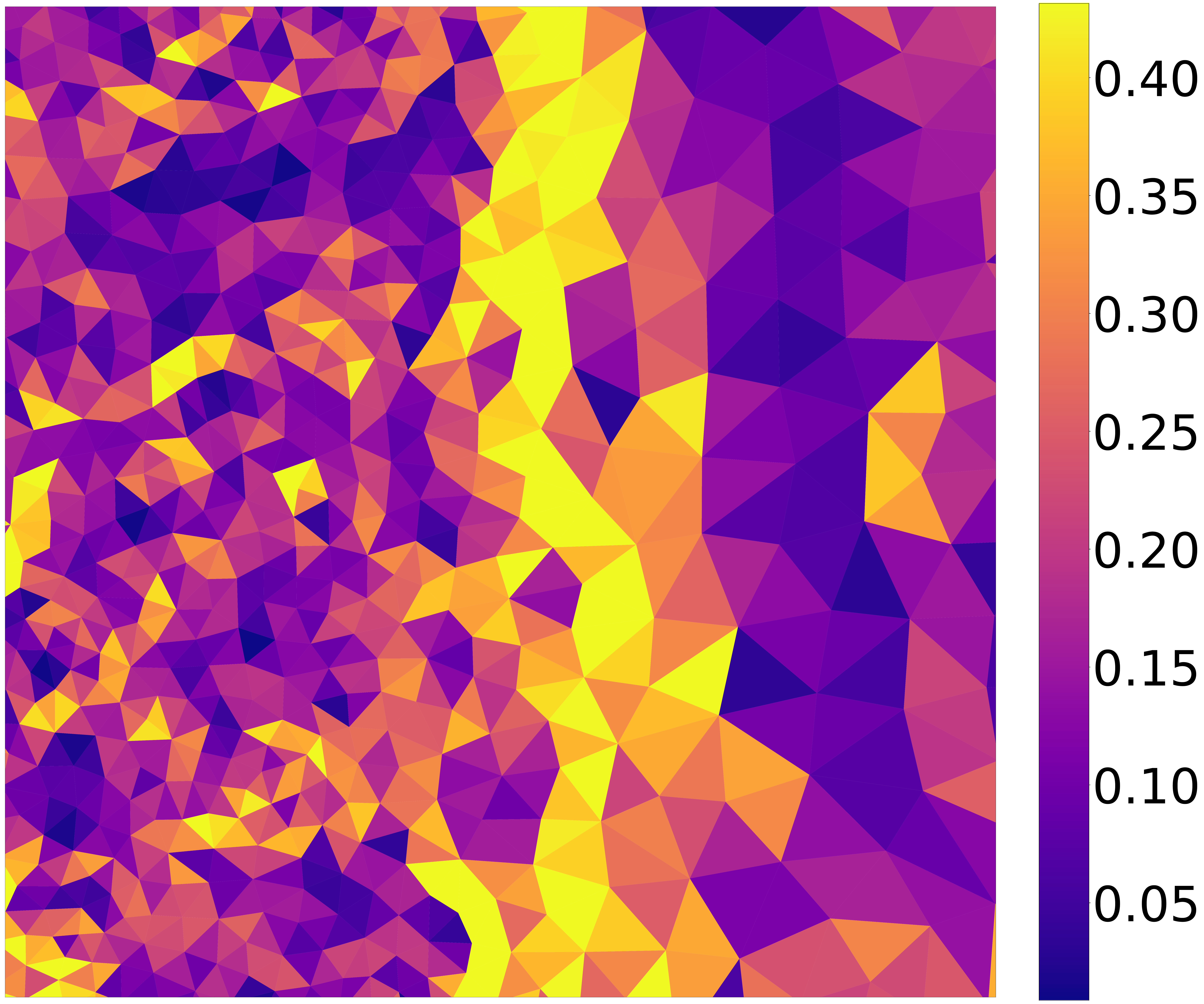}~~
 \includegraphics[height=.2\textwidth]{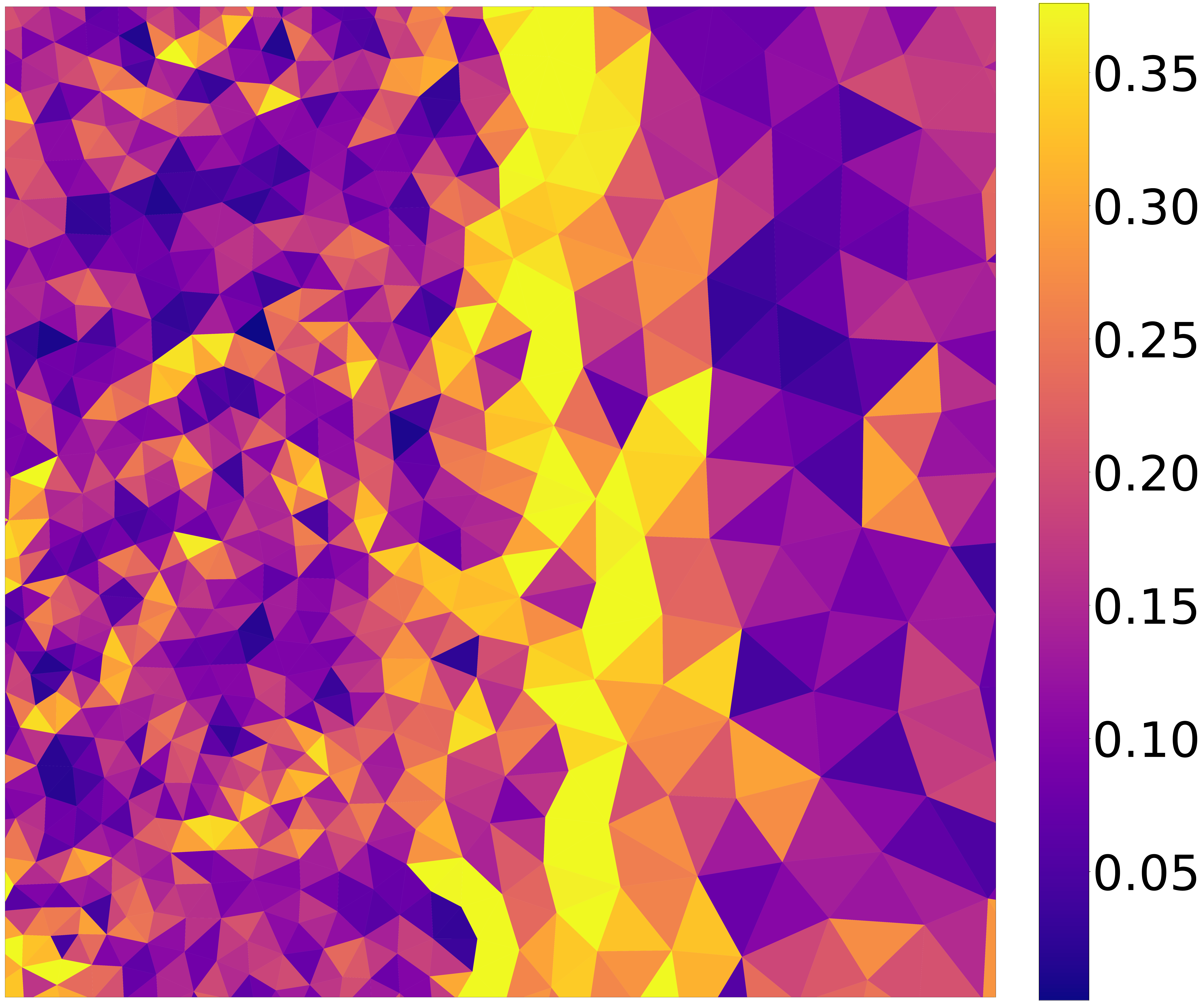}~~
 \includegraphics[height=.2\textwidth]{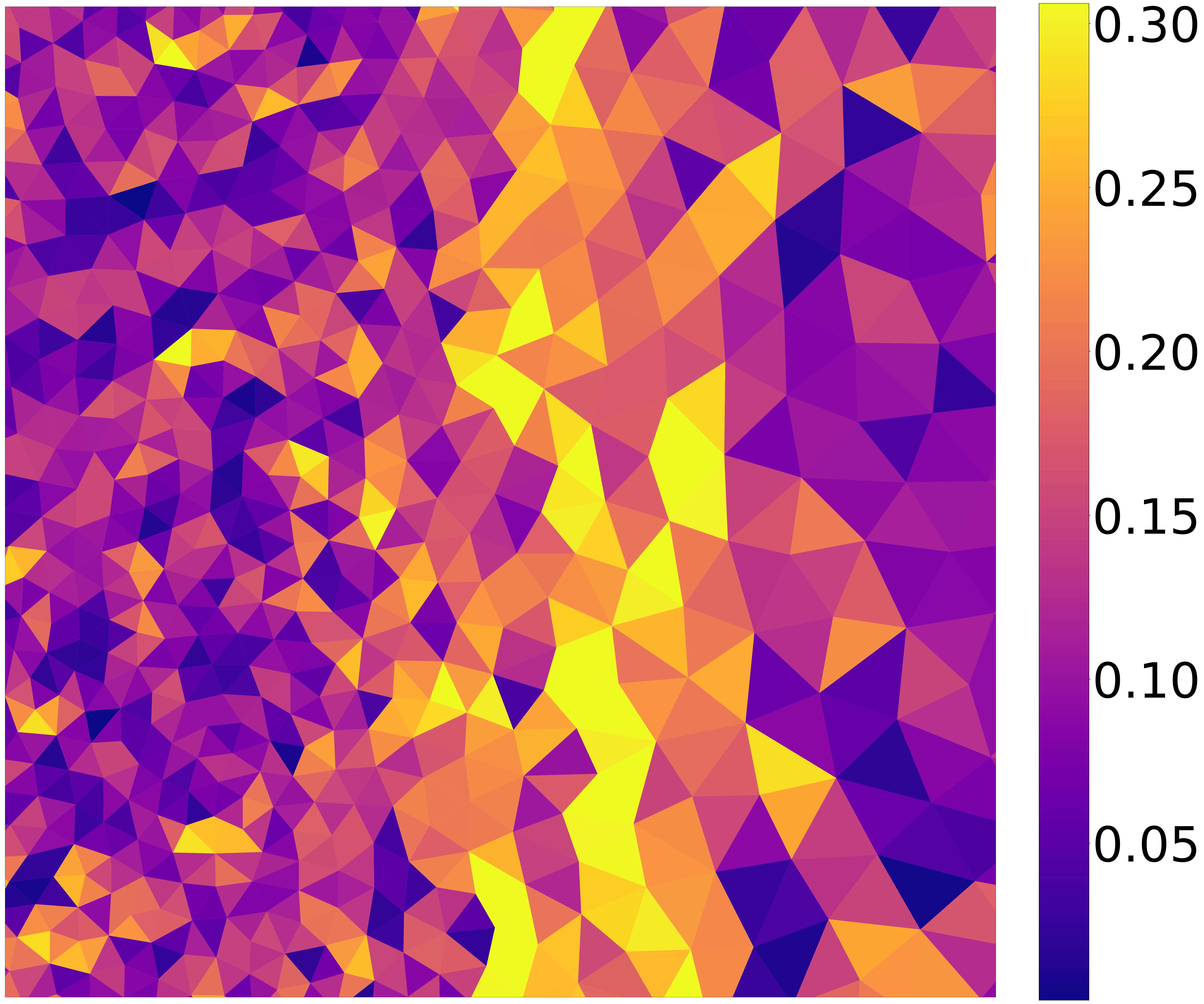}
\caption{\label{fig:lsu_lloyd} Delaunay triangulations with individual cells color-coded by their rescaled gradient magnitudes $\tilde G_\alpha$, after a certain number of Lloyd steps (from top left to bottom right): 0, 1, 2, 3, 5, 7, 10, 20. The color bar extends up to $60\%$ of the largest $\tilde G_\alpha$ value found in the data; any cell with a gradient magnitude above that threshold is colored yellow, essentially creating an overflow color bin. This was done in order to minimize the effect of outliers and better visualize the bulk of the cells with typical values.
}
\end{figure}

\begin{figure}[t]
 \centering
 \includegraphics[height=.23\textwidth]{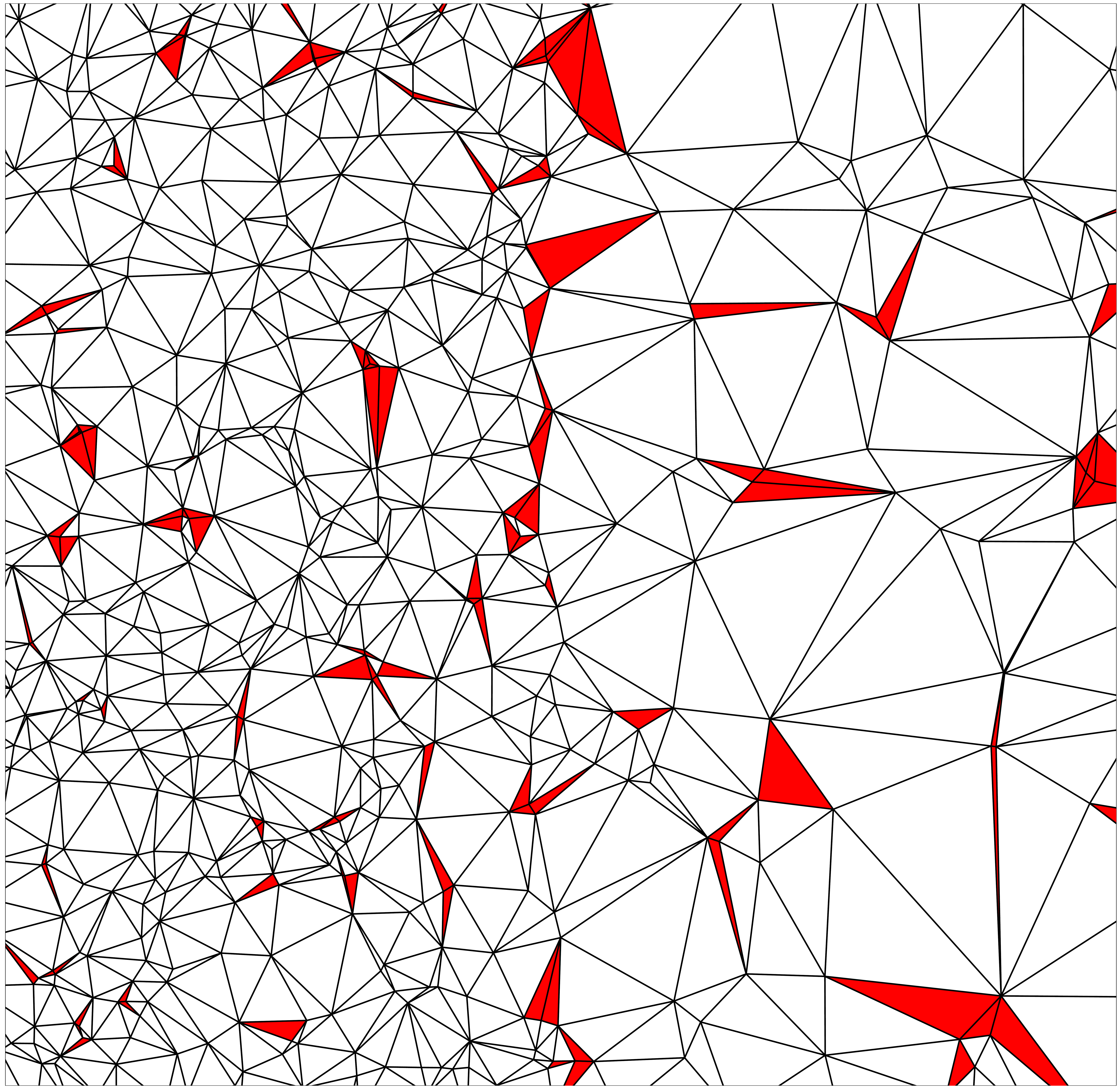}~~
 \includegraphics[height=.23\textwidth]{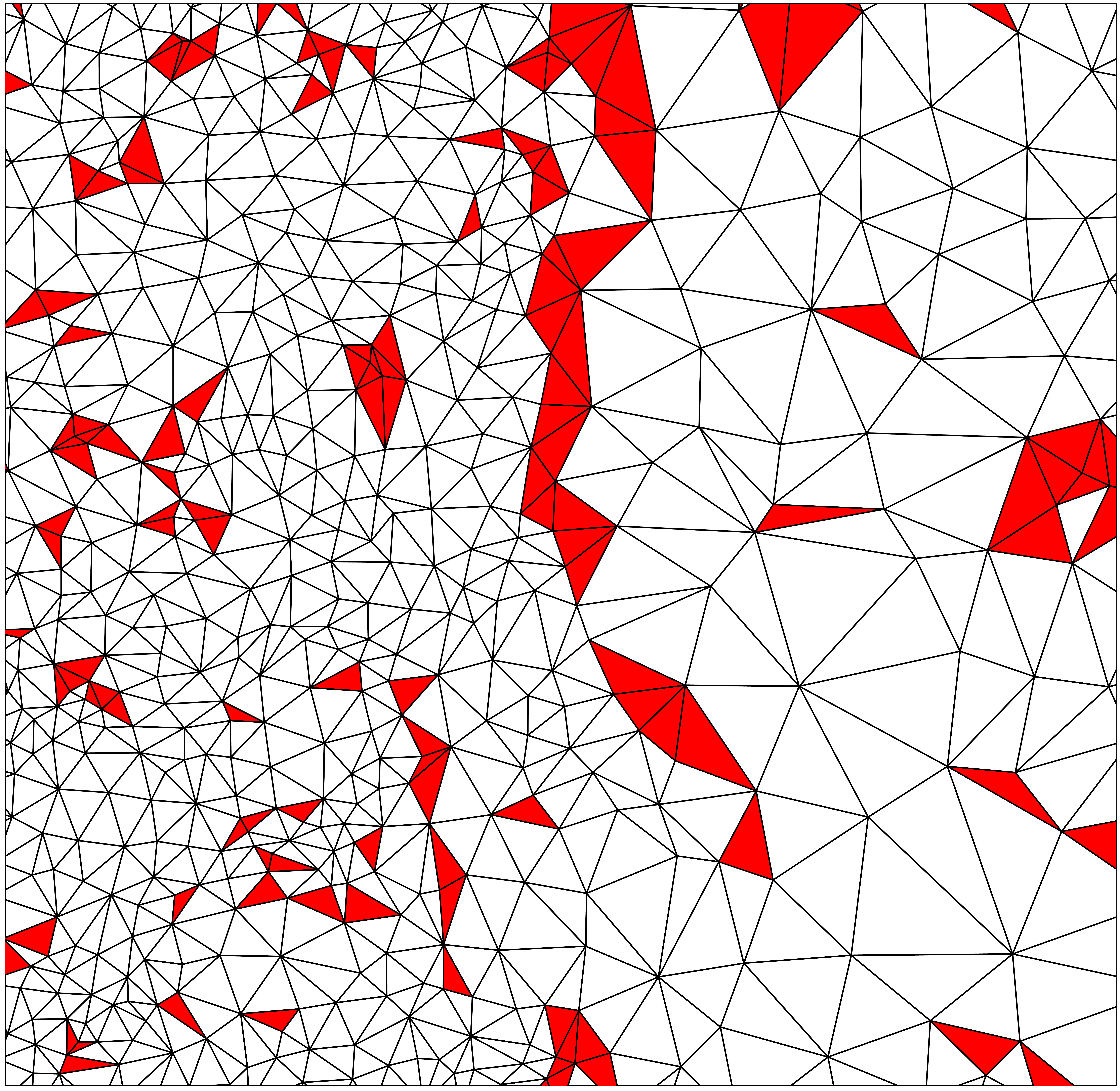}~~
 \includegraphics[height=.23\textwidth]{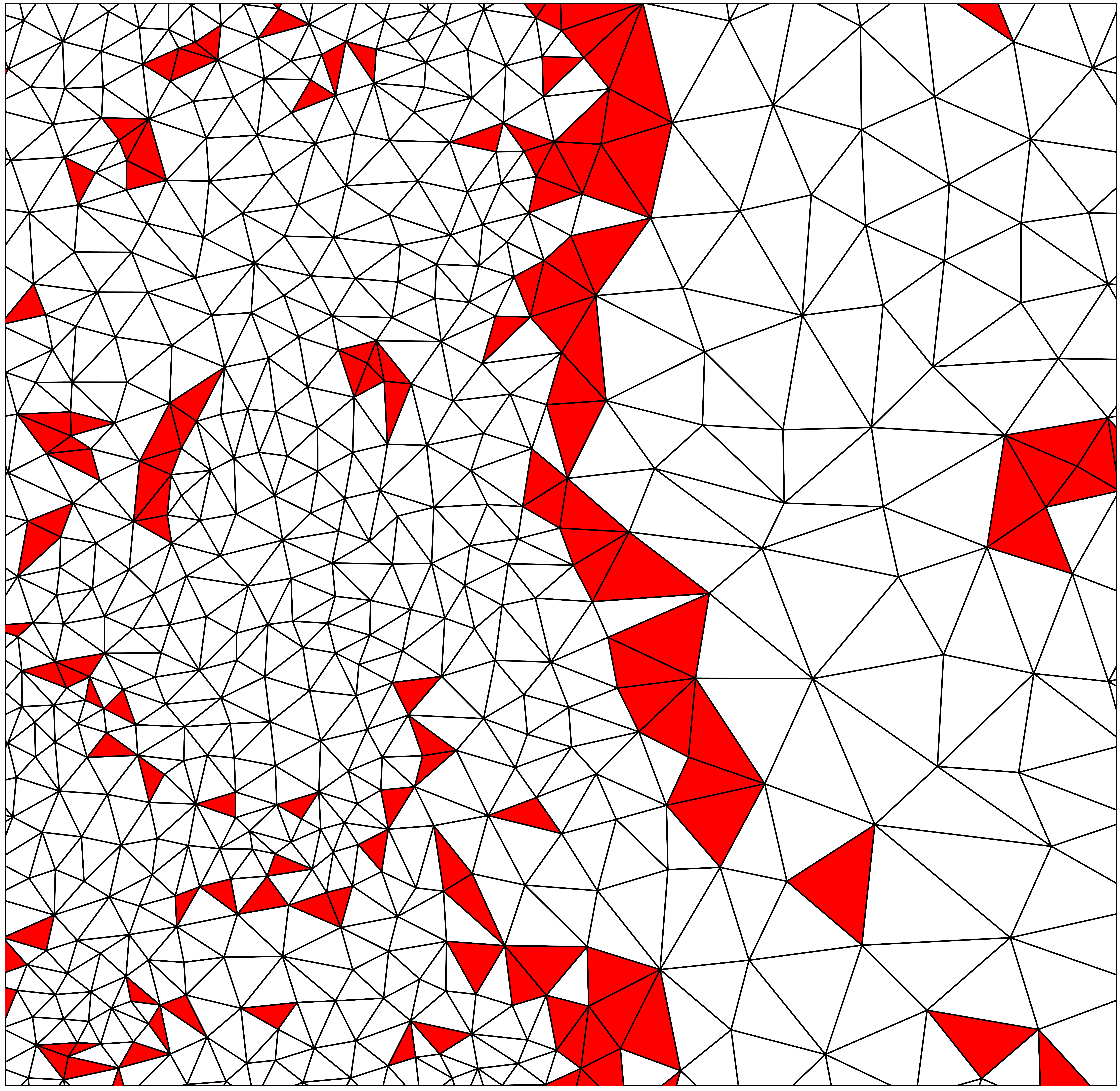}~~
\includegraphics[height=.23\textwidth]{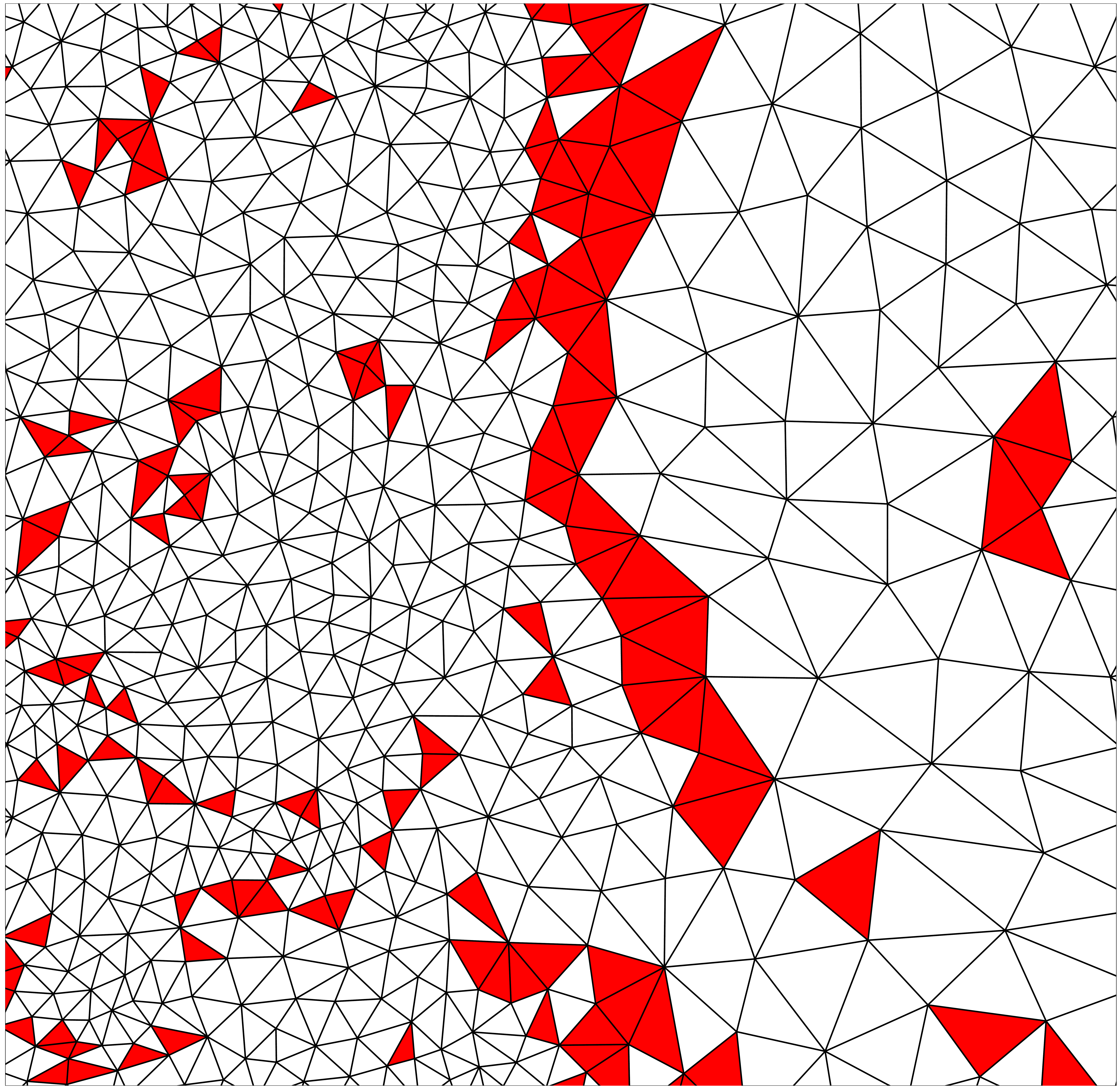}\\[5pt]
 \includegraphics[height=.23\textwidth]{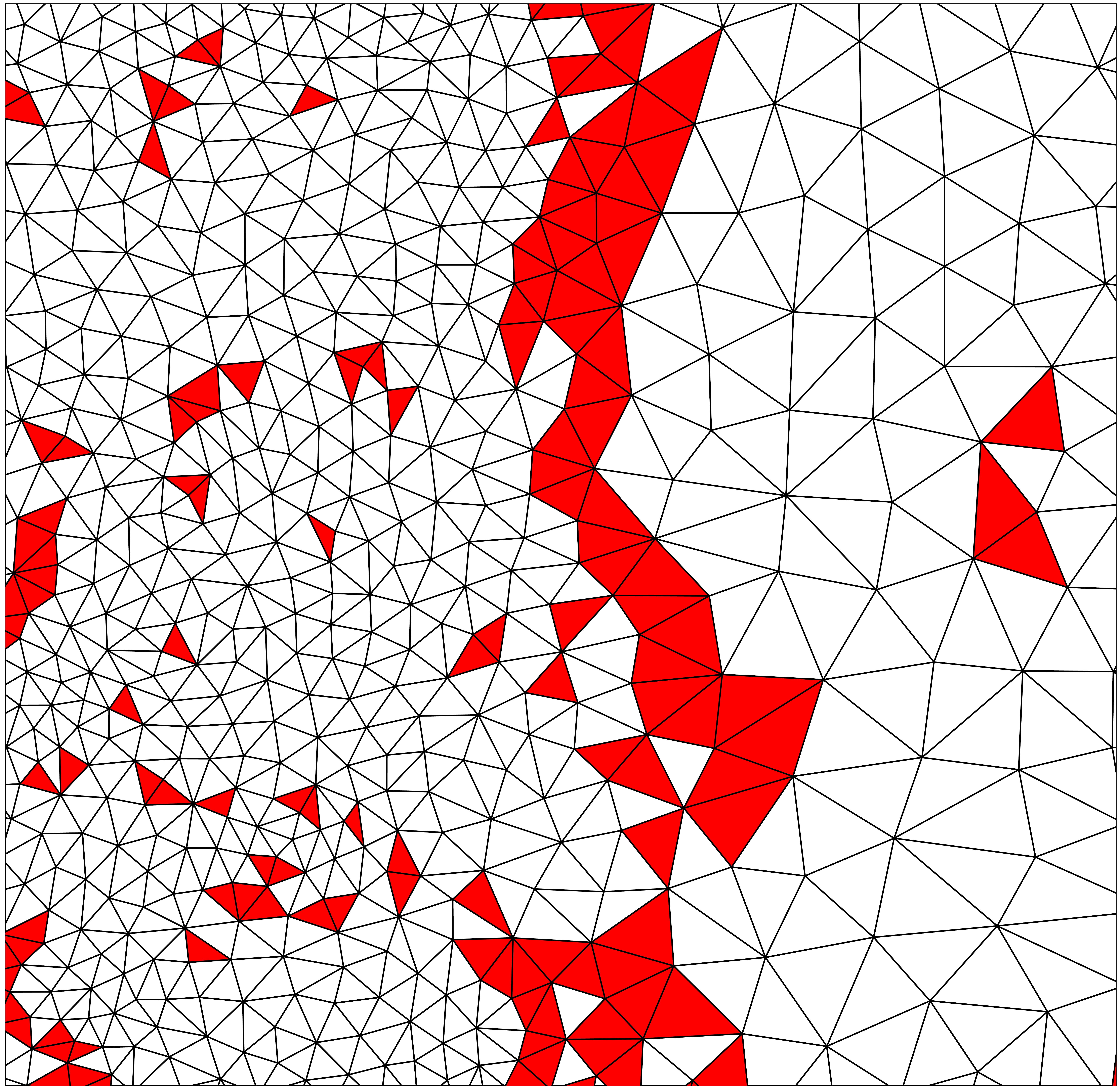}~~
 \includegraphics[height=.23\textwidth]{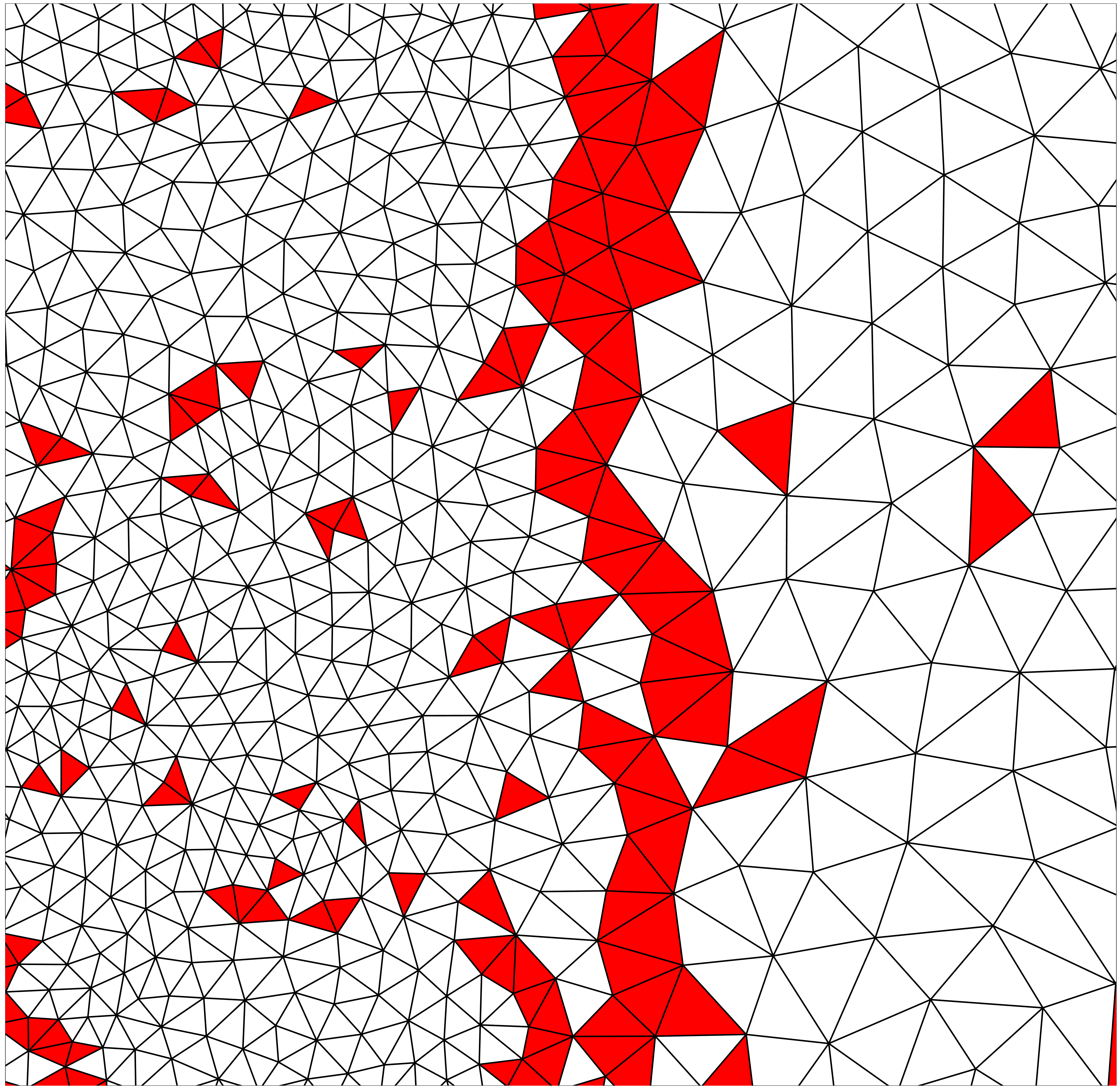}~~
 \includegraphics[height=.23\textwidth]{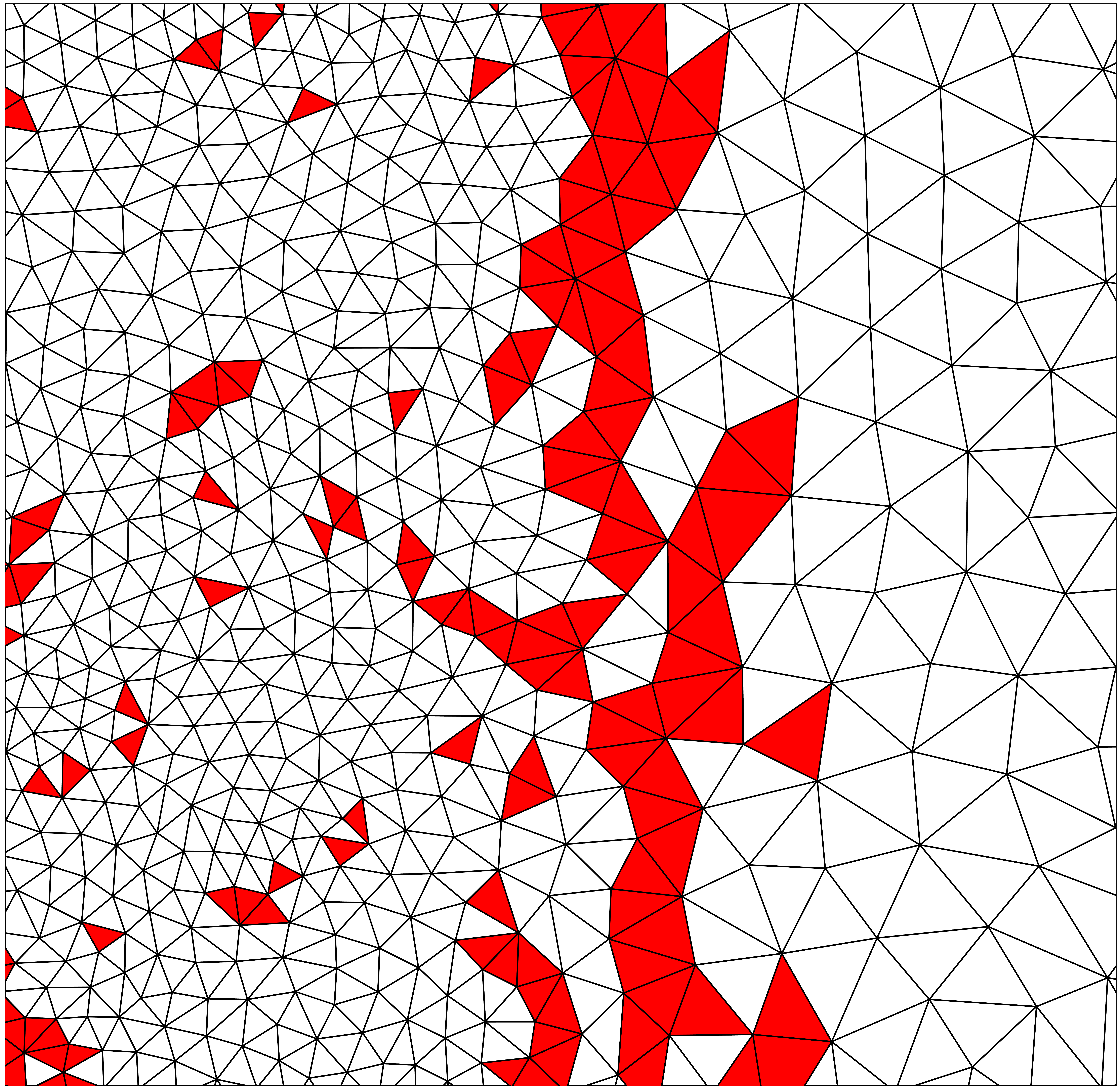}~~
 \includegraphics[height=.23\textwidth]{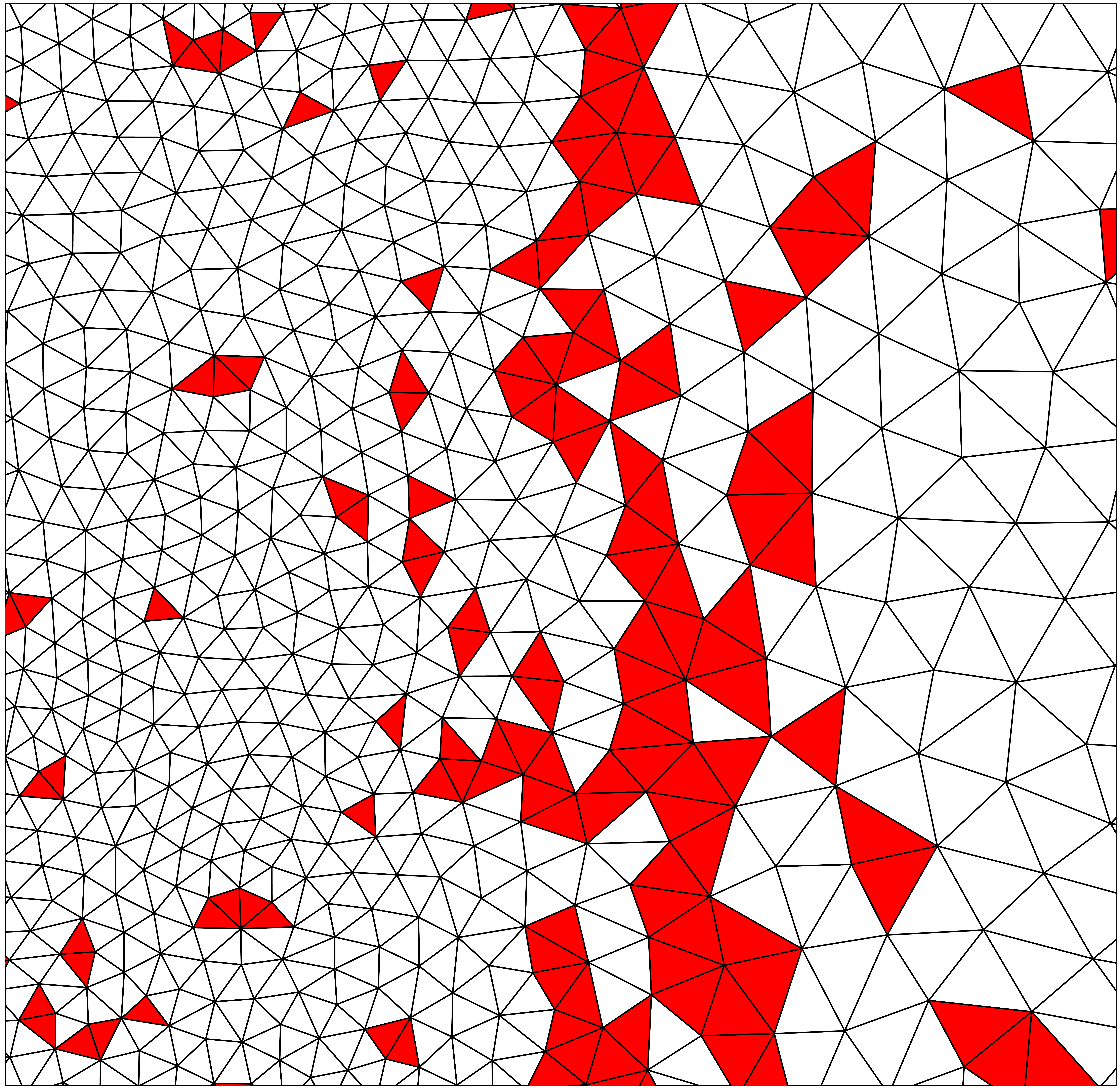}
\caption{\label{fig:after_cut_lloyd} The same as the right panel in Fig.~\ref{fig:scaled_gradients}, but showing the effect of applying the Lloyd algorithm with varying number of steps as in Fig.~\ref{fig:lsu_lloyd}.
}
\end{figure}

The effect of performing such Lloyd step uniformization (LSU) on our data is shown in Figs.~\ref{fig:lsu_lloyd} and \ref{fig:after_cut_lloyd}. Each figure has 8 panels, depicting the Delaunay tessellation\footnote{For similar plots illustrating the effect of LSU on the Voronoi tessellation, see \cite{Debnath:2015wra}. } of the data after a certain number of Lloyd iterations, starting with 0 (no Lloyd steps) in the upper left panels and going up to 20 Lloyd steps in the lower right panels. In addition, in Fig.~\ref{fig:lsu_lloyd} each Delaunay triangle $D_\alpha$ is color-coded by the rescaled magnitude $\tilde G_\alpha$ of the respective gradient (the gradients are recalculated after each step). Note that the color bar extends only up to $60\%$ of the largest gradient magnitude found in the data\footnote{For example, the left panel in Fig.~\ref{fig:scaled_gradients} reveals that the largest $\tilde G_\alpha$ value found in the data is around 75, while the color bar in the upper left panel of Fig.~\ref{fig:lsu_lloyd} only goes up to $75\times 60\%= 45$.}; any cell with a gradient magnitude above that threshold is colored yellow, essentially creating an overflow color bin. This was done in order to minimize the effect of outliers and better visualize the bulk of the cells with the more typical values. Fig.~\ref{fig:after_cut_lloyd}, on the other hand, marks the Delaunay triangles falling in the top $10^{\rm th}$ percentile of $\tilde G_\alpha$ values, just like the plot in the right panel of Fig.~\ref{fig:scaled_gradients}.


There are several lessons that can be drawn from Figs.~\ref{fig:lsu_lloyd} and \ref{fig:after_cut_lloyd}. First, as expected, the Lloyd relaxation causes the Delaunay triangles to become more regularly shaped. For example, notice that a large fraction of the triangles in the original Delaunay tessellation are obtuse (see the right panels in Figs.~\ref{fig:boundaryobjects}-\ref{fig:scaled_gradients}). However, within a few Lloyd steps, the fraction of obtuse triangles drops significantly and obtuse triangles are rather rare\footnote{Furthermore, the largest angle of any remaining obtuse triangle is typically not too far above $90^\circ$.} in the plots in the lower rows of Figs.~\ref{fig:lsu_lloyd} and \ref{fig:after_cut_lloyd}. More importantly, as shown in Figs.~\ref{fig:lsu_lloyd}, the LSU procedure also tends to wash out the noisy fluctuations in the calculated rescaled gradient magnitudes $\tilde G_\alpha$ within the bulk regions away from the boundary (note the decreasing range of $\tilde G_\alpha$ on the color bars). This further sharpens the contrast between the Delaunay triangles situated near the boundary versus those in the bulk. In particular, notice the gradual emergence of the boundary, which becomes quite pronounced and unmistakable after 5-7 Lloyd steps. However, the figures also show that the number of Lloyd steps should be chosen with care --- applying too few may not optimally showcase the boundary, while applying too many may cause the boundary to start disintegrating, as evidenced in the lower right panels after 20 iterations.

\subsubsection{Local averaging of gradient vectors}
\label{sec:averaging}

A third approach for smoothing out the local statistical fluctuations in the data is to perform some type of averaging procedure over a region extending beyond the individual Voronoi or Delaunay cells and their immediate neighbors. For example, Ref.~\cite{Debnath:2015wra} considered extending the calculation of $\bar\sigma_i$ in (\ref{defvar}) over several tiers of Voronoi neighbors (up to 5) and showed that this indeed produces the desired effect of reducing the fluctuations and sharpening the boundary identified by means of $\bar\sigma_i$.

\begin{figure}[t]
 \centering
 \includegraphics[width=.3\textwidth]{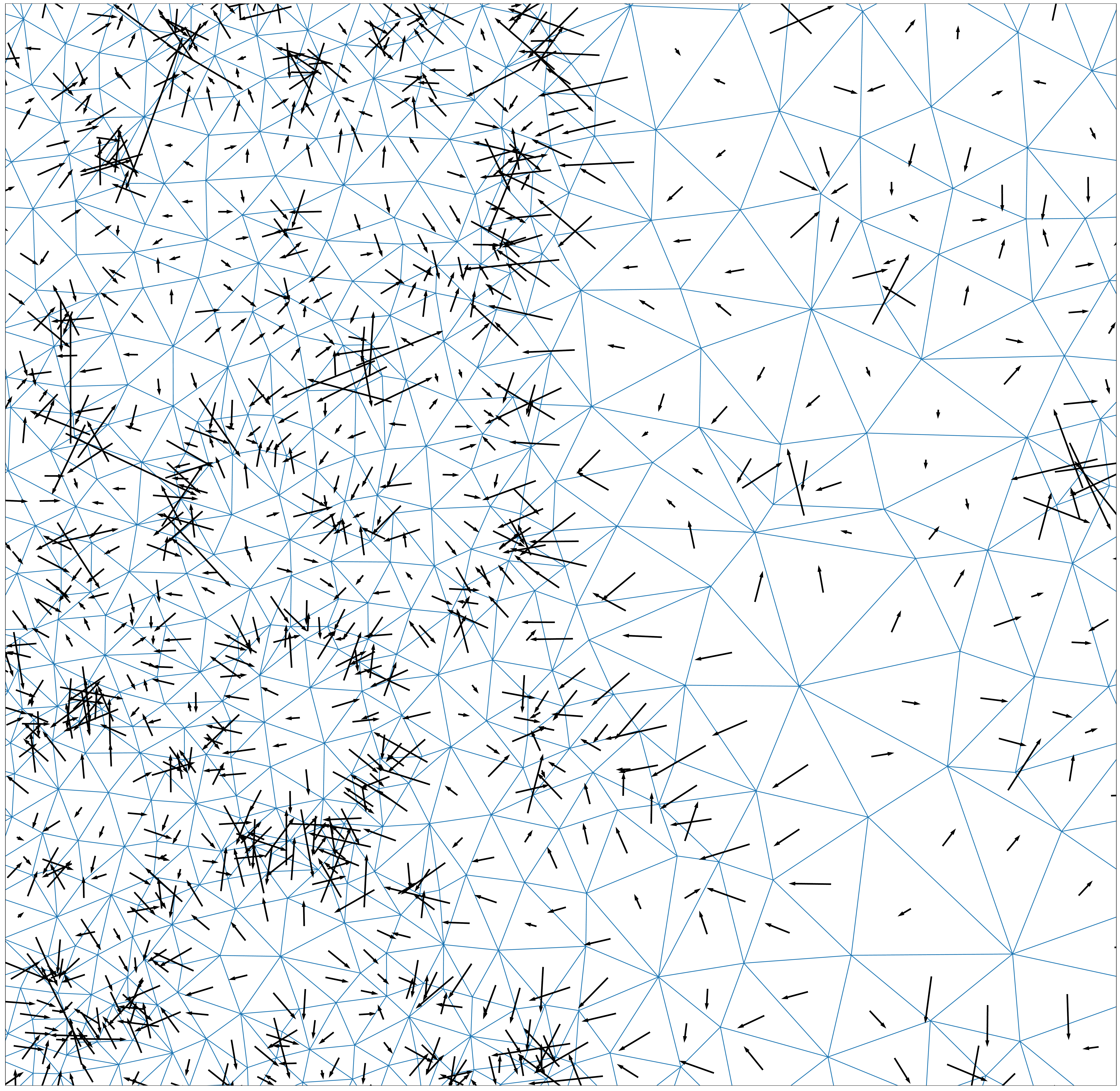}~~
 \includegraphics[width=.3\textwidth]{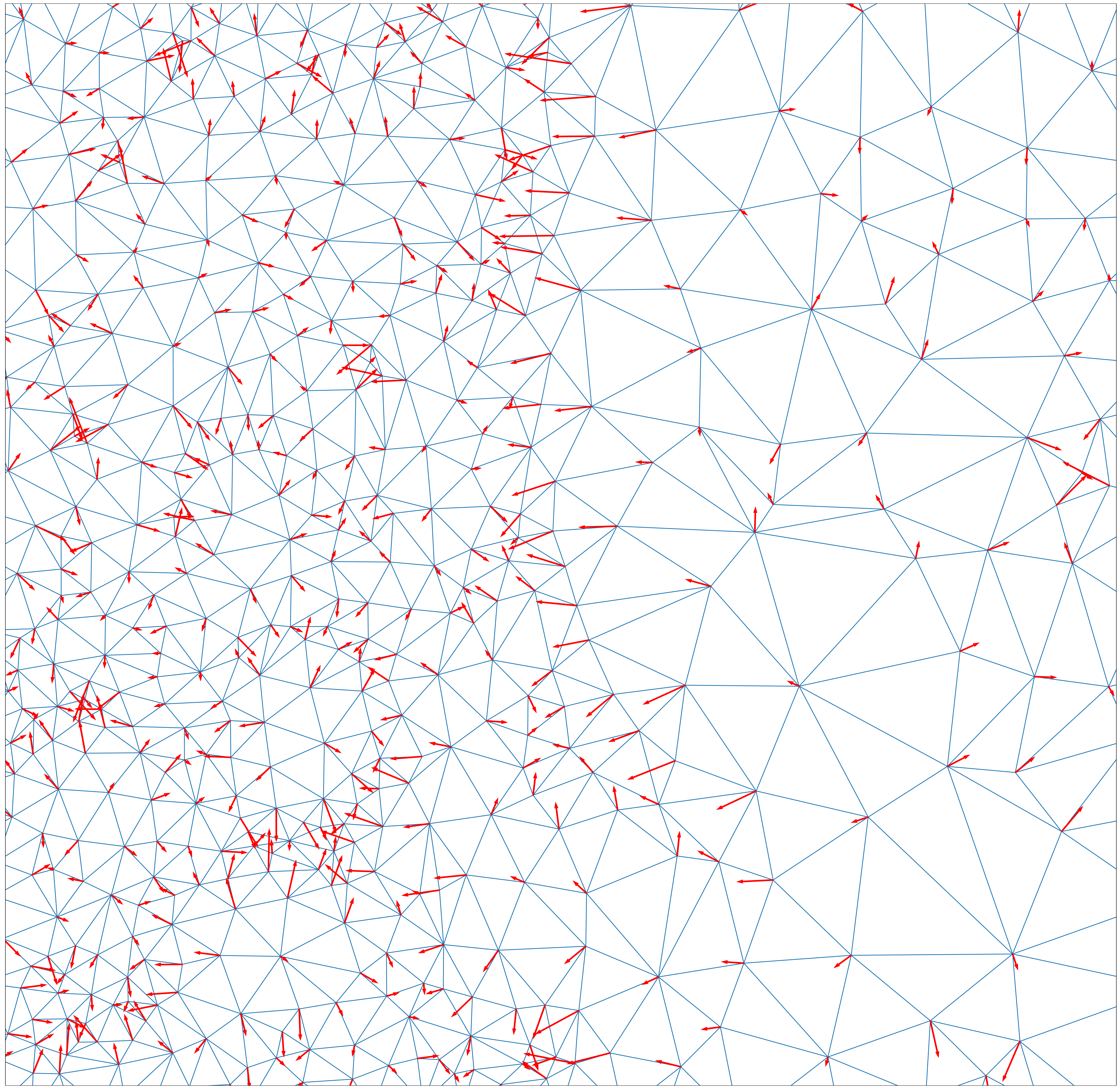}~~
 \includegraphics[width=.3\textwidth]{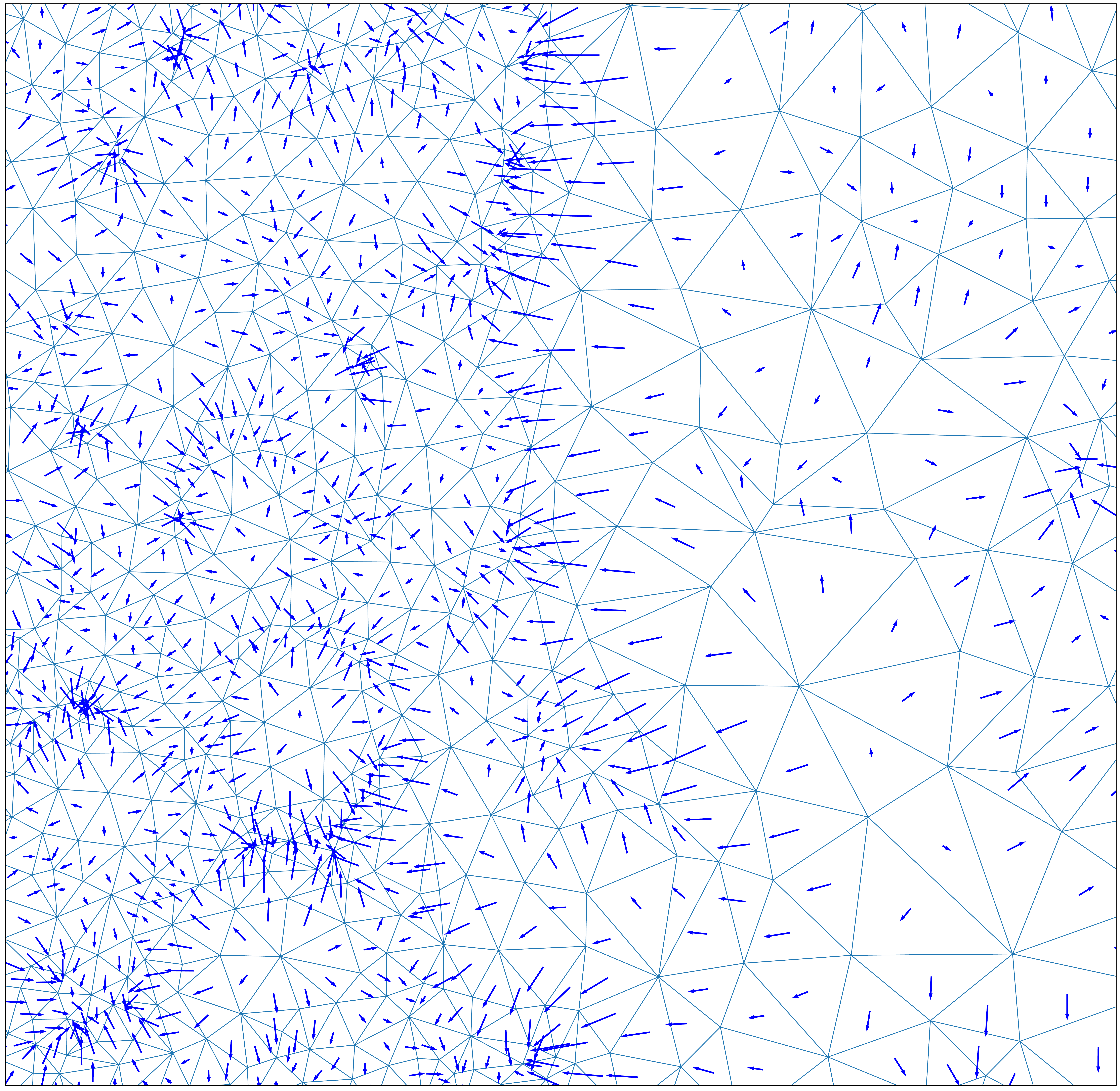}
\caption{\label{fig:vectors_averaged} Left panel: Rescaled gradient vectors $\tilde{\vec{G}}_\alpha$ computed via triangulation wombling, after applying one Lloyd iteration to the original data. Middle panel: the result $\langle{\vec{G}}\rangle_i$ from averaging the gradient vectors $\tilde{\vec{G}}_\alpha$ shown in the left panel at each data point $P_i$ according to eq.~(\ref{ave}). Right panel: the result $\langle{\vec{G}}\rangle_\alpha$ from further averaging the Voronoi-averaged gradient vectors $\langle{\vec{G}}\rangle_i$ shown in the middle panel at each Delaunay triangle according to eq.~(\ref{superave}).}
\end{figure}

\begin{figure}[t]
 \centering
 \includegraphics[height=.28\textwidth]{lsu_plot_4sbr_1lloyds}~~
 \includegraphics[height=.28\textwidth]{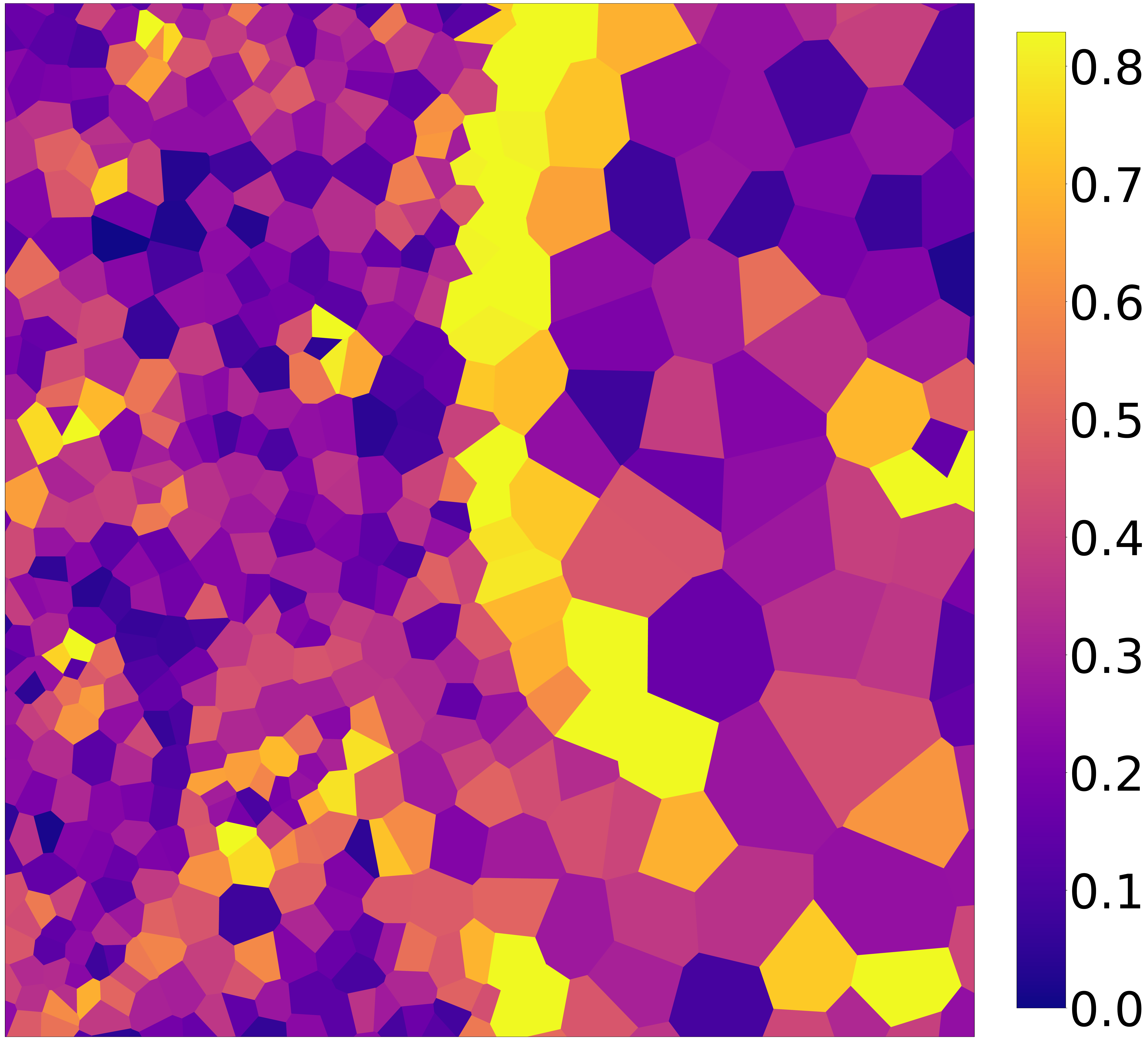}~~
 \includegraphics[height=.28\textwidth]{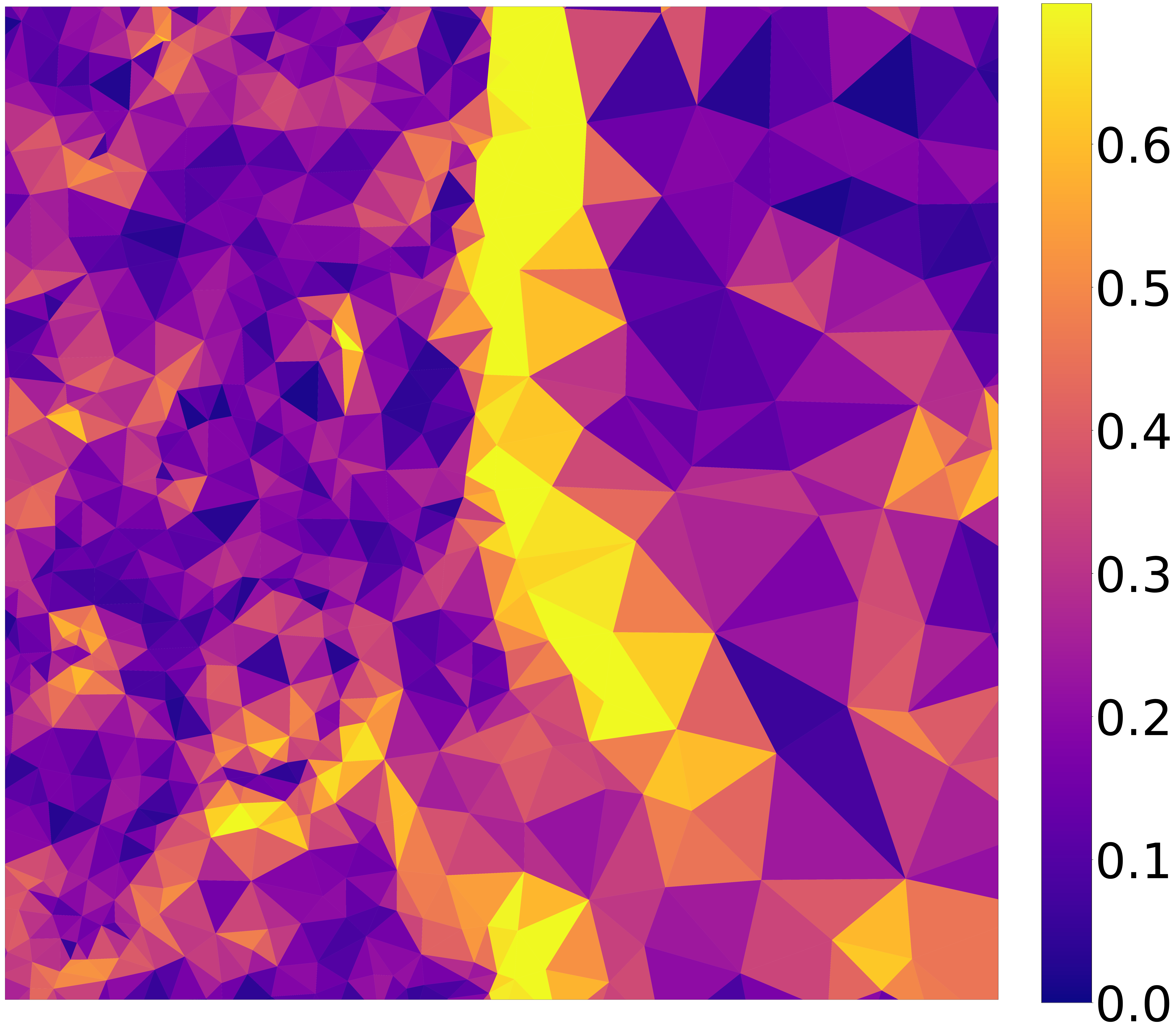}
\caption{\label{fig:lsu_averaged} Delaunay (left and right panels) or Voronoi (middle panel) triangulations of the data after applying one Lloyd iteration, with individual cells color-coded by the magnitudes of their respective rescaled gradient vectors computed from triangulation wombling (left panel), Voronoi-averaged according to eq.~(\ref{ave}) (middle panel), or Delaunay-averaged according to eq.~(\ref{superave}) (right panel).
}
\end{figure}

In our case here, we are dealing with the Delaunay tessellation instead, where the procedures of triangulation wombling (\ref{Gvec}) and rescaling (\ref{Grescaling}) allow us to directly compute the rescaled gradient vector $\tilde{\vec{G}}_\alpha$ associated with each Delaunay triangle $D_\alpha$. The result (after one Lloyd iteration) is shown in the left panel of Fig.~\ref{fig:vectors_averaged}, where we plot each vector $\tilde{\vec{G}}_\alpha$ at the centroid of the corresponding Delaunay triangle. This plot is the vector field analogue of the color map shown in the second panel of the upper row of Fig.~\ref{fig:lsu_lloyd}, which for ease of comparison we reproduce again in the left panel of Fig.~\ref{fig:lsu_averaged}. The difference is that the color map plots in Figs.~\ref{fig:lsu_lloyd} and \ref{fig:lsu_averaged} identify cells only by the magnitude  $\tilde{G}_\alpha$ of the gradient, while the vector field plots of Fig.~\ref{fig:vectors_averaged} include the directional information as well, which is useful in visualizing the spatial patterns and correlations of the gradient vectors.

Now, given the rescaled gradient vectors $\tilde{\vec{G}}_\alpha$ shown in the left panel of Fig.~\ref{fig:vectors_averaged}, there are two ways to perform local averaging of these vectors, depending on whether we want to associate the result from the averaging with a Voronoi cell (i.e., a Delaunay vertex) or with a Delaunay triangle:
\begin{itemize}
\item {\em Voronoi cell averaging.} Recall that according to the duality relation (\ref{dual1}), each Voronoi cell $V_i$ corresponds to a Delaunay vertex $D_{\alpha_1\alpha_2\alpha_3...\alpha_{N_i}}$, which is the common vertex of $N_i$ Delaunay triangles $D_{\alpha_1}, D_{\alpha_2}, D_{\alpha_3}, \ldots, D_{\alpha_{N_i}}$. Thus we can simply define the average gradient $\langle{\vec{G}}\rangle_i $ at any Voronoi cell $V_i$ to be\footnote{As an alternative to a simple sum as in (\ref{ave}), one could assign a weight for each vector $ \tilde {\vec{G}}_{\alpha_k}$, for example, the angular size of the $D_{\alpha_k}$ triangle as seen from the point $P_i$.}
\beq
\langle{\vec{G}}\rangle_i = \frac{1}{N_i}\sum_{k=1}^{N_i} \tilde {\vec{G}}_{\alpha_k}.
\label{ave}
\eeq
\item {\em Delaunay cell averaging.} Once we have the averaged vectors (\ref{ave}) at our disposal, we can go back to each Delaunay triangle and further average the three vectors (\ref{ave}) associated with its three vertices. 
\beq
\langle{\vec{G}}\rangle_\alpha = \frac{1}{3} \left(\langle{\vec{G}}\rangle_i + \langle{\vec{G}}\rangle_j + \langle{\vec{G}}\rangle_k \right), 
\label{superave}
\eeq
where the indices $i$, $j$ and $k$ label the data points at the vertices of $D_\alpha$ (recall the duality relation (\ref{dual3})). \end{itemize}

The vector fields resulting from the averaging prescriptions in eqs.~(\ref{ave}) and (\ref{superave}) are shown in the middle and right panels of Fig.~\ref{fig:vectors_averaged}, respectively. One can see that the statistical fluctuations are indeed getting suppressed as a result of the averaging, and furthermore, the directions of the gradient vectors near the boundary are becoming better correlated with each successive iteration. This directional correlation will become important in the next two sections where we shall compare the properties of {\em neighboring} cells in the tessellation. For now, in order to demonstrate the benefits from the averaging procedures (\ref{ave}) and (\ref{superave}) for the purposes of boundary detection, it is sufficient to update the color maps from Fig.~\ref{fig:lsu_lloyd} using the magnitudes of the {\em averaged} gradients instead. This is done in the middle and right panels of Fig.~\ref{fig:lsu_averaged}, where the individual cells in the tessellation have been color-coded by the magnitudes of the Voronoi-averaged gradient (\ref{ave}) and the Delaunay-averaged gradient (\ref{superave}), respectively. By comparing Fig.~\ref{fig:lsu_averaged} to Fig.~\ref{fig:lsu_lloyd}, we see that the local averaging procedures produce comparable benefits to Voronoi relaxation, so that we can view the two procedures as alternatives to the other. More specifically, local averaging seems to be at least equivalent to (if not better than) running on the order of 5-7 Lloyd iterations, which seemed to be the optimal choice in Figs.~\ref{fig:lsu_lloyd} and \ref{fig:after_cut_lloyd}. Of course, the two methods can also be applied simultaneously, so that their benefits can be optimally exploited. In our analysis of Sections \ref{sec:womblingdiscrete} and \ref{sec:significance}, unless specified otherwise, we shall choose to employ the Delaunay-averaged gradient vectors (\ref{superave}).

\section{Tagging elements of the tessellation as boundary candidates}
\label{sec:tagging}

Armed with the various estimates of the local gradient vectors discussed in the previous section, we are now ready for the next step of the wombling algorithm, namely, the tagging of Voronoi or Delaunay cells as boundary candidates. The standard approach is to place a lower cut on the relevant variable (typically the magnitude) which measures the size of the gradient. This selection singles out a certain set of candidate boundary cells as shown in Fig.~\ref{fig:after_cut_lloyd}. The purpose of this section is to study how effective\footnote{Selection efficiency is typically illustrated with ROC curves, where one varies the cut on the selection variable and plots the fraction of signal versus the fraction of background surviving the cut. This was also the procedure used in Refs.~\cite{Debnath:2015wra,Debnath:2016mwb}. Here, however, we prefer to simply show scatter plots of the tagging variable versus the distance to the boundary. In this way, we avoid the need to define what exactly is meant by a ``boundary'' object versus a ``bulk'' object.} this selection is and to suggest a potential improvement of the standard approach by utilizing the correlations between gradient vectors computed in {\em neighboring} cells. The idea will be to place a premium not just on cells whose own gradient $\tilde{\vec{G}}_\alpha$ has a large magnitude, but on cells where the neighboring gradients $\tilde{\vec{G}}_\beta$ have both a) large magnitudes and b) correlated directions with $\tilde{\vec{G}}_\alpha$. A convenient variable which captures the desired correlations between two vectors $\tilde{\vec{G}}_\alpha$ and $\tilde{\vec{G}}_\beta$ is their dot product,  $\tilde{\vec{G}}_\alpha\cdot \tilde{\vec{G}}_\beta$ \cite{Debnath:2015wra}.

\begin{figure}[t]
 \centering
 \includegraphics[width=.45\textwidth]{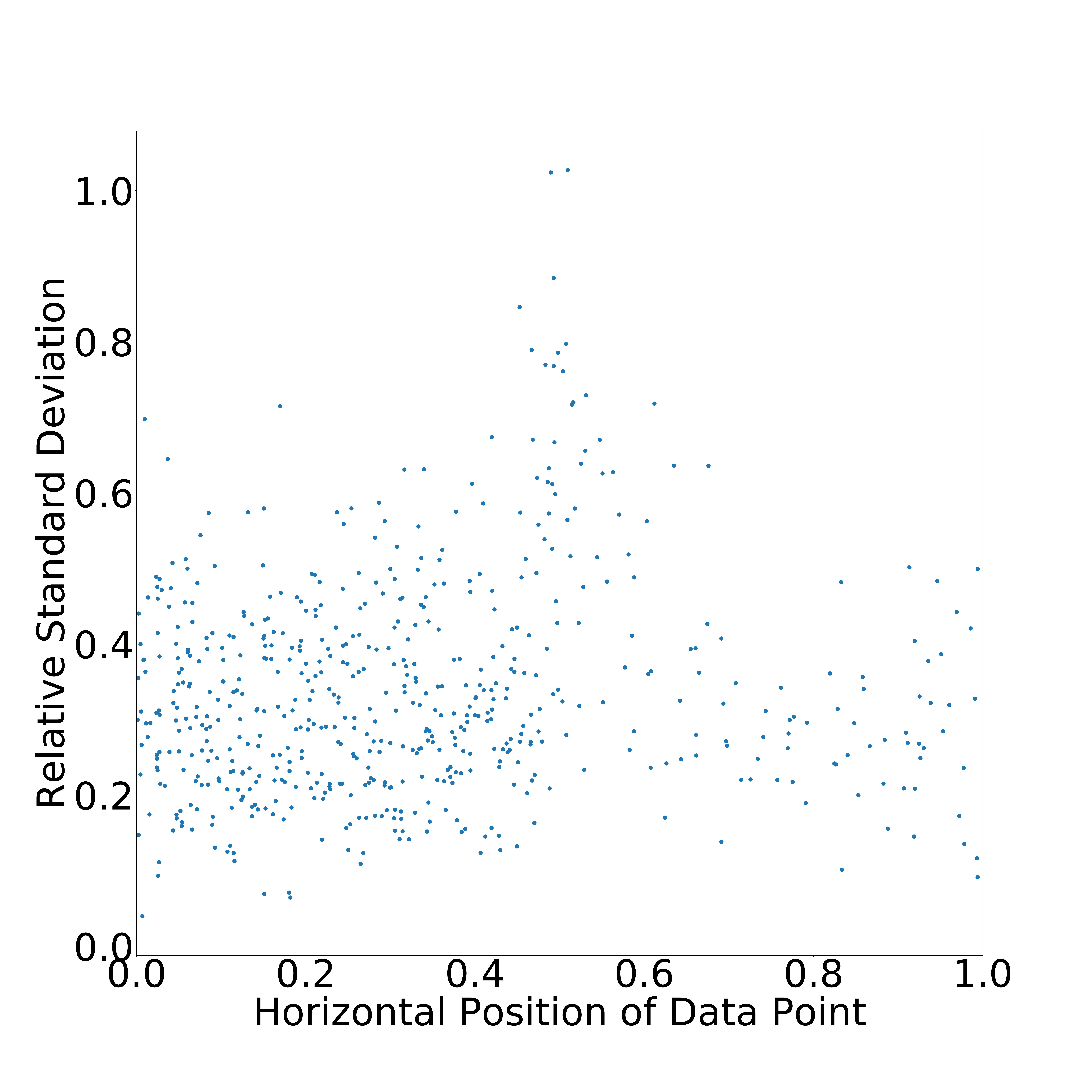}
 \includegraphics[width=.45\textwidth]{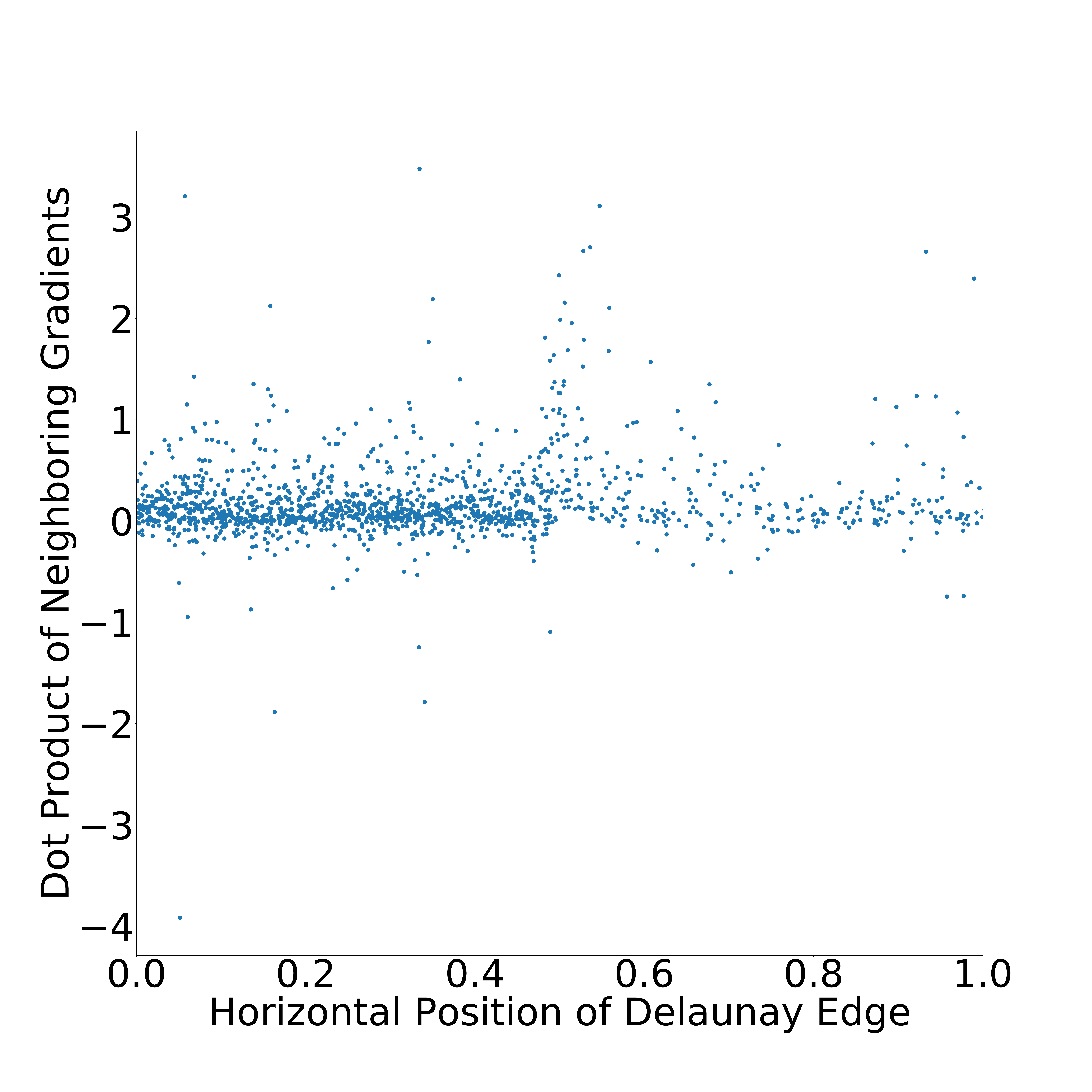}\\
 \includegraphics[width=.45\textwidth]{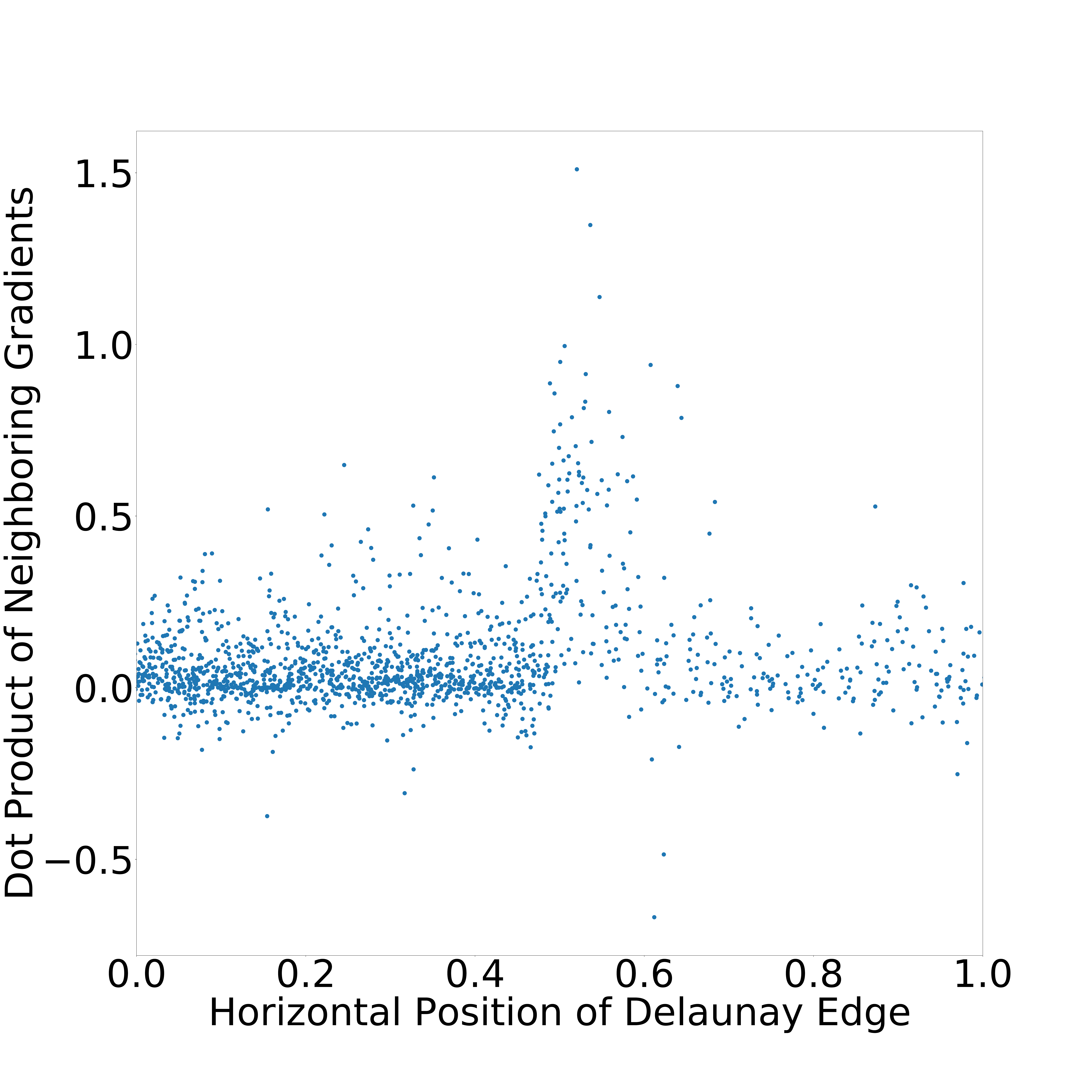}
 \includegraphics[width=.45\textwidth]{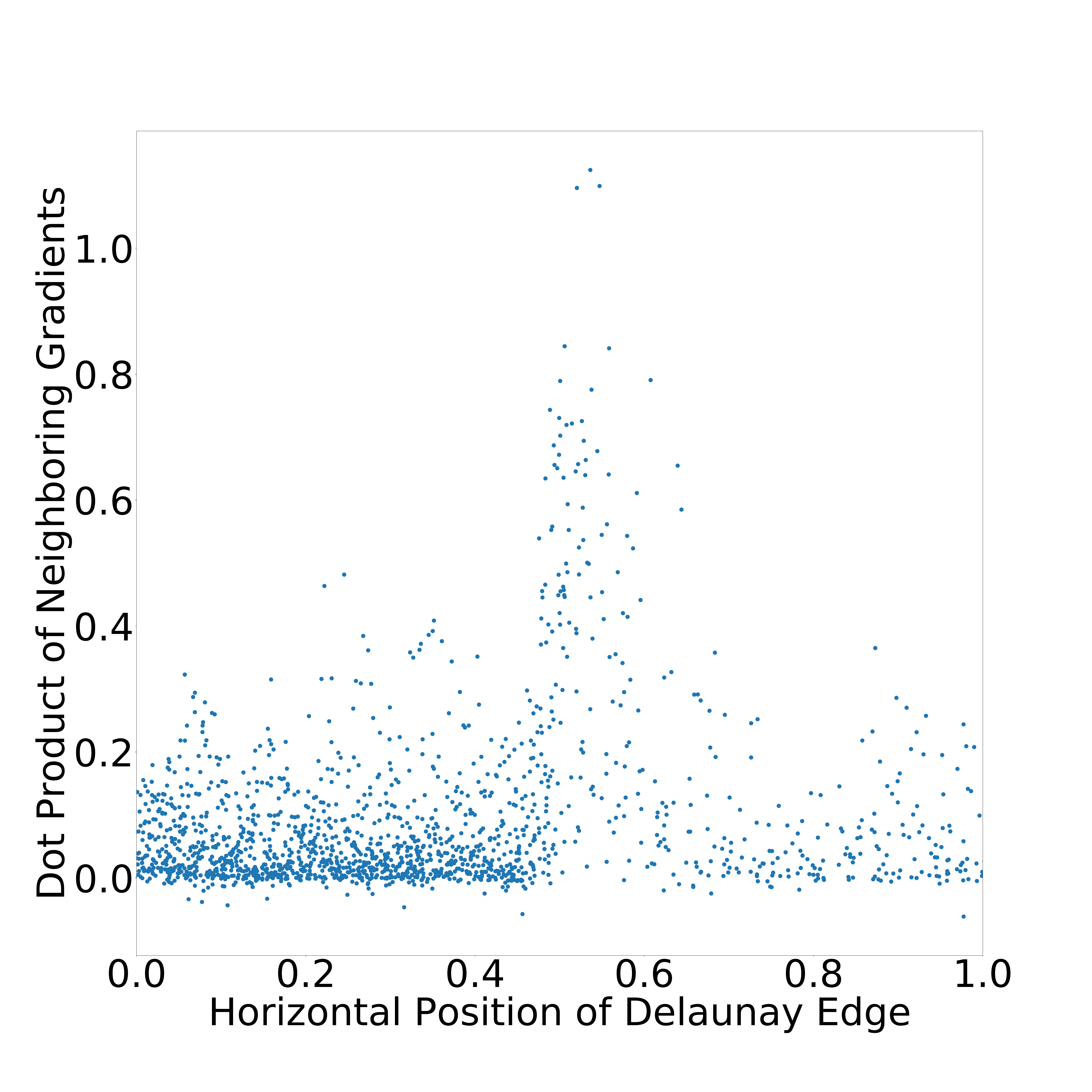}
\caption{\label{fig:taggers} Scatter plots of different tagging variables discussed in the text versus the horizontal position of the corresponding element in the tessellation. The upper left panel shows the relative standard deviation $\bar\sigma_i$ defined in (\ref{defvar}) versus the horizontal position $x_i$ of the generator point $P_i$. The other three panels show dot products of different  versions of neighboring gradient vectors versus the horizontal position $(x_i+x_j)/2$ of the midpoint of the respective Delaunay edge $D_{\alpha\beta}$: the rescaled gradients 
$\tilde{\vec{G}}_\alpha\cdot \tilde{\vec{G}}_\beta$ (upper right panel),
the Voronoi-averaged gradients $\langle{\vec{G}}\rangle_i\cdot \langle{\vec{G}}\rangle_j$ (lower left panel) and
the Delaunay-averaged gradients $\langle{\vec{G}}\rangle_\alpha\cdot \langle{\vec{G}}\rangle_\beta$ (lower right panel). }
\end{figure}

Fig.~\ref{fig:taggers} shows scatter plots of such dot products of neighboring vectors for the three types of gradients introduced in the previous section: $\tilde{\vec{G}}_\alpha\cdot \tilde{\vec{G}}_\beta$ (upper right panel), $\langle{\vec{G}}\rangle_i\cdot \langle{\vec{G}}\rangle_j$ (lower left panel) and $\langle{\vec{G}}\rangle_\alpha\cdot \langle{\vec{G}}\rangle_\beta$ (lower right panel). In each case, the result is plotted versus the horizontal position $(x_i+x_j)/2$ of the midpoint of the respective\footnote{In the case of $\tilde{\vec{G}}_\alpha\cdot \tilde{\vec{G}}_\beta$ and $\langle{\vec{G}}\rangle_\alpha\cdot \langle{\vec{G}}\rangle_\beta$, the relevant Delaunay edge is simply $D_{\alpha\beta}$, i.e., the edge separating the Delaunay triangles $D_\alpha$ and $D_\beta$, while in the case of $\langle{\vec{G}}\rangle_i\cdot \langle{\vec{G}}\rangle_j$ the relevant Delaunay edge is the one dual to the Voronoi edge $V_{ij}$, see eq.~(\ref{dual2}).} Delaunay edge $D_{\alpha\beta}$. For comparison, in the upper left panel we show a scatter plot of the relative standard deviation $\bar\sigma_i$ defined in (\ref{defvar}) versus the horizontal position $x_i$ of the corresponding generator point $P_i$. As explained in Sec.~\ref{sec:voronoigradients}, the relative standard deviation $\bar\sigma_i$ is constructed from the Voronoi tessellation and was found to perform best among several other Voronoi-constructed alternatives \cite{Debnath:2015wra}. The upper left panel in Fig.~\ref{fig:taggers} confirms that the highest values of $\bar\sigma_i$ are indeed found for cells near the boundary; in fact, the top 16 highest $\bar\sigma_i$ values belong to such cells. At the same time, we also observe a significant variation in the $\bar\sigma_i$ values for cells in the bulk; for $\bar\sigma_i < 0.7$ this starts introducing a certain number of false positives.

The remaining three panels of Fig.~\ref{fig:taggers} demonstrate that the corresponding dot products of gradients computed from the Delaunay tessellation are also efficient in identifying boundary objects (in this case, Delaunay edges). Among the three options illustrated in the plot, the dot product of the Delaunay-averaged gradients seems to perform the best --- the top 47 highest dot products of neighboring $\langle{\vec{G}}\rangle_\alpha$ vectors belong to Delaunay edges near the boundary; and there is a well defined cluster of points with large $y$ values in the boundary region $0.45<x<0.6$. Note the different $y$-axis range on these three plots --- the variation of dot product values is largest for $\tilde{\vec{G}}_\alpha\cdot \tilde{\vec{G}}_\beta$ and smallest for $\langle{\vec{G}}\rangle_\alpha\cdot \langle{\vec{G}}\rangle_\beta$, further demonstrating the beneficial effects from the averaging procedures discussed in Sec.~\ref{sec:averaging}.

\begin{figure}[t]
 \centering
 \includegraphics[width=.65\textwidth]{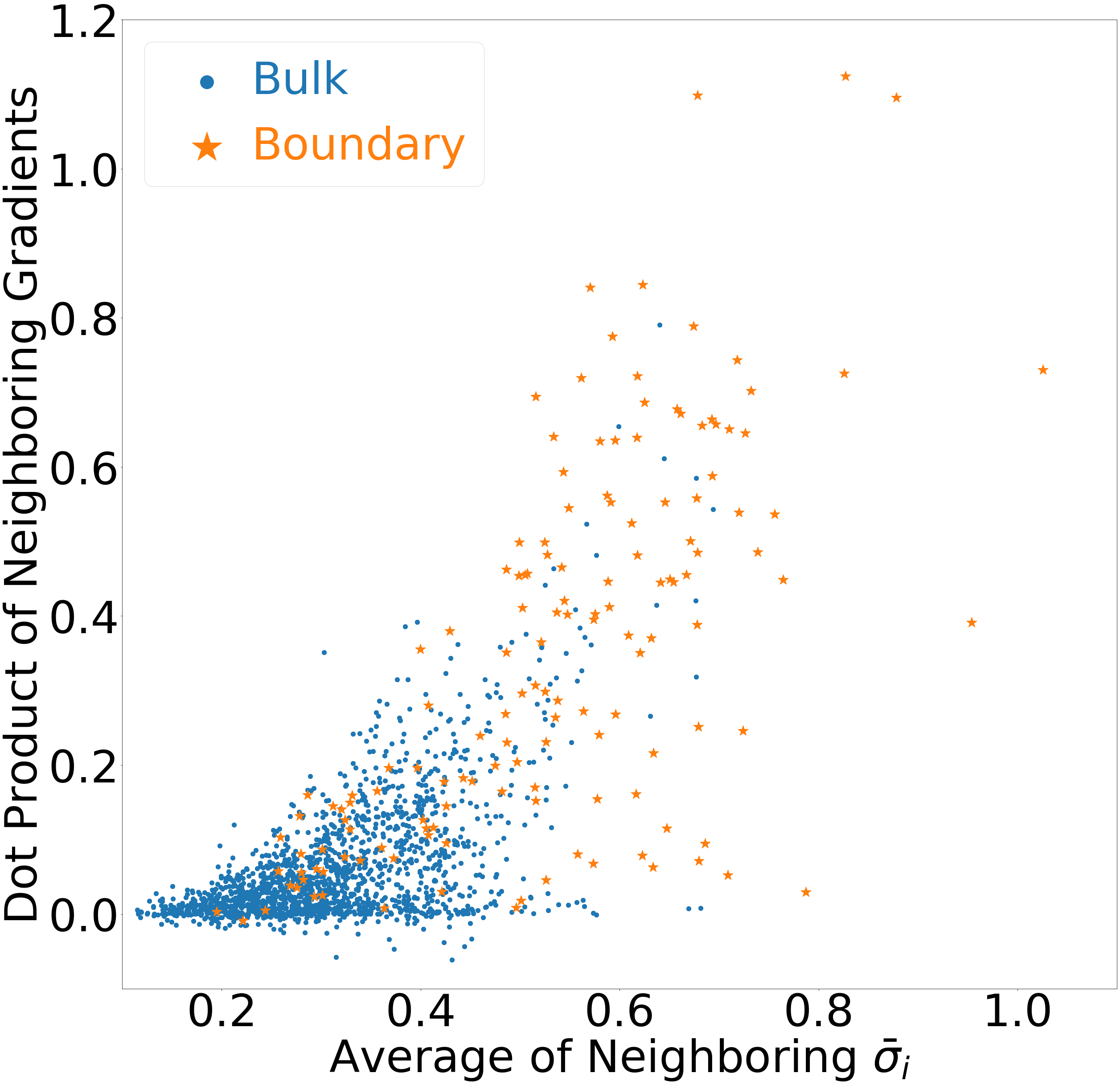}
\caption{\label{fig:dot_sigma} Scatter plot showcasing the correlation between two of the tagging variables from Fig.~\ref{fig:taggers} --- the relative standard deviation ($x$-axis) and the dot product $\langle{\vec{G}}\rangle_\alpha\cdot \langle{\vec{G}}\rangle_\beta$ of Delaunay-averaged gradients ($y$-axis). The orange stars correspond to Delaunay edges $D_{\alpha\beta}$ whose midpoint is located close to the boundary ($0.47 < x < 0.57$), while blue circles represent the remaining edges further away. Since each Delaunay edge $D_{\alpha\beta}$ is dual to two Voronoi points $P_i$ and $P_j$, the quantity shown on the $x$-axis is the average $(\bar\sigma_i+\bar\sigma_i)/2$ of the respective relative standard deviations $\bar\sigma_i$ and $\bar\sigma_j$. }
\end{figure}

Comparing the top left panel in Fig.~\ref{fig:taggers} to the other three panels, we conclude that the gradient dot products, which take advantage of the correlations between neighboring gradient vectors in terms of both direction and magnitude, are able to identify the boundary better than $\bar\sigma_i$ and similar variables (Fig.~\ref{fig:dot_sigma} shows a direct comparison between the relative standard deviation and the dot product $\langle{\vec{G}}\rangle_\alpha\cdot \langle{\vec{G}}\rangle_\beta$ of Delaunay-averaged gradients). As a byproduct of this new method of tagging, we have also automatically built up a network of associations among the Delaunay cells in the triangulation, which is readily available for use in the next step (agglomeration), where one attempts to construct the actual boundary. This is the subject of the next section.

\section{Agglomeration of tagged boundary elements}
\label{sec:agglomeration}

\begin{figure}[t]
 \centering
 \includegraphics[width=.3\textwidth]{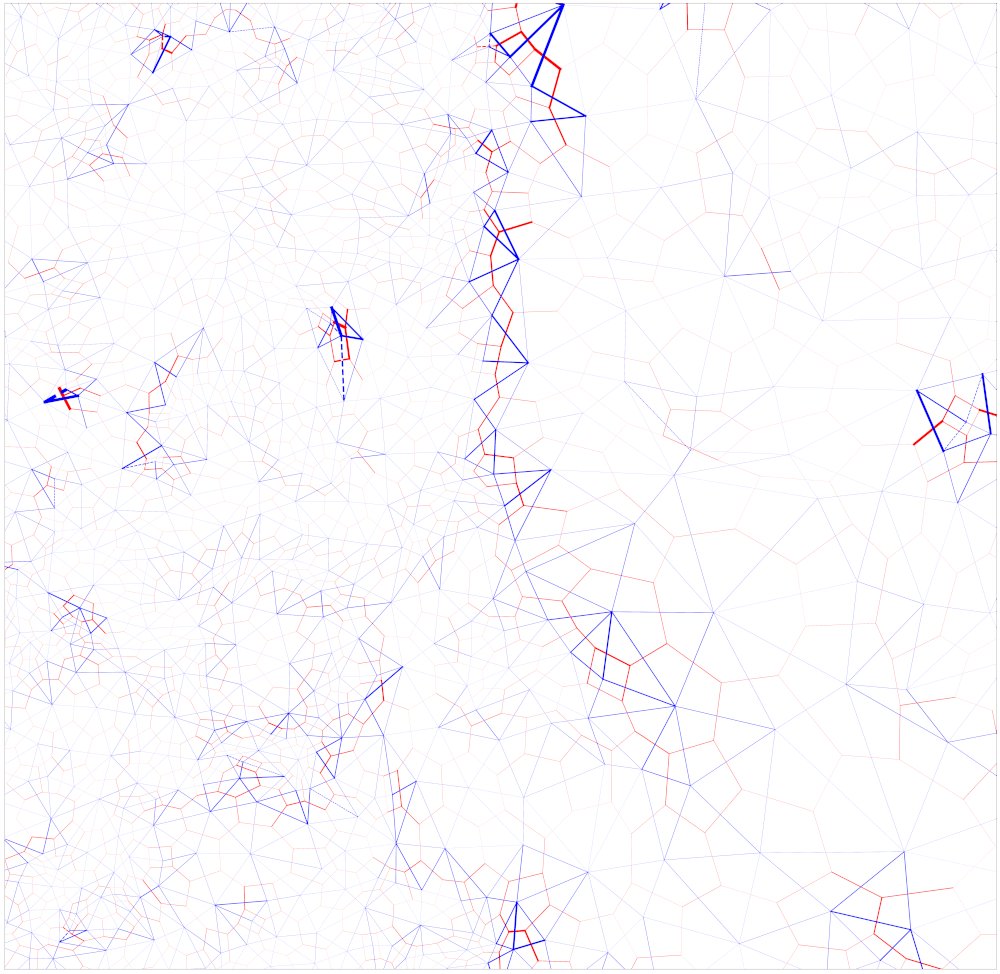}~~
 \includegraphics[width=.3\textwidth]{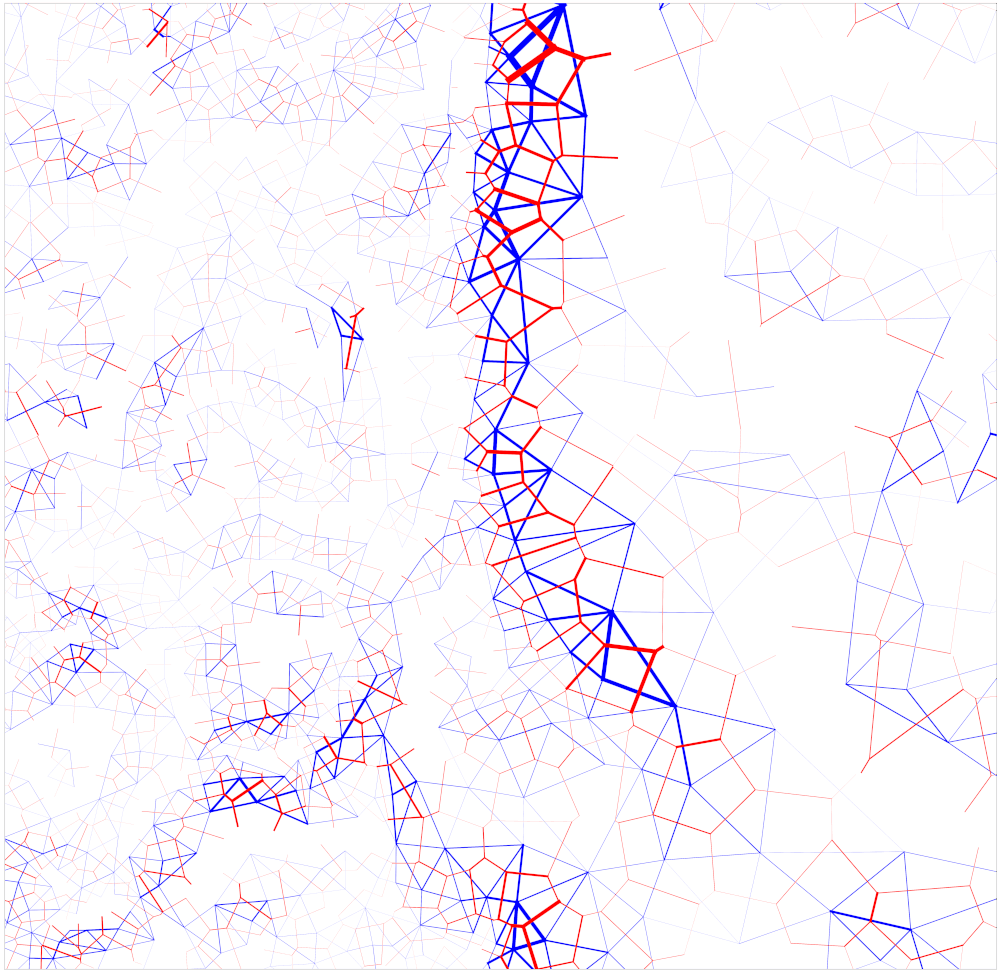}~~
 \includegraphics[width=.3\textwidth]{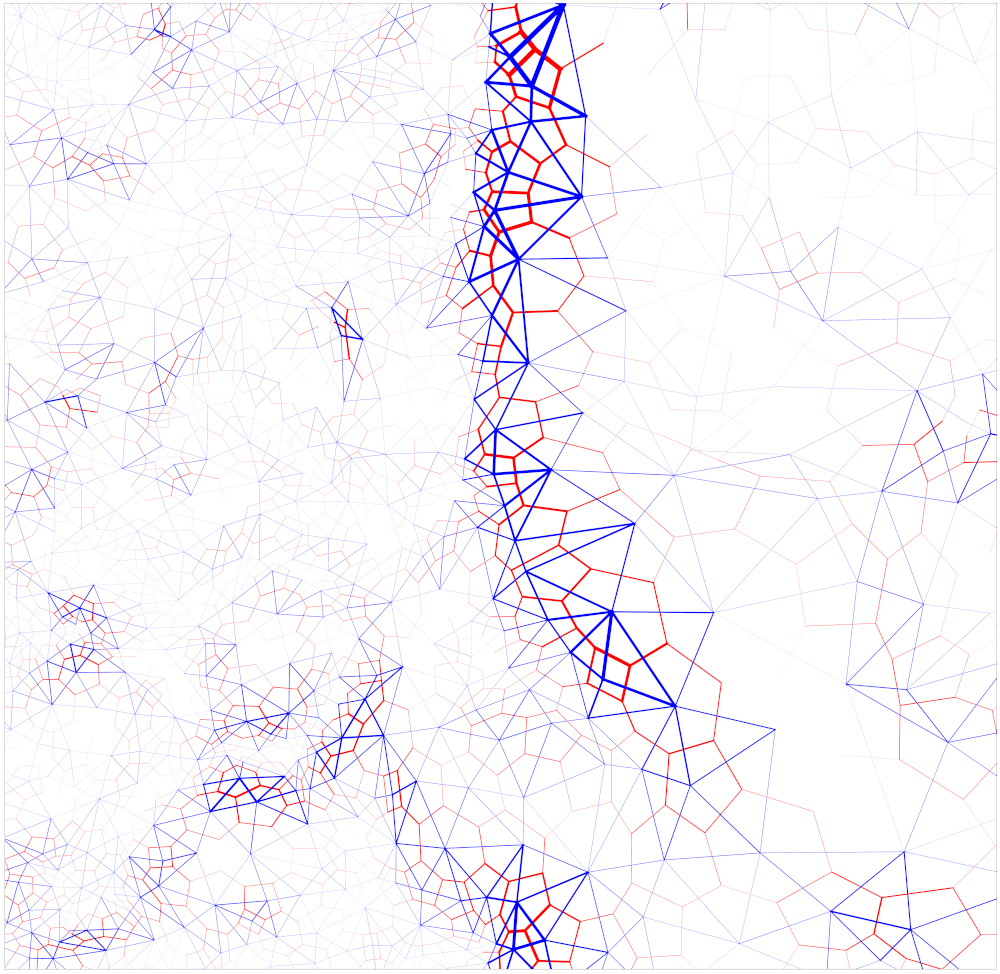}
\caption{\label{fig:spiderwebs} Weighted network representations of the Voronoi tessellation (red lines) and the Delaunay triangulation (blue lines). The line thickness is proportional to the weight, which is given by $\tilde{\vec{G}}_\alpha\cdot \tilde{\vec{G}}_\beta$ (left panel), $\langle{\vec{G}}\rangle_i\cdot \langle{\vec{G}}\rangle_j$ (middle panel) and $\langle{\vec{G}}\rangle_\alpha\cdot \langle{\vec{G}}\rangle_\beta$ (right panel). Edges with negative weights are not shown.}
\end{figure}

The previous step (the tagging of boundary elements) typically fails to result in a continuous boundary, especially in case of weak signals. Instead, the algorithm produces a collection of scattered ``islands'' of tagged cells, as seen in the right panels of Figs.~\ref{fig:unscaled_gradients} and \ref{fig:scaled_gradients} and in the top panels of Fig.~\ref{fig:after_cut_lloyd}. This necessitates the next step of agglomerating the individual tagged cells into subgraphs and evaluating whether the resulting pattern is consistent with a linear boundary \cite{DaleFortin}. The downside of this approach is that it does not take into further consideration the cells which have failed the tagging cut --- those cells are simply ignored from this point on. Another potential drawback is that the tagging and agglomeration steps are done independently from each other, so that the existence of any spatial correlations among neighboring cells is not being used during the tagging. As already mentioned in the previous section, both of these problems are avoided when we use the dot products of neighboring gradients as tagging variables. As illustrated in the last three panels of Fig.~\ref{fig:taggers}, each dot product of gradients can be uniquely associated with a Delaunay edge $D_{\alpha\beta}$ or with its dual Voronoi edge $V_{ij}$, see eq.~(\ref{dual2}). We can then treat the original Voronoi and Delaunay tessellations as {\em weighted networks}, where each edge is assigned a weight equal to the dot products of the corresponding two gradients. This weighted network representation is illustrated in Fig.~\ref{fig:spiderwebs}, where we superimpose the Voronoi tessellation (red lines) and the Delaunay triangulation (blue lines). The weight of an edge is indicated by the line thickness --- thicker lines imply higher weights and vice versa. The three panels show three different ways to compute the weights from the gradient dot products, depending on which set of gradient vectors from Fig.~\ref{fig:vectors_averaged} we choose to use: $\tilde{\vec{G}}_\alpha\cdot \tilde{\vec{G}}_\beta$ (left panel), $\langle{\vec{G}}\rangle_i\cdot \langle{\vec{G}}\rangle_j$ (middle panel) or $\langle{\vec{G}}\rangle_\alpha\cdot \langle{\vec{G}}\rangle_\beta$ (right panel). As seen in Fig.~\ref{fig:taggers}, a certain fraction of dot products are negative; if that is the case, the corresponding edge is not plotted.

The three panels in Fig.~\ref{fig:spiderwebs} can be contrasted with the corresponding results in Fig.~\ref{fig:lsu_averaged}, where we used just the magnitudes of the individual gradient vectors, without any reference to their neighbors. The boundary seems to be better outlined in Fig.~\ref{fig:spiderwebs}, particularly when we make use of the averaging procedures from Sec.~\ref{sec:averaging}. We also note the benefit of plotting the Voronoi and Delaunay tessellations simultaneously --- the orientation of the edges with respect to the boundary is random, so whenever a given edge happens to be orthogonal to the boundary, its dual tends to be parallel to it, so taken together, they trace out the correct shape of the boundary.  
 
Fig.~\ref{fig:spiderwebs} also elucidates the results from Fig.~\ref{fig:taggers}, where we observed that while many edges situated close to the boundary enjoyed relatively large values of their gradient dot products, there was also a non-negligible fraction of edges near the boundary with rather low values of the gradient dot products. Fig.~\ref{fig:spiderwebs} now reveals that these two populations are spatially correlated --- note how the edges with large values of the gradient dot products are linked together, as are their counterparts. This confirms that using the gradient dot products as tagging variables automatically also takes care of the agglomeration. 

Until now we have been following the standard steps of a wombling analysis. As already mentioned, the last remaining step is to evaluate the statistical significance of the observed pattern of tagged boundary candidates. Note that the last three figures illustrate tagging procedures for each of the four types of boundary candidate objects defined in Sec.~\ref{sec:boundaryobjects}: Voronoi cells (middle panel in Fig.~\ref{fig:lsu_averaged} and upper left panel in Fig.~\ref{fig:taggers}), Delaunay cells (left and right panels in Fig.~\ref{fig:lsu_averaged}), Delaunay edges (upper right and lower panels in Fig.~\ref{fig:taggers} and all three panels in Fig.~\ref{fig:spiderwebs}) and Voronoi edges (Fig.~\ref{fig:spiderwebs}). Of course, since the Voronoi and Delaunay edges are dual to each other, any procedure which can tag one edge type can also be applied to tag the other.

In the next three sections we shall outline an alternative approach originally proposed in \cite{Banerjee2006}, which allows us to perform the tagging, agglomeration and statistical evaluation steps in one go. In order to introduce and illustrate the method, in the next section Sec.~\ref{sec:womblingcontinuous} we shall start with the case of a continuously defined function $f(x,y)$ and then proceed to analyze the case of point data in Secs.~\ref{sec:womblingdiscrete} and \ref{sec:significance}.

\section{Finding wombling boundaries: analytical examples}
\label{sec:womblingcontinuous}

In order to bypass the tagging and agglomeration steps, Ref.~\cite{Banerjee2006} proposed to directly consider various curves $C$ in the $(x,y)$ plane, and to associate a ``wombling measure'' $\Gamma$ with each one, so that true wombling boundaries can be identified by their large values of $\Gamma$. Since a wombling boundary is supposed to represent a zone of rapid change in the function $f$, it is natural to define the wombling measure in terms of the local gradient $\vec{\nabla} f$ suitably integrated along $C$. In particular, Ref.~\cite{Banerjee2006} defined $\Gamma$ to be the total gradient flux through $C$
\beq
\Gamma[C] \equiv \int_C \left( \vec\nabla f \cdot \hat{n}_C \right) d\ell,
\label{eq:Gammadef}
\eeq
where $\hat{n}_C$ is a unit vector normal to the curve $C$ and $d\ell$ is the infinitesimal length along $C$. Additionally, Ref.~\cite{Banerjee2006} also considered the average gradient flux 
\beq
\bar\Gamma[C] \equiv \frac{\int_C \left( \vec\nabla f \cdot \hat{n}_C \right) d\ell}{\int_C d\ell},
\label{eq:Gammaavedef}
\eeq
where using the total length of the curve as a normalization factor eliminates the unfair advantage of curves which happen to be too long. For this reason, in what follows we shall make use of (\ref{eq:Gammaavedef}) and not (\ref{eq:Gammadef}).\footnote{An additional variation mentioned in \cite{Banerjee2006} was to consider integrating the flux in absolute value, e.g., 
$$
\Gamma[C] \equiv \int_C \left| \vec\nabla f \cdot \hat{n}_C \right| d\ell,
$$
in order to avoid cases where a large positive flux over one section of the curve $C$ is cancelled by a large negative flux over another section. However, in our case such cancellations are welcome since the noise fluctuations are random and we would like to allow them to cancel out each other as much as possible. We have confirmed numerically in our examples that the absolute value alternative leads to lower sensitivity.} 

Without any further constraints on the type of curves $C$ that we are allowed to consider, this method would be rather impractical. To make further progress, two approaches are possible. The first one is the model-dependent route --- if we specify exactly what kind of new physics model generates the wombling boundary, then $C$ can be specified by only a handful of parameters (typically the masses of the new particles). Then the problem of maximizing the functional (\ref{eq:Gammaavedef}) over all possible curves $C$ reduces to a simple global maximization problem in the parameter space describing $C$ \cite{Debnath:2016gwz}. Here, however, we would like to stay as model-independent as possible, so we shall not assume any specific parametrization of the boundary. At the same time, we do not want to consider arbitrarily general curves $C$ either. 

An intermediate compromise approach is the following. Note that any curve $C$ can locally be approximated by a straight line segment. Therefore, we can perform a scan of the $(x,y)$ plane where at each point $P$ we try a line segment (centered on $P$) of fixed length $L$ and arbitrary angular orientation $\varphi$. Each point in the so-defined 4-dimensional parameter space $(x,y,L,\varphi)$ corresponds to a well-defined line segment, for which the wombling measure (\ref{eq:Gammaavedef}) can be calculated. The regions in $(x,y,L,\varphi)$ parameter space with large values for $\bar\Gamma[C]$ would then identify (segments of) the wombling boundary. Since this procedure involves optimization in a 4-dimensional parameter space, it will be difficult to illustrate here. This is why from now on we choose to focus on the $(L,\varphi)$ subspace --- one can think of this as first zooming in on an interesting region of the $(x,y)$ plane and then testing for the presence of a linear wombling line segment.

\begin{figure}[t]
 \centering
 \includegraphics[width=.8\textwidth]{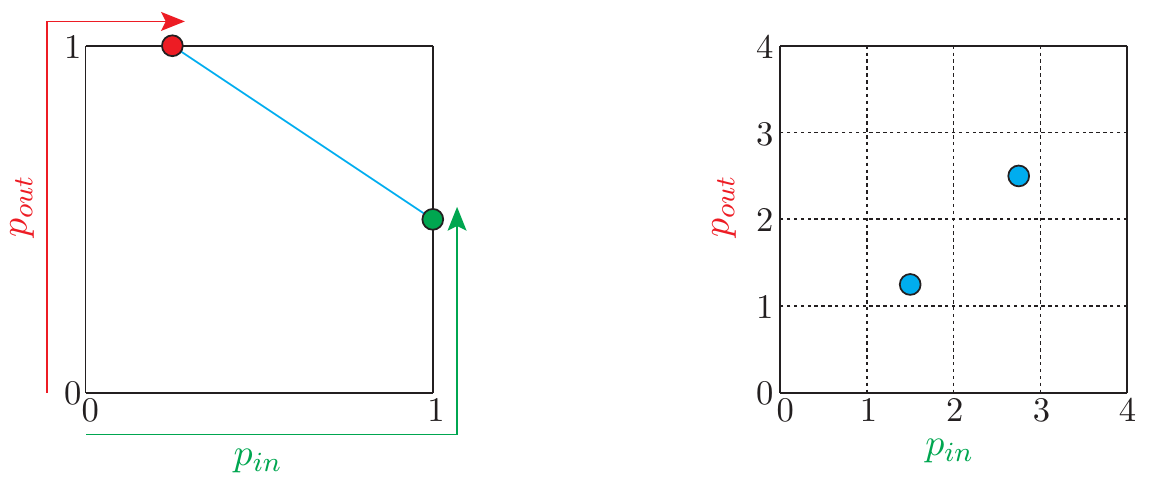}
\caption{\label{fig:perimeter} Our parametrization of all possible straight lines intersecting the unit square in terms of the coordinate $p_{in}$ of the entry point  and the coordinate $p_{out}$ of the exit point, where $p_{in}$ ($p_{out}$) is measured counterclockwise (clockwise) along the perimeter. The right panel shows the complete $(p_{in},p_{out})$ parameter space, which is a square of side length 4. The blue dot represents the blue example line shown in the left panel and its partner (when the line is traversed in reverse).}
\end{figure}

Our reparametrization of the remaining two degrees of freedom describing the line segment is illustrated in Fig.~\ref{fig:perimeter}. As before, we retain the unit square as our field of view. We then consider all possible straight lines crossing the unit square --- each such line can be identified by the point where it enters the square and the point where it exits the square. We shall identify the locations of those two points by their respective coordinates $p_{in}$ and $p_{out}$ measured along the perimeteter, as shown in the left panel of Fig.~\ref{fig:perimeter}. Since the perimeter of a unit square is equal to 4, the parameter space $(p_{in}, p_{out})$ spans the $4\times4$ square shown in the right panel of Fig.~\ref{fig:perimeter} --- any point within that $4\times4$ square can be uniquely associated with a straight line crossing the field of view in one of the two possible directions. For example, $(p_{in}, p_{out})=(0,2)$ describes a diagonal line traversed from the lower left corner to the upper right corner, while $(p_{in}, p_{out})=(2,0)$ describes the same diagonal line covered in reverse. In our previous examples, the true boundary was located at $x=0.5$, and corresponds to either\footnote{In our conventions, reversing the direction of a given line $(p_{in}, p_{out})$ implies $(p_{in}, p_{out})\to (4-p_{out}, 4-p_{in})$. } $(p_{in}, p_{out})=(0.5,1.5)$ or  $(p_{in}, p_{out})=(2.5,3.5)$.

In the remainder of this section and in the next Sec.~\ref{sec:womblingdiscrete}, our main goal will be to compute the wombling measure $\bar\Gamma[C]$ in the $(p_{in}, p_{out})$ parameter space and identify the relevant wombling boundary segment(s). First we shall illustrate this procedure with the example of a continuous function $f(x,y)$ before tackling the case of point datasets in the next section.

\subsection{A straight line boundary}
\label{sec:continuousline}

In this subsection we shall revisit the vertical straight line boundary example from the previous sections. However, we shall not use the original distribution (\ref{fSMfLine}), for two reasons: first, the discontinuity at $x=0.5$ would generate an infinite gradient when computed analytically, and second, the distribution (\ref{fSMfLine}) corresponds to an idealized situation where the effects of particle widths and detector smearing are ignored. In any realistic experimental analysis the sharp step at $x=0.5$ will be smeared and the boundary will be characterized by a large but finite gradient.

\begin{figure}[t]
 \centering
 \includegraphics[width=.45\textwidth]{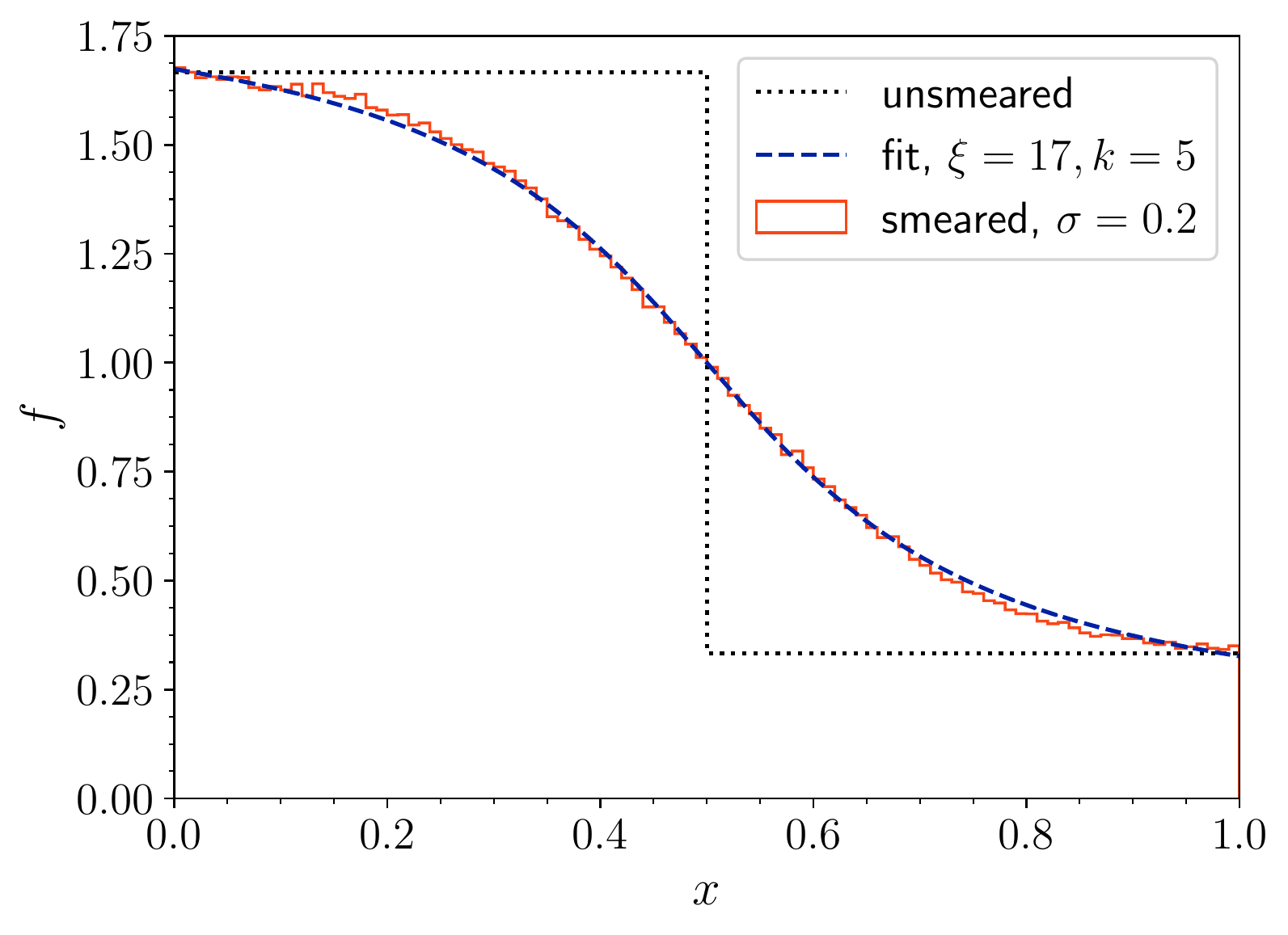}~~~~
 \includegraphics[width=.45\textwidth]{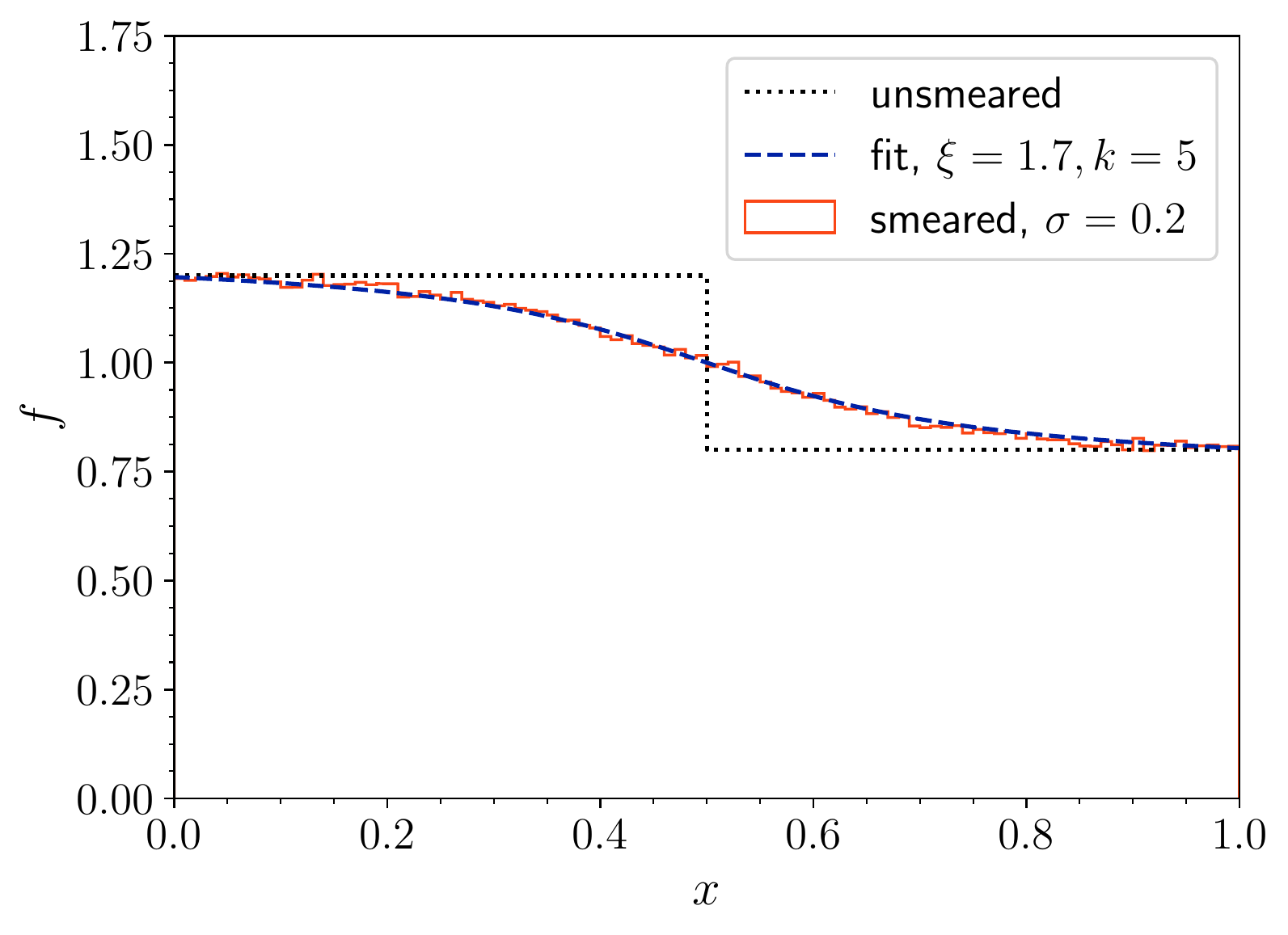}
 \caption{\label{fig:lineexample} The black dotted line shows the original distribution (\ref{fSMfLine}) for $\rho=5$ (left panel) and $\rho=1.5$ (right panel) before any smearing. The red solid histogram is the result from applying Gaussian smearing with $\sigma=0.2$. The blue dashed line is the result from fitting to the ansatz (\ref{eq:smearingfunction}).}
\end{figure}

This is illustrated in Fig.~\ref{fig:lineexample}, where the black dotted lines show (the $x$-dependence of) the original distribution (\ref{fSMfLine}) before any smearing, for $\rho=5$ (left panel) and $\rho=1.5$ (right panel). We then apply Gaussian smearing with $\sigma=0.2$, resulting in the red histograms, which have the typical shapes expected in a realistic experimental analysis. In particular, notice how the gradient at the boundary is significantly reduced as a result of the smearing, making the task of finding the wombling boundary quite challenging. At this point we fit an analytical function to the so obtained smeared distributions. For the fit, we choose to utilize the (unit-normalized) ArcTan sigmoid function 
\beq
g_{res}(x; a, \xi, k) = 1 - \frac{2}{\pi} \frac{\xi-1}{\xi+1} \arctan \Bigl(k(x-a)\Bigr),
\label{eq:arctansigmoid}
\eeq
whose derivative is maximal (in absolute value) at $x=a$, the sharpness of the transition being controlled by the parameter $k$. The remaining parameter $\xi$ is analogous to $\rho$ in the sense that (compare to (\ref{rhodef}))
$$
\xi=\frac{g(x=-\infty)}{g(x=\infty)}.
$$
Since our field of view is limited to $x\in [0,1]$ and the boundary is at $x=0.5$, for the actual fit we choose the parametrization 
\beq
f(x,y) = g_{res}(x; a=0.5, \xi, k)
\label{eq:smearingfunction}
\eeq
and then adjust $\xi$ and $k$ to match the smeared distributions shown by the red histograms. As seen in Fig.~\ref{fig:lineexample}, the fit reproduces the effects of smearing rather well, so in the rest of this subsection we shall use (\ref{eq:smearingfunction}) as our analytically defined distribution.

\begin{figure}[t]
 \centering
 \includegraphics[height=.4\textwidth]{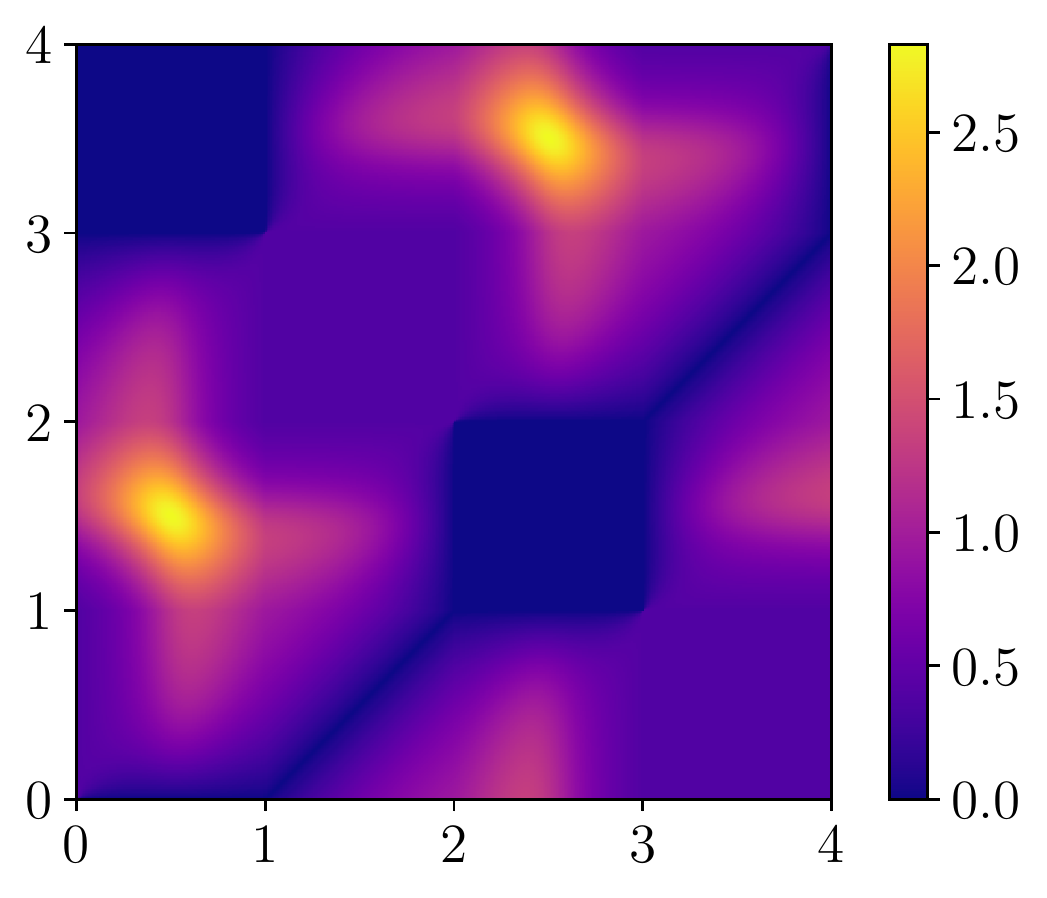}~~~~
 \includegraphics[height=.4\textwidth]{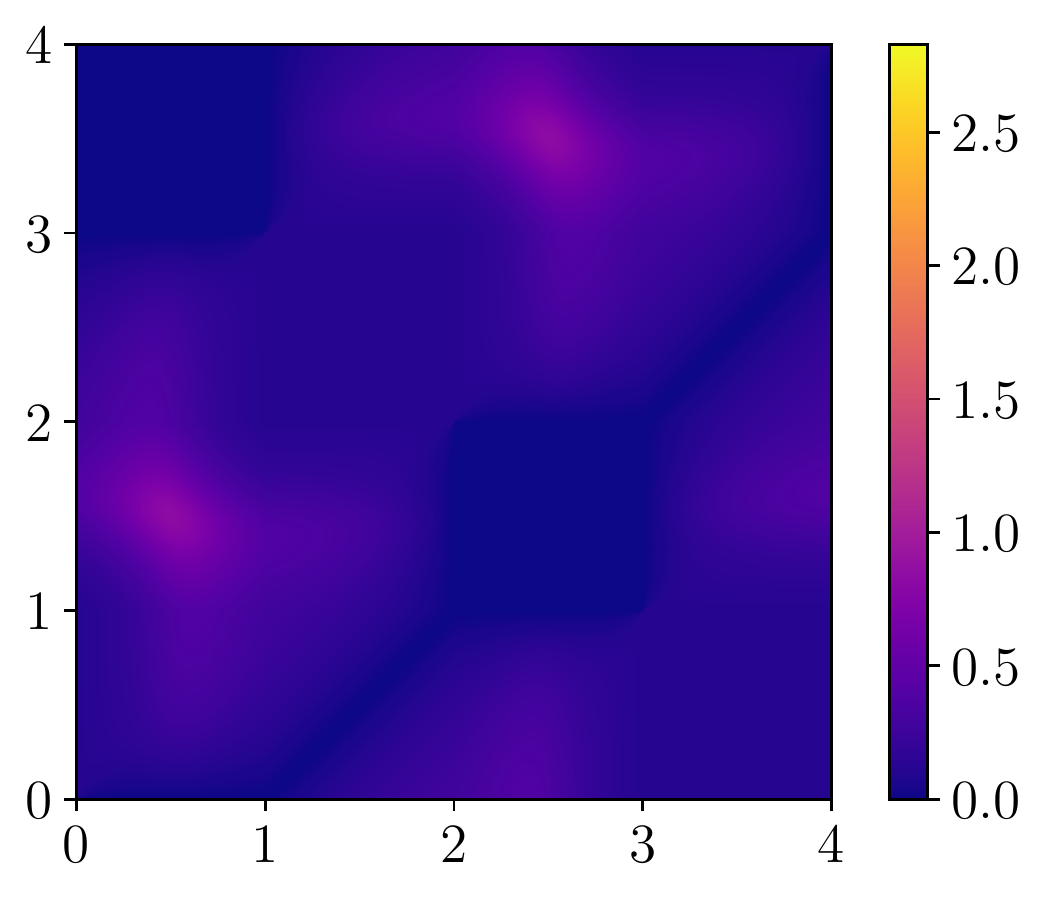}
 \caption{\label{fig:lineresult} The result from the linear wombling procedure described in the text on a data set with $\rho=5$ (left) and $\rho=1.5$ (right). The heat map is color coded according to the value of $|\bar\Gamma[C]|$ and exhibits two degenerate maxima since each line is traversed twice (once in each direction).}
\end{figure}

Using the respective fit (\ref{eq:smearingfunction}) as our proxy, we can now compute the wombling measure $\bar\Gamma[C]$ in the $(p_{in}, p_{out})$ space of all lines intersecting our unit square. The result for $\rho=5$ ($\rho=1.5$) is shown in the left (right) panel of Fig.~\ref{fig:lineresult}. We choose to plot the absolute value of $\bar\Gamma[C]$, since the sign of $\bar\Gamma[C]$ is determined by the direction in which we traverse the line, and does not have any bearing on whether the line is a wombling boundary or not. Fig.~\ref{fig:lineresult} reveals that, as expected, there are two locations with maximal wombling measures: at $(p_{in}, p_{out})=(0.5,1.5)$ and at $(p_{in}, p_{out})=(2.5,3.5)$. Both of those correspond to the same vertical line at $x=0.5$ which was the true boundary. This demonstrates that the method is indeed able to find the correct boundary. The significance of these findings, however, is sensitive to the amount of signal present --- in the left panel, where $\rho=5$, the two 	winning answers are very clearly identified, while in the right panel (using the same color scale) they appear to be less noticeable, which is a hint that the effect might be in danger of being washed out once we include the statistical fluctuations present in the point data --- this issue will be investigated in detail in Sec.~\ref{sec:womblingdiscrete} below.

\subsection{A circular boundary line}
\label{sec:continuouscircle}

\begin{figure}[t]
 \centering
 \includegraphics[height=.4\textwidth]{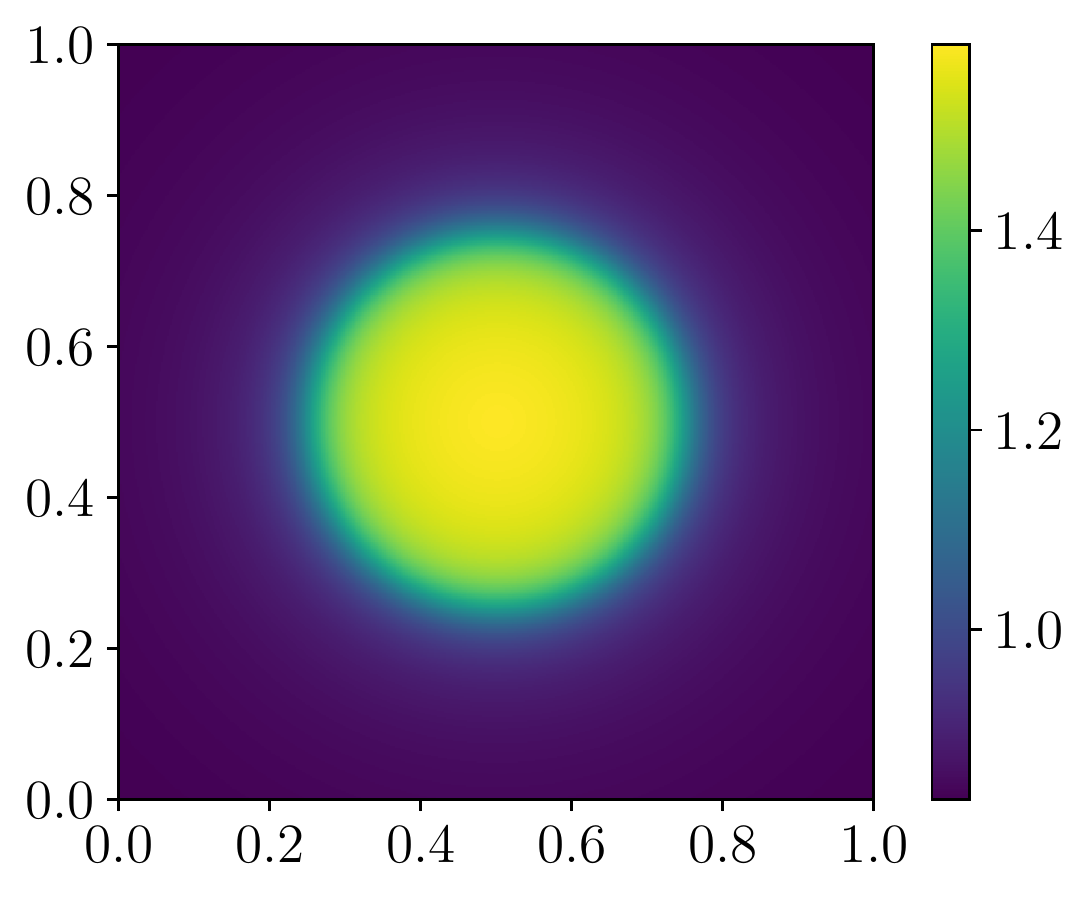}~~~~
 \includegraphics[height=.4\textwidth]{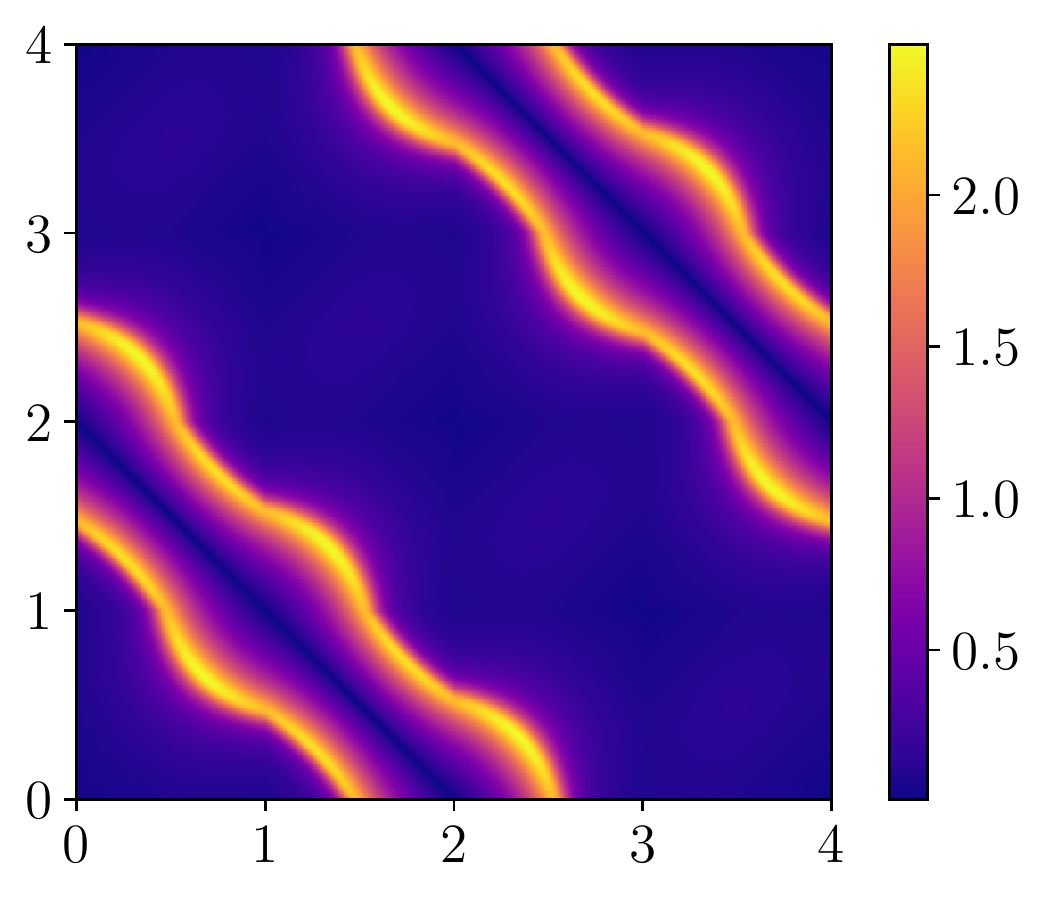}
 \caption{\label{fig:circleexample} Color maps of the probability distribution (\ref{eq:fcircle}) used for the circular boundary example (left panel) and of the corresponding $|\bar\Gamma[C]|$ in the $(p_{in}, p_{out})$ parameter space (right panel).}
\end{figure}

The example in the previous Sec.~\ref{sec:continuousline} might appear somewhat contrived since the true boundary was a straight line and at the same time, we also used straight lines in computing the wombling measure $\bar\Gamma[C]$. Since the shapes of the lines match, it was inevitable to find a unique best match, as shown in Fig.~\ref{fig:lineresult}. To be fair, we shall now consider a less trivial example where the true boundary has a different shape from the line segments which we use to test for the presence of a wombling boundary. In particular, we shall revisit the case of a circular boundary introduced in Sec.~\ref{sec:simdetails}, where the probability distribution was given by (\ref{fSMfCircle}). Once again, we shall not rely directly on (\ref{fSMfCircle}), but in order to account for the detector resolution, we shall sample the events according to 
\beq
f(r,\varphi) \sim r g_{res}(r; a=0.25, \xi=2, k=30),
\label{eq:fcircle}
\eeq
where $r$ and $\varphi$ are the polar coordinates in the plane, measured from an origin at the center of the circle, and the smearing function $g_{res}$ was already defined in (\ref{eq:arctansigmoid}). The resulting probability distribution is plotted in the left panel of Fig.~\ref{fig:circleexample}. Note that the densities on the two sides of the circular boundary differ by no more than a factor of 2, so in this sense this example is analogous to $\rho\sim2$ in our usual notation. The corresponding heat map of $|\bar\Gamma[C]|$ in the $(p_{in}, p_{out})$ parameter space is shown in the right panel of Fig.~\ref{fig:circleexample}. The most striking difference from the previous results in Fig.~\ref{fig:lineresult} is that now we find not just a single wombling boundary candidate, but a whole class of wombling boundaries, identified by the two\footnote{We obtain two stripes because of the double counting $(p_{in}, p_{out})\longleftrightarrow (4-p_{out}, 4-p_{in})$ due to the possibility to traverse a line segment in each of the two opposite directions, as shown in the right panel of Fig.~\ref{fig:perimeter}.} bright yellow stripes running diagonally across the plot. A careful inspection of the right panel in Fig.~\ref{fig:circleexample} reveals that each of the identified wombling boundary candidates is tangential to the circular boundary, which suggests that the dominant contribution to the integral comes from the region in the vicinity of the circular boundary, where the gradient is largest. The slight difference in the brightness along the stripes can then be attributed to the different orientation of the lines, which leads to differences in the (sub-dominant) flux contributions in the regions away from the true boundary. 

\section{Finding wombling boundaries: point data examples}
\label{sec:womblingdiscrete}

Having illustrated the basic idea of Refs.~\cite{Banerjee2006,Gelfand2015} with the continuous examples from the previous section, we shall now apply it to point data. Following the outline of Sec.~\ref{sec:womblingcontinuous}, we shall first consider the case of a straight line boundary in Sec.~\ref{sec:discretestraight} and then the case of a circular boundary in Sec.~\ref{sec:discretecircular}.

\subsection{A straight line boundary}
\label{sec:discretestraight}

\subsubsection{An example with $\rho=5$}
\label{sec:rho5example}

Our first straight-line boundary example will be the same point data example which we have been using so far throughout the paper to illustrate the various methods and techniques of a wombling analysis, see Figs.~\ref{fig:boundaryobjects}, \ref{fig:correlations}, \ref{fig:unscaled_gradients}, \ref{fig:scaled_gradients}, \ref{fig:lsu_lloyd}, \ref{fig:after_cut_lloyd}, \ref{fig:vectors_averaged}, \ref{fig:lsu_averaged}, \ref{fig:taggers} and \ref{fig:spiderwebs}. (As a reminder, we used $N=500$ points generated according to the distribution (\ref{fSMfLine}) with $\rho=5$.) In particular, we shall repeat the procedure from Sec.~\ref{sec:womblingcontinuous} and compute a wombling measure $\bar\Gamma[C]$ for each possible straight line $C$ crossing the field of view. However, since we are now dealing with discretely sampled point data instead of a continuous function $f(x,y)$, we need to adapt the definition (\ref{eq:Gammaavedef}) as follows
\beq
\bar\Gamma[C] \equiv \frac{\sum_\alpha \left( \langle\vec{G} \rangle_\alpha \cdot \hat{n}_C \right) \Delta\ell_\alpha}{\sum_\alpha \Delta\ell_\alpha},
\label{eq:Gammaavediscrete}
\eeq
where each sum runs over all Delaunay triangles $D_\alpha$ which are crossed by the straight line $C$ and, as before, $\langle\vec{G} \rangle_\alpha$ is the Delaunay-averaged\footnote{In principle, we can define the wombling measure (\ref{eq:Gammaavediscrete}) in terms of the original gradient vectors $\tilde{\vec G}_\alpha$ or in terms of the Voronoi-averaged gradient vectors $\langle \vec G\rangle_i$ defined in (\ref{superave}). However, as argued in Secs.~\ref{sec:tagging} and \ref{sec:agglomeration}, the Delaunay-averaged vectors offer the best option for our purposes.} gradient vector (\ref{superave}) associated with $D_\alpha$. Since the Delaunay tessellation gives complete coverage of the field of view, the line $C$ necessarily gets fragmented into individual line segments of length $\Delta \ell_\alpha$, defined so that each segment is contained within a single Delaunay cell $D_\alpha$. Finally, $\hat n_C$ is a unit vector orthogonal to the straight line $C$ and therefore, to each individual line segment $\Delta \ell_\alpha$ as well, so that an index $\alpha$ on it is unnecessary.

\begin{figure}[t]
 \centering
 \includegraphics[width=.65\textwidth]{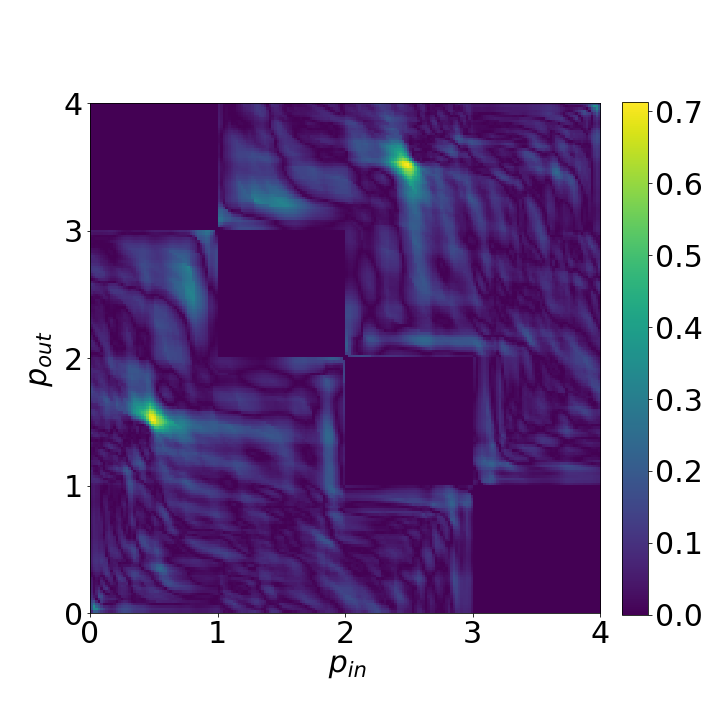}
 \caption{\label{fig:oceanplot5} Results from applying the wombling procedure of Sec.~\ref{sec:womblingcontinuous} to the illustrative point data example used in the previous sections ($N=500$ points generated from (\ref{fSMfLine}) with $\rho=5$). For concreteness, at the pre-processing stage we applied one Lloyd iteration and then used the Delaunay-averaged gradient vectors which were defined in (\ref{superave}). In constructing the heat map, we sampled the $(p_{in}, p_{out})$ parameter space on a 200 by 200 grid with step size 0.02 and then computed $\bar\Gamma[C]$ from eq.~(\ref{eq:Gammaavediscrete}). The four dark squares in the plot correspond to line segments which overlap with one of the edges of the unit square which is our field of view - those segments were assigned $\bar\Gamma[C]=0$ by default and excluded from further consideration. For reference, the true boundary is located at $(p_{in},p_{out})=(0.5,1.5)$, or equivalently, at $(p_{in},p_{out})=(2.5,3.5)$. }
\end{figure}

Fig.~\ref{fig:oceanplot5} shows a heat map of the wombling measure $\bar\Gamma[C]$ computed from eq.~(\ref{eq:Gammaavediscrete}) throughout the parameter space $(p_{in},p_{out})$ of all possible straight lines passing through our data. As a pre-processing step, we applied one Lloyd iteration and then used the Delaunay-averaged gradient vectors illustrated in the right panel of Fig.~\ref{fig:vectors_averaged}. The heat map in Fig.~\ref{fig:oceanplot5} contains four dark squares situated along the diagonal from the upper left corner to the lower right corner. All points within those four squares define line segments $C$ which do not enter the field of view at all, and instead run along one of the edges of the field of view. Clearly, such line segments are irrelevant for our wombling boundary analysis, so they have been assigned $\bar\Gamma[C]=0$ by default and are excluded from further consideration.

\begin{figure}[t]
 \centering
 \includegraphics[width=.5\textwidth]{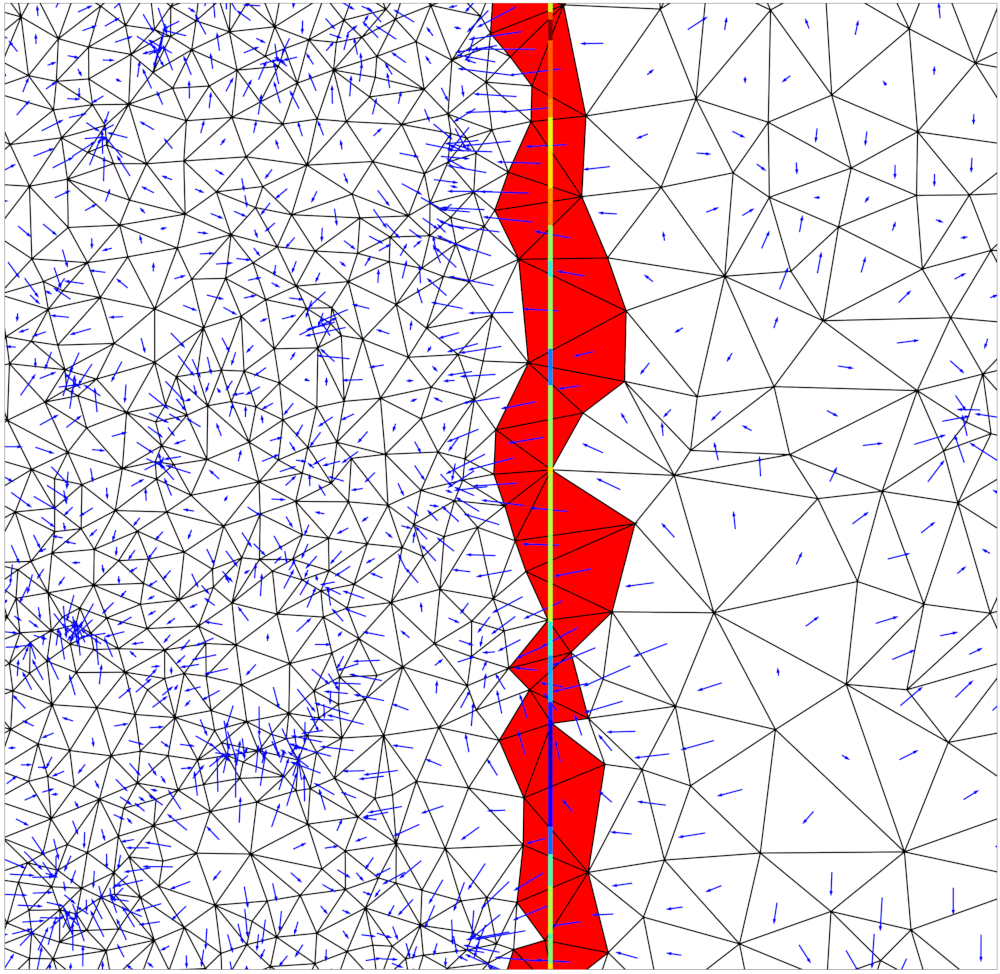}
 \caption{\label{fig:lineresult5} The line $C_w$ with the largest wombling measure $\bar\Gamma[C]$ in Fig.~\ref{fig:oceanplot5}, plotted over the Delaunay-averaged gradient vector field from the right panel of Fig.~\ref{fig:vectors_averaged}. The individual line segments $\Delta\ell_\alpha$ have been color-coded by their respective individual contributions $\langle\vec{G} \rangle_\alpha \cdot \hat{n}_C$ to the total flux. The red-shaded Delaunay cells are those which are crossed by the line $C_w$ and are therefore included in the two summations in (\ref{eq:Gammaavediscrete}).}
\end{figure}

Fig.~\ref{fig:oceanplot5} reveals that there is a unique line with the largest possible wombling measure --- let us denote this winning line with $C_w$:
\beq
C_w \equiv \argmax_{C} \left( \bar\Gamma[C]  \right).
\label{Cwdef}
\eeq 
By construction, the winning line $C_w$ is {\em the best} wombling boundary candidate among the set of all straight-line boundary candidates. Is it the correct wombling boundary though? According to the result from Fig.~\ref{fig:oceanplot5}, the answer in this example is yes: $C_w$ is found at the exact location $(0.5,1.5)$ (or equivalently, $(2.5,3.5)$) of the true theoretical boundary $x=0.5$. This can be verified explicitly in Fig.~\ref{fig:lineresult5}, where we plot $C_w$ overlaid on top of the Delaunay-averaged gradient vectors $\langle\vec{G} \rangle_\alpha$ from the right panel of Fig.~\ref{fig:vectors_averaged}. Fig.~\ref{fig:lineresult5} not only confirms that $C_w$ is the correct wombling boundary, but also helps us understand why $C_w$ was chosen by the algorithm. Note that for any given line $C$, the calculation of its wombling measure $\bar\Gamma[C]$ depends only on the Delaunay cells $D_\alpha$ which happen to be crossed by the line $C$ --- in Fig.~\ref{fig:lineresult5} those cells are shaded in red. A careful inspection of Fig.~\ref{fig:lineresult5} reveals that in the large majority of red-shaded Delaunay cells the gradient vectors are both large and (roughly) orthogonal to the line $C_w$, thus maximizing the average flux (\ref{eq:Gammaavediscrete}) through it. To better visualize this, we have color-coded the individual line segments $\Delta\ell_\alpha$ of $C_w$ according to their individual contributions $\langle\vec{G} \rangle_\alpha \cdot \hat{n}_C$ to the total flux, with warm (cold) colors corresponding to large (small) values. We see that $C_w$ is predominantly colored with warm colors, indicating large fluxes all the way throughout. On the other hand, it is not difficult to convince oneself that this will not be true for any other randomly chosen line $C$ crossing the field of view --- the gradient vectors may happen to be relatively large and perhaps even roughly orthogonal to it purely by chance in some restricted region, but this will not occur consistently along the full length of the line as was the case with $C_w$. In short, the wombling procedure applied here automatically takes into account the spatial correlations of the gradient vectors along a wombling boundary. 

\subsubsection{An example with $\rho=1.5$}
\label{sec:rho1p5}

\begin{figure}[t]
 \centering
 \includegraphics[width=.4\textwidth]{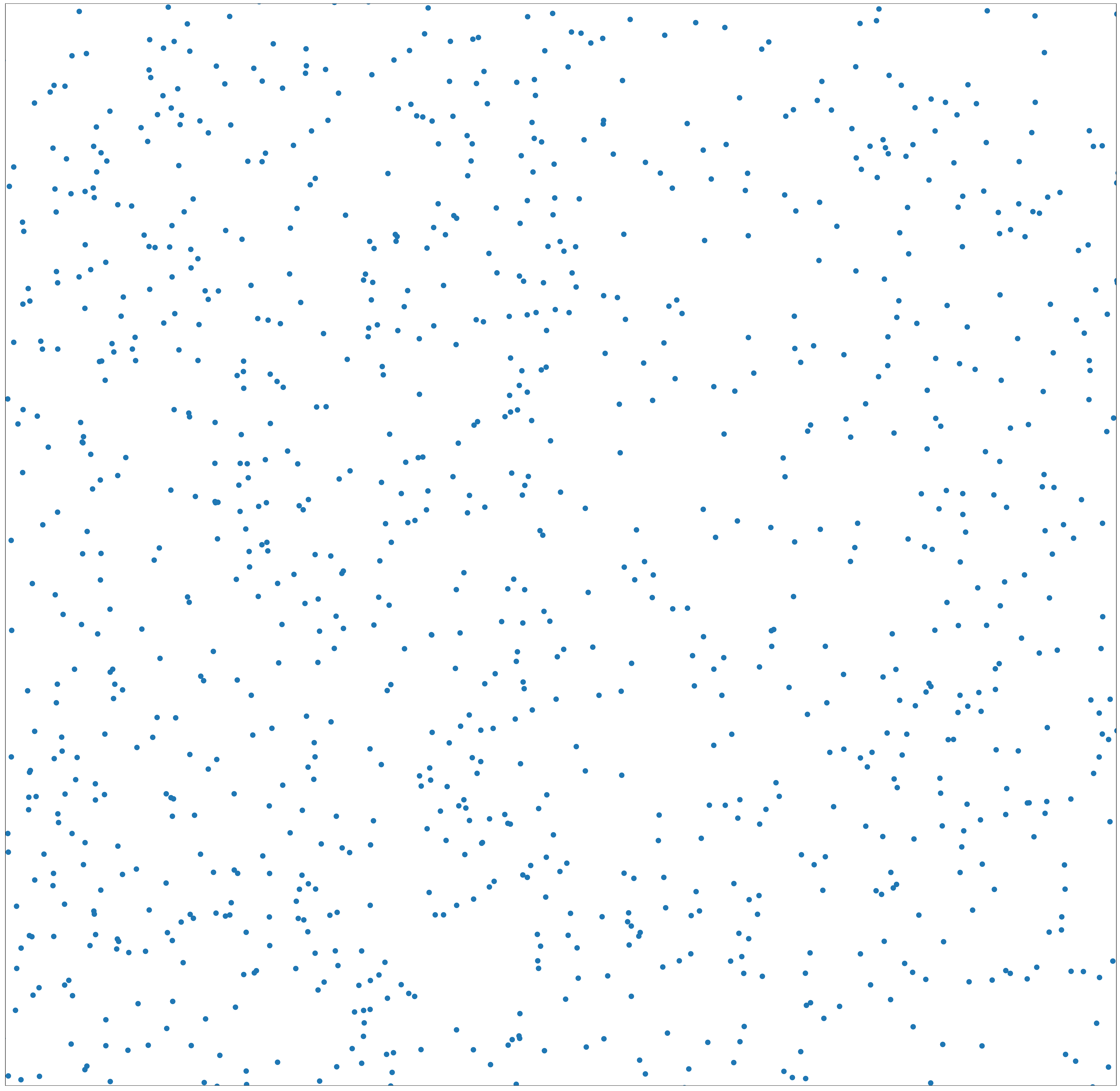}~~~
 \includegraphics[width=.4\textwidth]{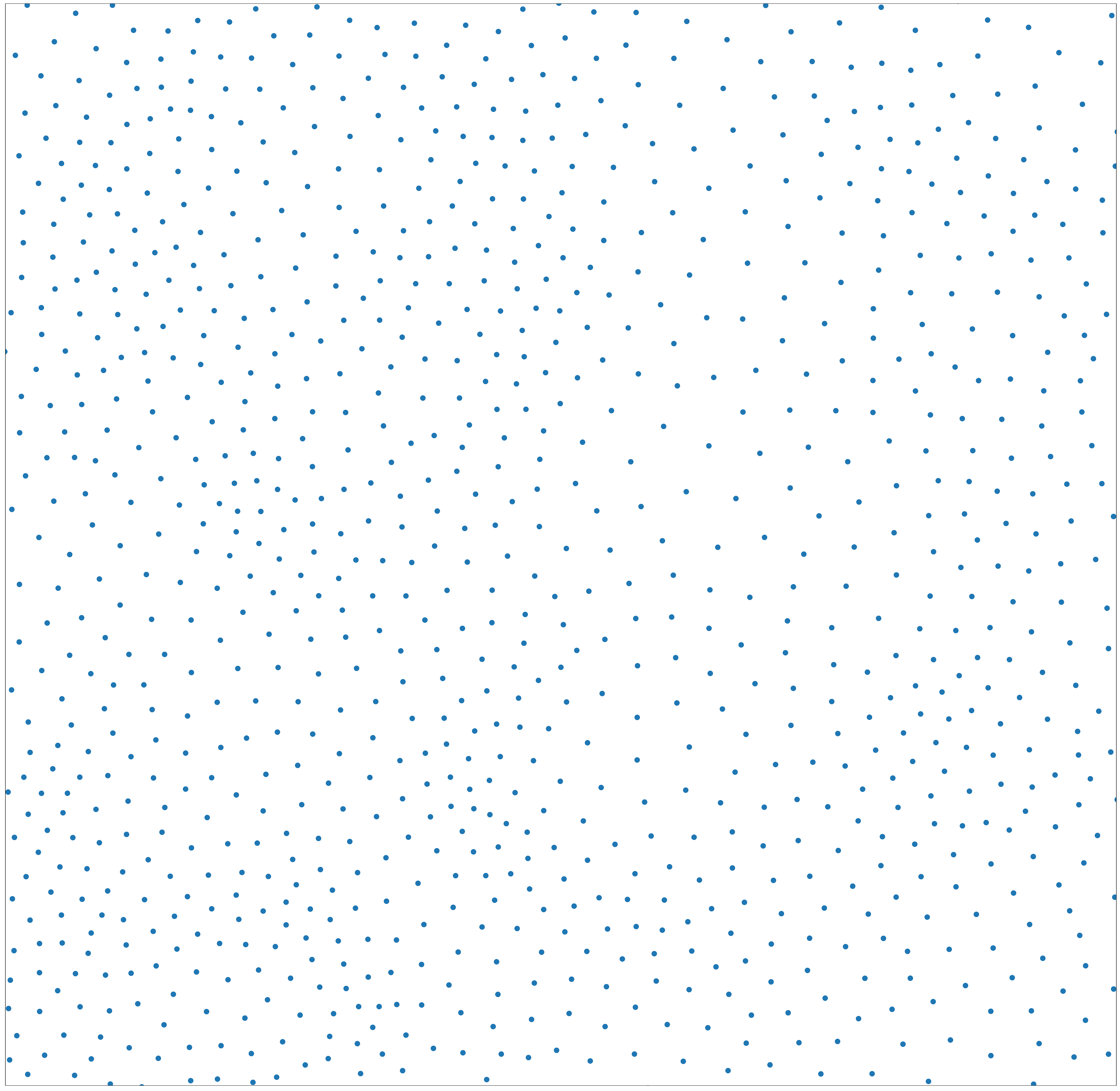}
 \caption{\label{fig:linedata1p5} The point data for the straight line boundary example considered in Sec.~\ref{sec:rho1p5}. The left panel shows the original $N=1000$ points distributed according to (\ref{fSMfLine}) with $\rho=1.5$. The right panel shows the same data after applying 5 Lloyd iterations. }
\end{figure}

We are now ready to tackle a more difficult case, with a smaller signal to background ratio. In this subsection we consider $N=1000$ data points, still distributed according to (\ref{fSMfLine}), but with the much smaller value of $\rho=1.5$. This data is shown in the left panel of Fig.~\ref{fig:linedata1p5}. Unlike the previous example, this time the wombling boundary is not as easy to identify visually in the data. As already discussed in Ref.~\cite{Debnath:2015wra}, the weaker the signal, the more Lloyd steps are needed for optimal results. Correspondingly, here we apply 5 Lloyd iterations at the preprocessing stage and obtain the data shown in the right panel of Fig.~\ref{fig:linedata1p5}, on which the subsequent wombling analysis is done. 

\begin{figure}[t]
 \centering
 \includegraphics[width=.65\textwidth]{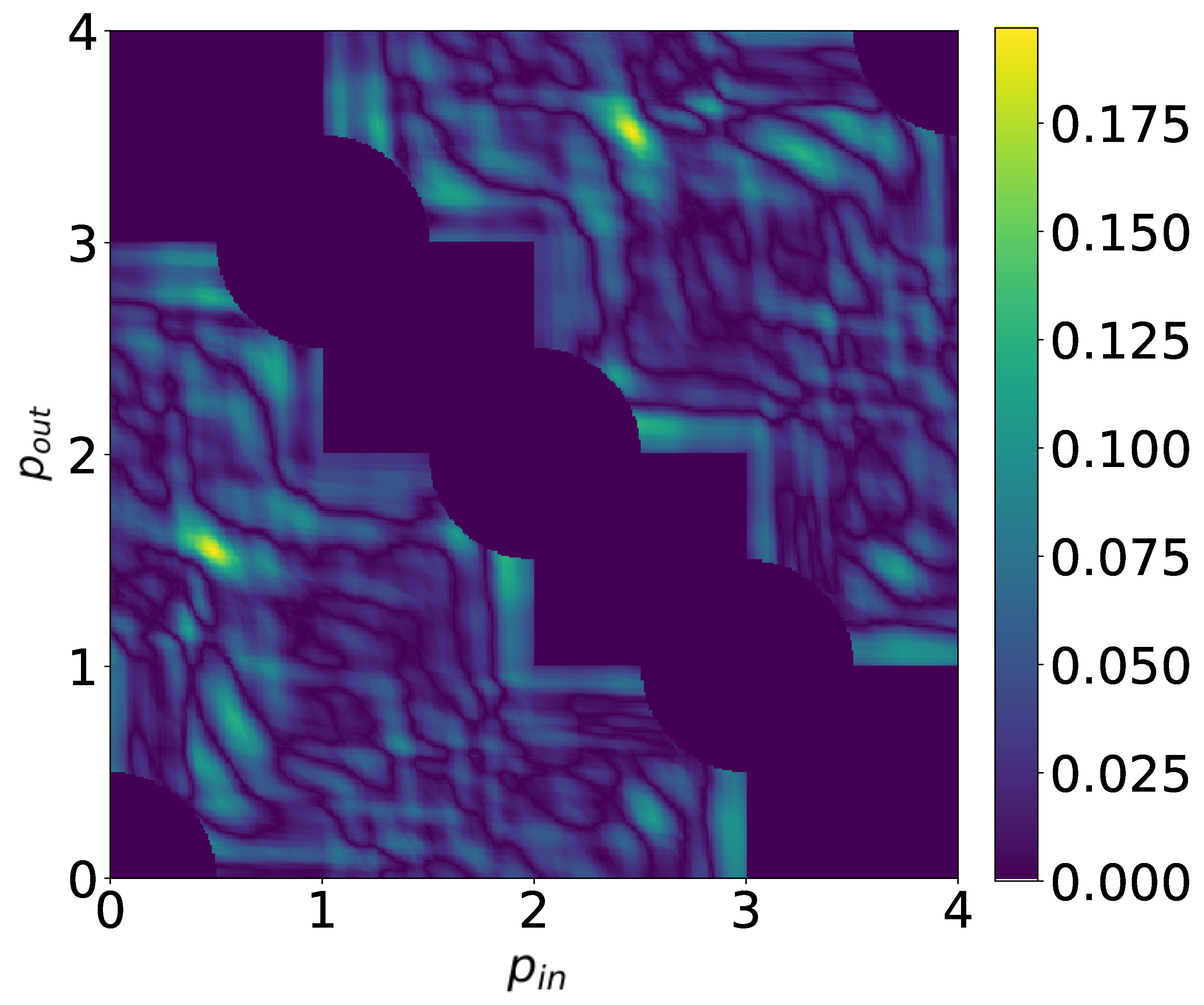}
 \caption{\label{fig:oceanplot1p5} The same as Fig.~\ref{fig:oceanplot5} but for the dataset  with $\rho=1.5$ shown in the right panel of Fig.~\ref{fig:linedata1p5}. Here we exclude from the analysis not only lines along the perimeter of the field of view but also any line of length less than 0.5; in the heatmap such lines are assigned $\bar\Gamma[C]=0$ by default.  }
\end{figure}
\begin{figure}[h]
 \centering
 \includegraphics[width=.5\textwidth]{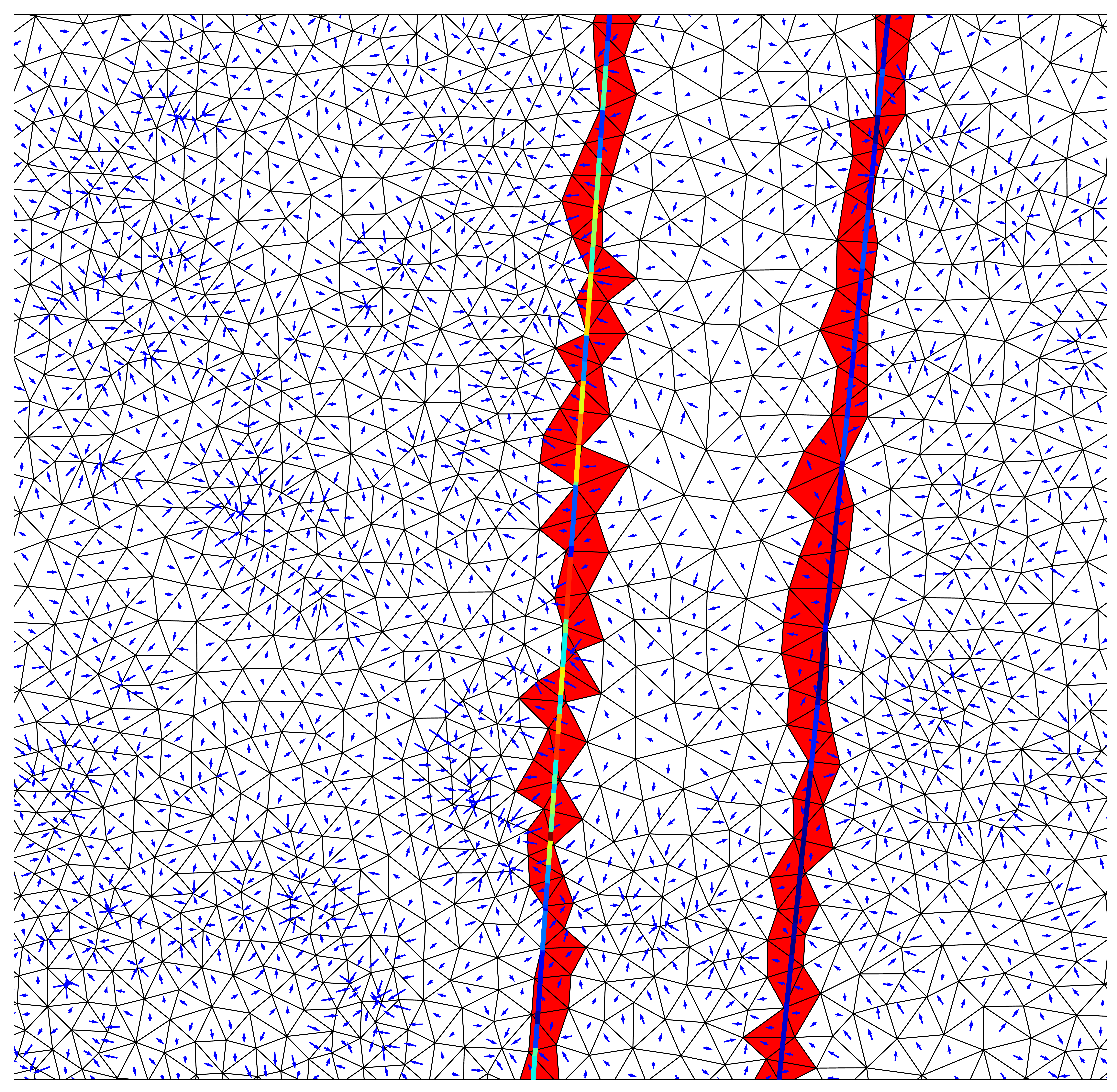}
 \caption{\label{fig:lineresult1p5} The same as Fig.~\ref{fig:lineresult5} but for the dataset  with $\rho=1.5$ shown in the right panel in Fig.~\ref{fig:linedata1p5}. In addition to the line $C_w$ with the largest value of the wombling measure $\bar\Gamma$, for comparison we also show a generic line with a more typical value of $\bar\Gamma[C]$; for concreteness, we chose the line at $(p_{in},p_{out})=(0.7,0.8)$ which happens to have $\bar\Gamma[C]=0.08$.}
\end{figure}

The end result from our wombling procedure is shown in Fig.~\ref{fig:oceanplot1p5}, which is the analogue of Fig.~\ref{fig:oceanplot5}. We notice that the typical values obtained for $\bar\Gamma[C]$ are now much lower than what we saw in Fig.~\ref{fig:oceanplot5}. This is to be expected due to the smaller value of $\rho$ --- the boundary is less pronounced, and the magnitudes of the gradient vectors are generally reduced as well. Despite these difficulties, the boundary is still correctly identified --- notice the two bright spots in the heatmap located near $(0.5,1.5)$ and $(2.5,3.5)$, which is the right answer. The corresponding winning line $C_w$ has $\bar\Gamma[C_w]=0.19$ and is plotted in Fig.~\ref{fig:lineresult1p5}, where for illustration we also show a second, rather generic, line $C$ at $(p_{in},p_{out})=(0.7,0.8)$ which has a more typical value of the wombling parameter, $\bar\Gamma[C]=0.08$. The individual line segments of each of the two lines are color-coded similar to Fig.~\ref{fig:lineresult5}, i.e., proportional\footnote{In order to highlight the differences between the lines segments of the two lines, we additionally apply the scaling (\ref{eq:rescaling}) motivated below in Sec.~\ref{sec:shape}.} to their individual contributions to the total flux. We see that the winner $C_w$ is again colored with mostly warm colors, indicating large flux contributions everywhere (except for just a few spots where the gradient vectors happen to be either too small or oriented along the line), while the generic line $C$ is colored with mostly cold colors, confirming that its wombling measure $\bar\Gamma$ is indeed rather small.

By comparing Figs.~\ref{fig:oceanplot5} and \ref{fig:oceanplot1p5}, one can notice that in Fig.~\ref{fig:oceanplot1p5} we have excluded from consideration not only lines which run purely along the perimeter of the field of view but also any line of length less than 0.5; this additional constraint results in the elimination of several quadrant sectors on the plot --- one at the lower left corner, one at the upper right corner, and six along the diagonal running from the upper left corner to the lower right corner. Since our wombling measure is the {\em average} flux, short lines which literally ``cut a corner'' of the square field of view, can potentially pick up large gradients due to local fluctuations in the bulk, without an opportunity to cancel those fluctuations elsewhere. In other words, if we encounter two candidate lines with the same value of $\bar\Gamma$, and one is much longer than the other, then we would treat the longer line as the more likely wombling boundary. Thus eliminating very short lines from consideration early on would go a long way in simplifying the significance estimation procedure, see Sec.~\ref{sec:significance} below.

\subsection{A circular boundary}
\label{sec:discretecircular}

\subsubsection{The results from a wombling analysis with straight line segments}
\label{sec:circlerho5}

\begin{figure}[t]
 \centering
 \includegraphics[width=.4\textwidth]{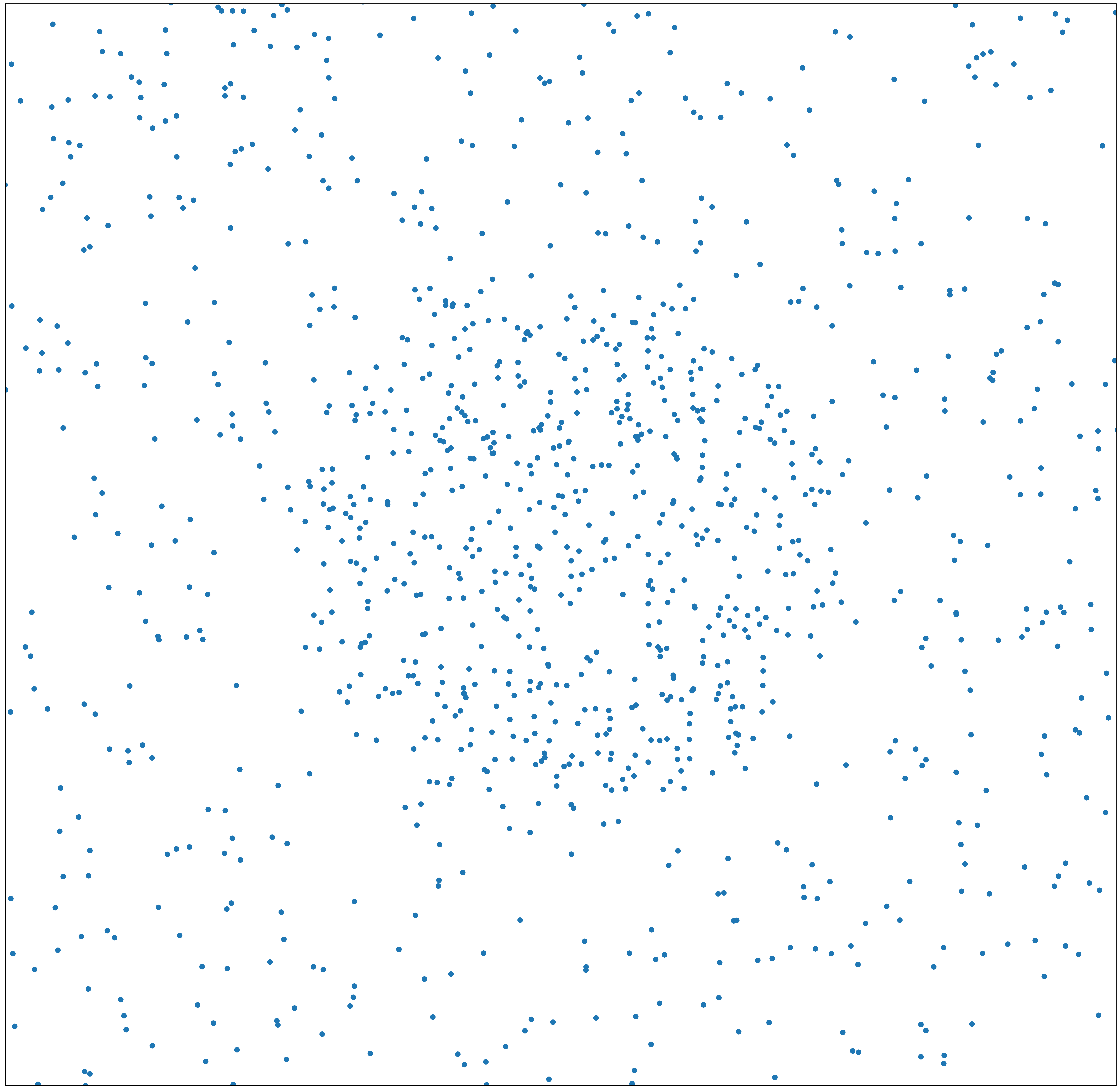}~~~
 \includegraphics[width=.4\textwidth]{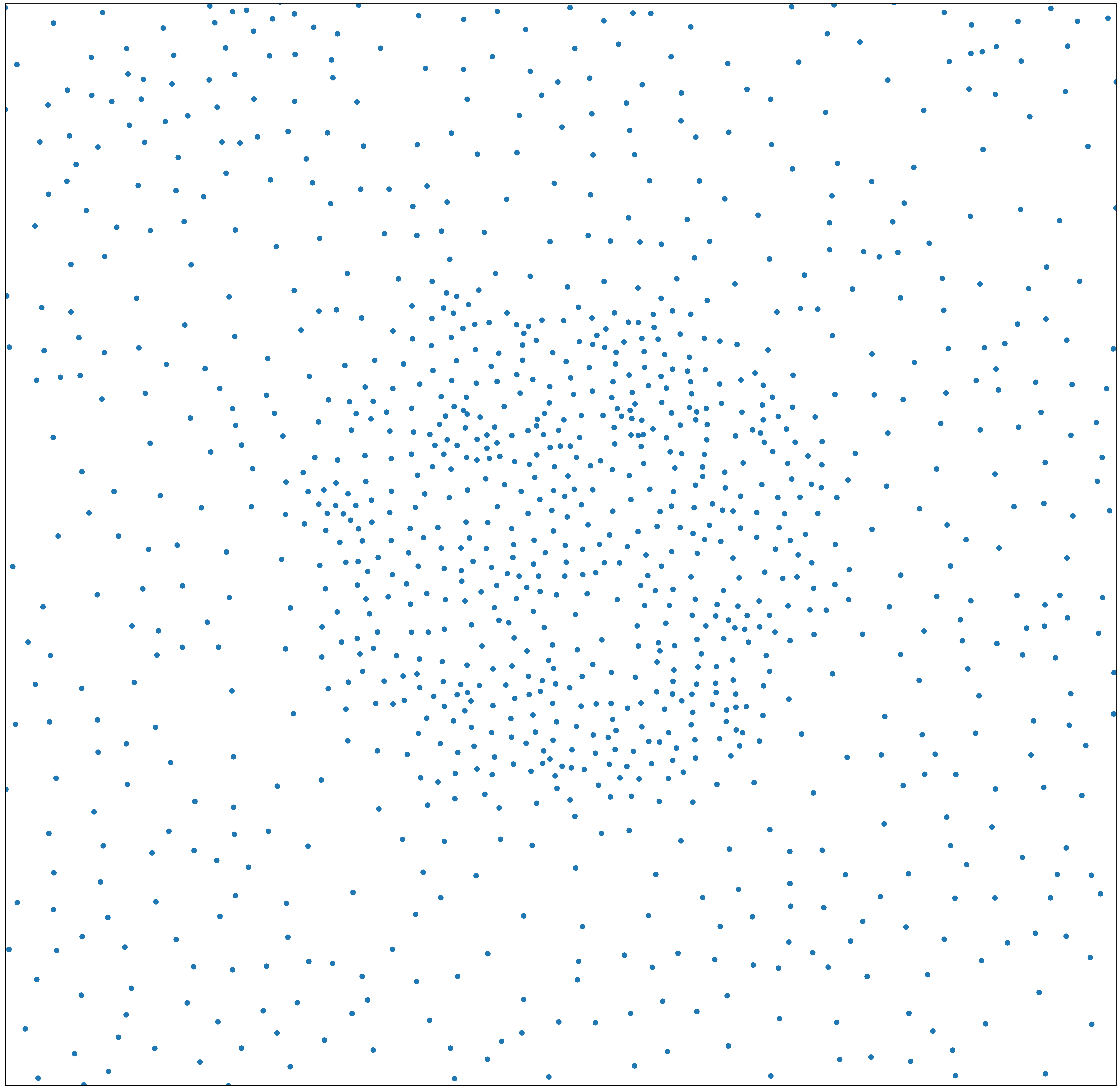}
 \caption{\label{fig:circledata2} The point data for the circular boundary example considered in Sec.~\ref{sec:circlerho5}. The left panel shows the original $N=1000$ points distributed according to (\ref{fSMfCircle}) with $\rho=5$. The right panel shows the same data after 1 Lloyd iteration.}
\end{figure}

We now proceed with the discrete version of the circular boundary example considered in Sec.~\ref{sec:continuouscircle}. The point dataset is shown in the left panel of Fig.~\ref{fig:circledata2} and consists of $N=1000$ points distributed according to the probability distribution with a circular boundary (\ref{fSMfCircle}) with $\rho=5$. As a preprocessing step, we then apply a single Lloyd iteration, obtaining the data shown in the right panel of Fig.~\ref{fig:circledata2}, on which the subsequent wombling analysis is performed.

\begin{figure}[t]
 \centering
 \includegraphics[width=.65\textwidth]{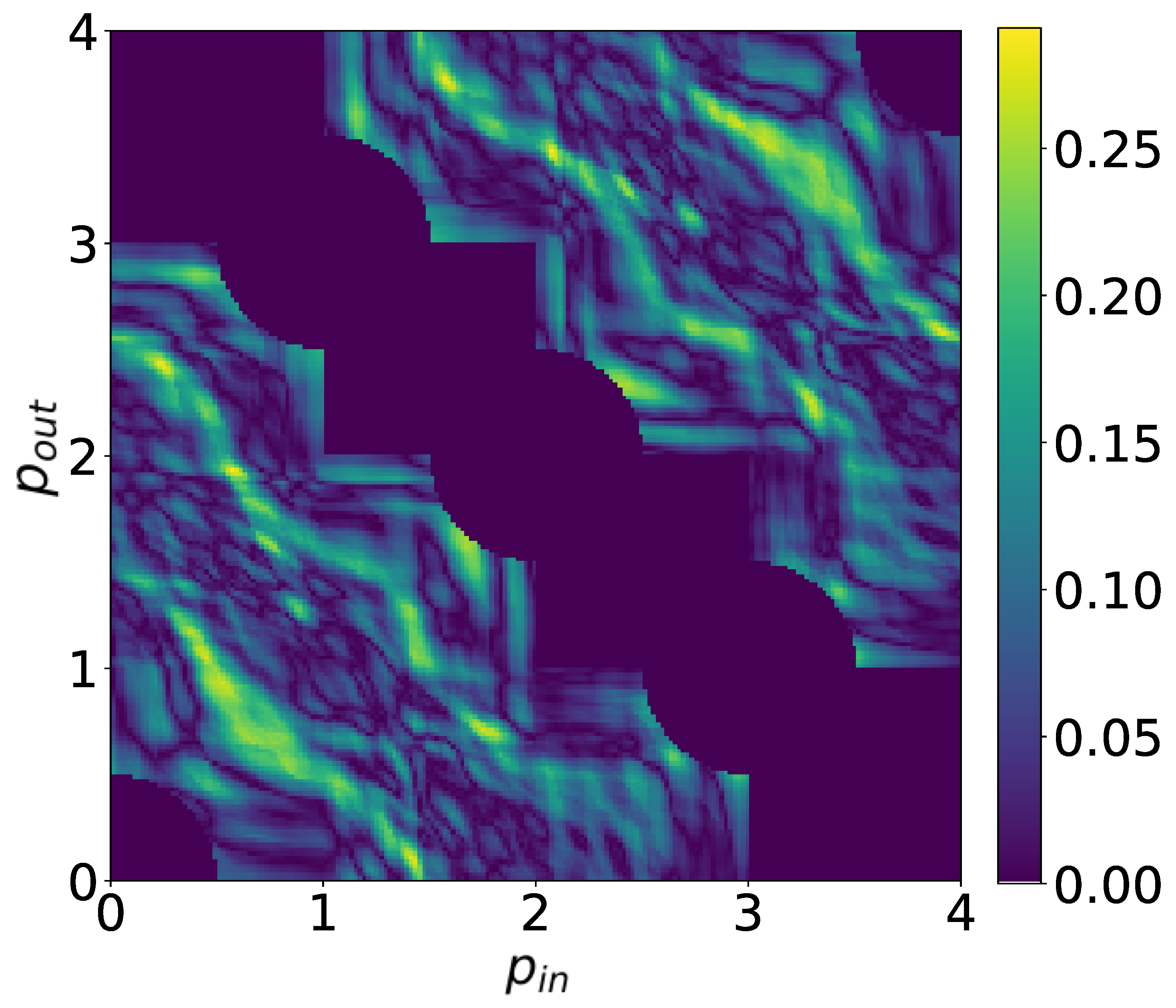}
 \caption{\label{fig:oceancircle5} The same as Fig.~\ref{fig:oceanplot1p5} but for the dataset  with a circular boundary shown in the right panel of Fig.~\ref{fig:circledata2}.}
\end{figure}
\begin{figure}[t]
 \centering
 \includegraphics[height=.4\textwidth]{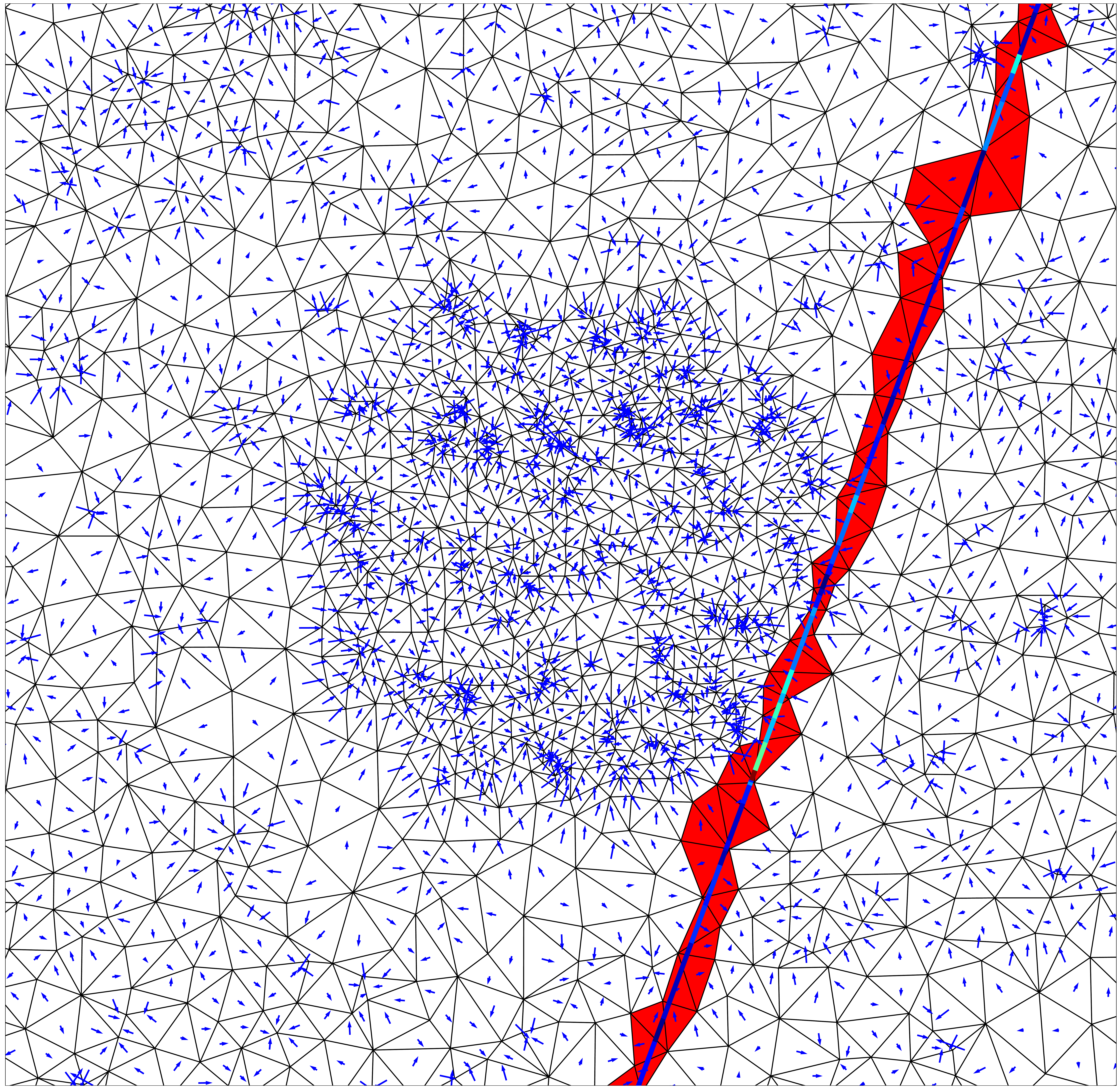}
 ~~~~~
 \includegraphics[height=.4\textwidth]{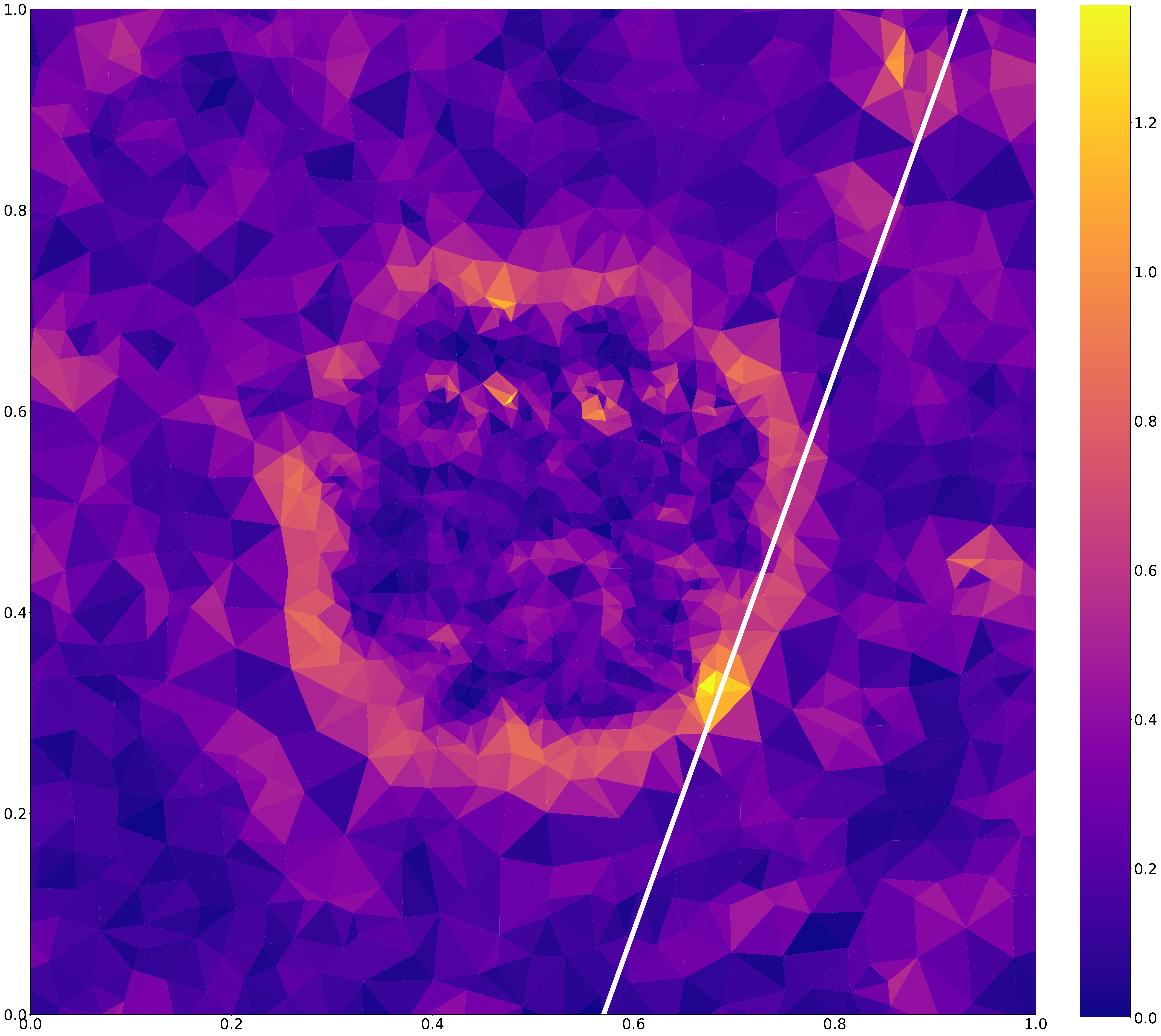}
 \caption{\label{fig:circleresult5} 
 The line $C_w$ with the largest wombling measure in the circular boundary example of Sec.~\ref{sec:discretecircular}, overlaid on the Delaunay-averaged gradient vectors $\langle \vec{G} \rangle_\alpha$ (left panel) or on the heatmap of their magnitudes (right panel).  }
\end{figure}

The results from the wombling procedure are displayed in Figs.~\ref{fig:oceancircle5} and \ref{fig:circleresult5}. Fig.~\ref{fig:oceancircle5} is the analogue of Figs.~\ref{fig:oceanplot5} and \ref{fig:oceanplot1p5}, but for the circular boundary example considered in this subsection. Comparing to those previous figures, we notice that the largest absolute values for $\bar\Gamma$ obtained here are lower than those in Fig.~\ref{fig:oceanplot5} but higher than those in Fig.~\ref{fig:oceanplot1p5}. The former is due to the fact that we are using straight line segments to test for a curvilinear wombling boundary, and no single line segment can capture the full extent of a circular boundary, while the latter is due to the fact that the value of $\rho$ is higher for the dataset used to produce Fig.~\ref{fig:oceancircle5}.

Fig.~\ref{fig:oceancircle5} exhibits the typical pattern observed in the continuous version of this example (the right panel of Fig.~\ref{fig:circleexample}). The locations with the largest values of $|\bar\Gamma|$ trace out the two stripes seen in Fig.~\ref{fig:circleexample}, which indicates that in the discrete version of the example it is still the line segments tangential to the circular boundary which tend to have large values of $|\bar\Gamma|$. The line $C_w$ with the largest wombling measure happens to be at $(p_{in}, p_{out})=(0.55,1.9)$ and is plotted in Fig.~\ref{fig:circleresult5}, together with the set of Delaunay-averaged gradient vectors $\langle \vec{G} \rangle_\alpha$ (left panel) or the heatmap of their magnitudes (right panel). As anticipated, $C_w$ is tangential to the circular boundary, and the largest contributions to its wombling measure indeed come from the (warm-colored) segments in the vicinity of the boundary. One should keep in mind that the particular line $C_w$ shown in Fig.~\ref{fig:circleresult5} is a very close winner among several other worthy challengers with similar values for the wombling measure --- as Fig.~\ref{fig:oceancircle5} showed, there are several locations along the two bright stripes with similarly large values of $|\bar\Gamma|$.

\subsubsection{Identifying the true shape of the boundary}
\label{sec:shape}

The analysis from the previous subsection \ref{sec:circlerho5} demonstrated that even in the case of a curvilinear boundary, our wombling procedure produces reasonable results --- it is able to identify a class of line segments, each of which already contains a portion of the true boundary. Unfortunately, none of the identified line segments is able to reproduce the full boundary all by itself. In this subsection, we shall therefore address the question of being able to globally reconstruct the wombling boundary, regardless of its shape,  from the results presented so far.

\begin{figure}[t]
 \centering
  \includegraphics[height=.5\textwidth]{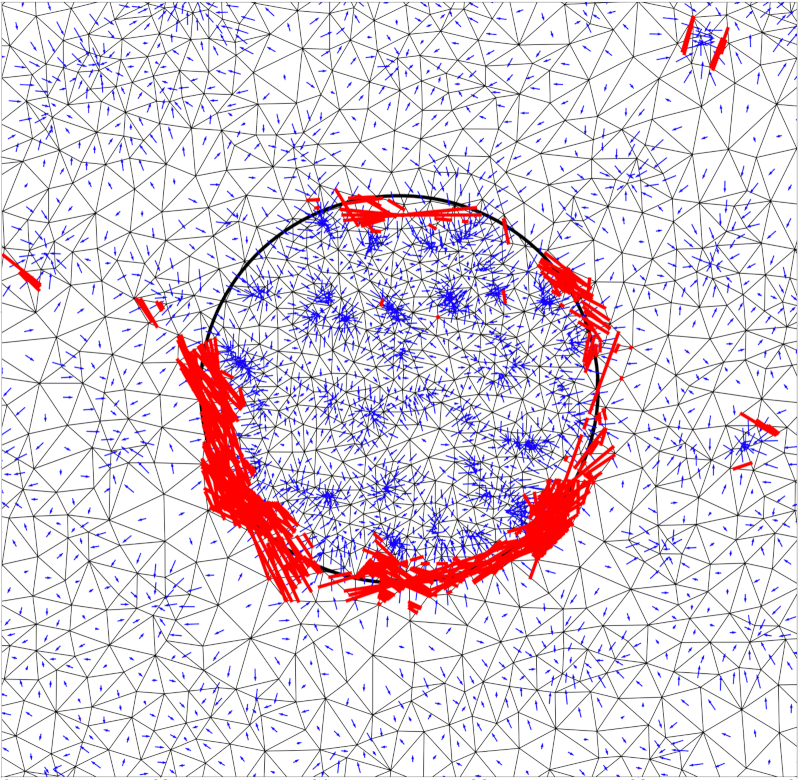}
 \caption{\label{fig:circleresult6} The individual segment selection procedure described in Sec.~\ref{sec:shape}. After scanning the $(p_{in}, p_{out})$ parameter space on an $80\times80$ grid and computing the wombling measures $\bar\Gamma[C]$, the average flux for each individual line segment $\Delta\ell_\alpha$ has been rescaled according to (\ref{eq:rescaling}) with $\gamma=4$ and the plot shows (in red) the segments with rescaled flux values in the top 1 percentile.}
\end{figure}

In principle, there can be several different approaches to this problem. 
\begin{itemize}
\item {\em Algorithmic wombling.} The standard approach is the algorithmic wombling procedure outlined in the introduction \cite{DaleFortin}. One applies a lower cut (tagging) on the magnitudes of the locally estimated gradient vectors and then suitably connects them (agglomeration). Although this approach has been subject to criticism \cite{Banerjee2006}, its modern implementation can perhaps benefit from some of the improvements which we have introduced here, in particular gradient rescaling (Sec.~\ref{sec:gradrescaling}), Voronoi relaxation (Sec.~\ref{sec:LSU}), local averaging of gradient vectors (Sec.~\ref{sec:averaging}), utilizing improved tagging variables which account for the spatial correlations among neighboring gradient vectors (Sec.~\ref{sec:tagging}), etc. 
\item {\em Use the correct ansatz for the shape of the boundary.} At the cost of giving up model-independence, one could focus on a particular theory model, derive the parametric form of the expected boundary shape, and then use that parametrization to test for the presence of such boundaries in the data. This approach was proved successful in specific event topologies motivated by supersymmetry \cite{Debnath:2016mwb,Debnath:2016gwz,DebnathPhD}, but relies on the experimenter being able to make the correct theory model assumption.
\item {\em Construct the envelope of the line segments with the largest wombling measures.} As shown by the results in Figs.~\ref{fig:circleexample} and \ref{fig:oceancircle5}, when probing a curvilinear boundary with straight line segments, we obtain a whole family of wombling boundary candidates which can be tagged with their relatively large values of $\bar\Gamma[C]$. We saw that each of these candidates is tangential to the true boundary, therefore, the task of constructing the true boundary reduces to the task of finding a planar curve which is tangential to each of the tagged straight line candidate segments at some point. The answer to this problem is precisely the envelope curve \cite{wikienvelope}.
\item {\em Identify and agglomerate ``the best'' line segments $\Delta\ell_\alpha$.} A specific realization of an approximate piece-wise reconstruction of the envelope curve mentioned above is offered by the following procedure. Note that our analysis not only tags straight lines $C$ with large values of the wombling measure, but it also identifies which individual portions $\Delta\ell_\alpha$ of those lines are most likely to be tangential to a true boundary, as indicated by the rainbow-color coding of the line $C_w$ in Figs.~\ref{fig:lineresult5}, \ref{fig:lineresult1p5} and \ref{fig:circleresult5}. This suggests that instead of working with the full line $C$ tagged by the algorithm, we can instead focus our attention on the individual elements $\Delta\ell_\alpha$ from it with the largest contributions to the average flux through $C$. The straightforward application of this method, however, will reintroduce sensitivity to local gradient fluctuations. In order to avoid this, we propose to rank the individual elements $\Delta\ell_\alpha$ not by their average flux $\langle \vec{G} \rangle_\alpha\cdot \hat n_C$, as was done in Figs.~\ref{fig:lineresult5}, \ref{fig:lineresult1p5} and \ref{fig:circleresult5}, but by the {\em rescaled} average flux 
\beq
\left( \langle \vec{G} \rangle_\alpha\cdot \hat n_C \right) \times \left| \bar\Gamma[C]\right|^\gamma,
\label{eq:rescaling}
\eeq
where the power $\gamma$ is a suitably chosen positive parameter, which interpolates smoothly between ``complete locality'' ($\gamma=0$) and ``complete globality'' ($\gamma\to\infty$). The scaling by a power of $\bar\Gamma[C]$ in eq.~(\ref{eq:rescaling}) ensures that a given element $\Delta\ell_\alpha$ is judged not only by the local flux going through it, but also by its association with a suitable wombling boundary candidate line $C$. By increasing the power $\gamma$, we can suppress the effects of local statistical fluctuations, and in the limit of $\gamma\to\infty$ we eventually recover our previous results where one would select {\em all} the segments  $\Delta\ell_\alpha$ belonging to the best wombling candidate line $C_w$. However, using finite values of $\gamma$ allows us to 1) let in ``the best'' individual segments $\Delta\ell_\alpha$ from other candidate lines $C$ which are not $C_w$, but have comparably large values of $\bar\Gamma[C]$, and 2) eliminate from consideration those individual segments $\Delta\ell_\alpha$ from the winning line $C_w$ whose local flux values are too low --- presumably such elements belong to $C_w$ only because we have used the wrong ansatz for the shape of the boundary. This approach strikes the right balance between the ``locality'' of the flux through an individual element $\Delta\ell_\alpha$ and the ``globality''  (in the method of calculation) of the wombling measure $\bar\Gamma[C]$. The procedure is illustrated in Fig.~\ref{fig:circleresult6}, where we show the individual line segments $\Delta\ell_\alpha$ whose rescaled flux values (\ref{eq:rescaling}) are in the top 1 percentile, after scanning the $(p_{in}, p_{out})$ parameter space on an $80\times80$ grid. We see that, despite using the wrong ansatz (straight lines), the circular boundary is reconstructed quite well, with only a few stragglers showing up in the bulk.
\end{itemize}

\section{Significance estimation}
\label{sec:significance}

In the previous sections we discussed different techniques for identifying wombling boundaries in point datasets. In applications to collider event data in high-energy physics, the presence of a wombling boundary could be indicative of new physics, if its location is in a region of phase space which is unremarkable from the point of view of the SM background. However, before claiming a discovery, one must be confident that such wombling boundaries cannot be accidentally generated by SM data alone. For this purpose, it is necessary to supplement any proposed wombling technique with a corresponding prescription for assessing the significance of any reconstructed wombling boundary. Since previous work \cite{Debnath:2015wra,Debnath:2016mwb,Debnath:2016gwz} did not address this issue, we shall now do so using a frequentist approach.

\begin{figure}[t]
 \centering
 \includegraphics[width=.65\textwidth]{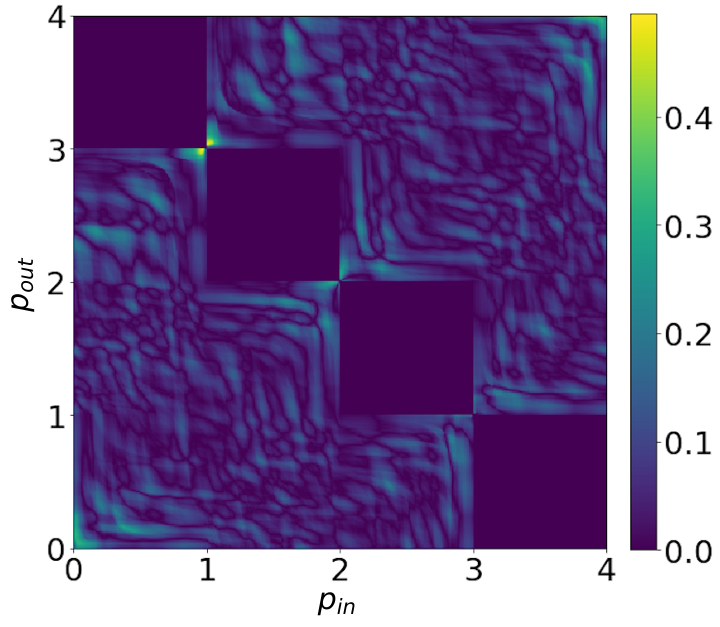}
 \caption{\label{fig:bkndresult} A heat map of wombling measures $|\bar\Gamma[C]|$ analogous to Figs.~\ref{fig:oceanplot5}, \ref{fig:oceanplot1p5} and \ref{fig:oceancircle5}, but for a typical background pseudo-experiment where the point data is generated from the pure-background distribution (\ref{fSMeqC}) and then a single Lloyd iteration is applied.   }
\end{figure}

The end result of the wombling method described in Sec.~\ref{sec:womblingdiscrete} was the selection of the best possible wombling line candidate $C_w$, together with its corresponding wombling measure
\beq
\bar \Gamma_w \equiv \bar\Gamma[C_w].
\label{eq:Gammawdef}
\eeq
In order to use $|\bar \Gamma_w|$ as our test statistic, we need to know its distribution under the pure-background hypothesis. For this purpose, we generate a number of pseudo-experiments where the point data is generated from the pure-background distribution (\ref{fSMeqC}), and for each pseudo-experiment, we repeat the wombling analysis from the previous section. A typical result from one such pseudo-experiment is shown in Fig.~\ref{fig:bkndresult}, where, in analogy to Figs.~\ref{fig:oceanplot5}, \ref{fig:oceanplot1p5} and \ref{fig:oceancircle5}, we show a heat map of $|\bar\Gamma[C]|$ in the $(p_{in},p_{out})$ parameter space.
We see that, as expected, in the absence of a real signal the typical values for the wombling measure are relatively low almost everywhere in the $(p_{in},p_{out})$ parameter space, except at one very special location near $(1,3)$. It is not difficult to realize that this location corresponds to very short candidate lines $C$ which ``cut'' the lower right corner of our field of view. We already alluded to this problem at the very end of Sec.~\ref{sec:rho1p5}, and our proposed solution was simply to exclude such very short lines from consideration\footnote{Another possibility could be to use the unnormalized wombling measure (\ref{eq:Gammadef}). We shall leave this option open for a future investigation \cite{MPR}.}. Therefore, when we derive the $|\bar\Gamma_w|$ distributions below, we shall apply a minimum cut on the allowed length of any candidate line $C$. In order to make sure this problem does not reappear, we shall conservatively increase our previous minimum length cut from $0.5$ to $\sqrt{2}/2$, which is the length of the line connecting the midpoints of any two neighboring edges of our field of view.

\begin{figure}[t]
 \centering
 \includegraphics[width=.5\textwidth]{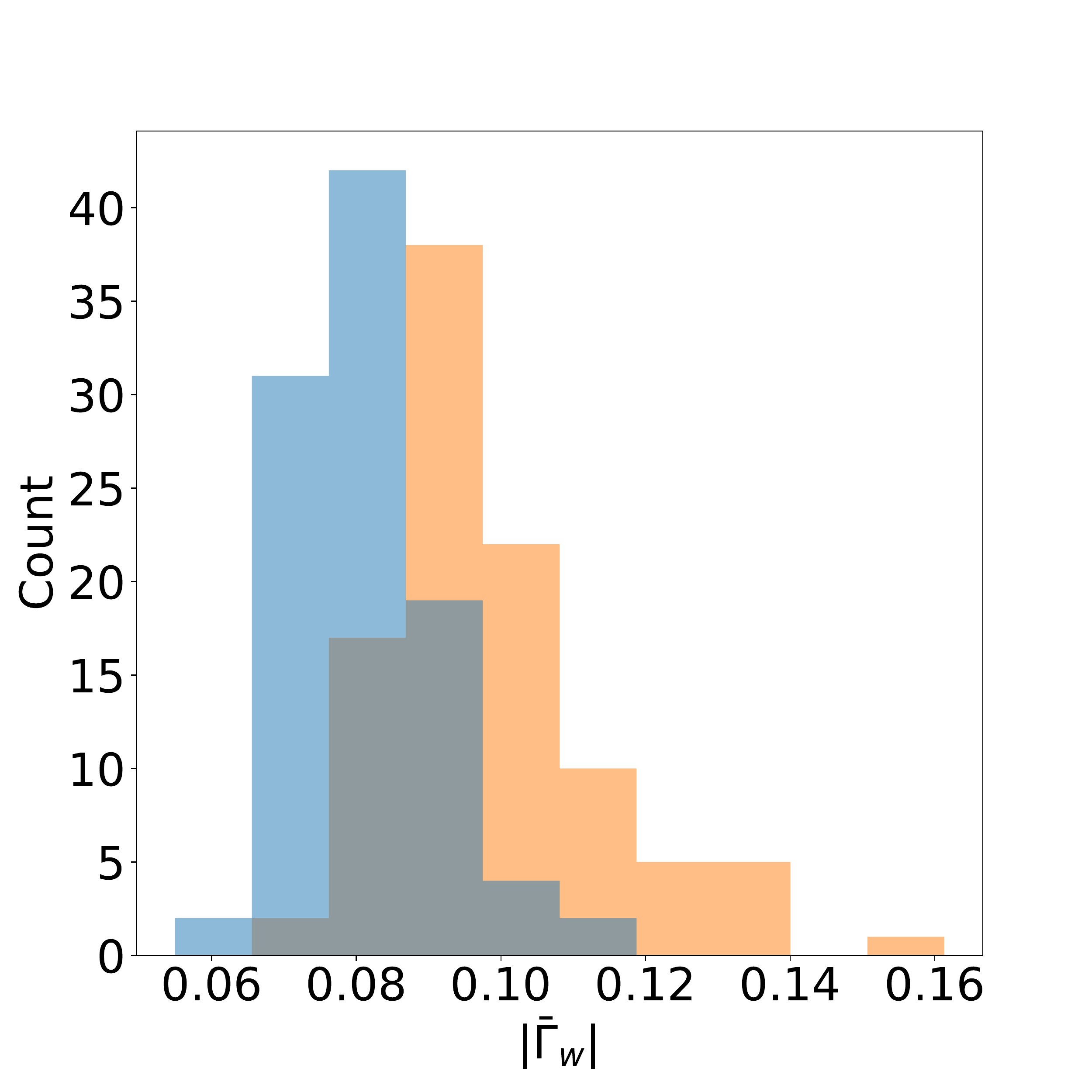}
 \caption{\label{fig:distributions} Distributions of $|\bar\Gamma_w|$ for two populations of 100 pseudo-experiments each, and with $N=1000$ points per pseudo-experiment. The blue histogram corresponds to data generated from the pure-background distribution (\ref{fSMeqC}) as in the left panel of Fig.~\ref{fig:datasetsFlat}. The orange histogram represents pseudo-experiments with added signal with a straight-line boundary as in the middle panel of Fig.~\ref{fig:datasetsFlat}, with the points sampled from (\ref{fSMfLine}) with $\rho=1.5$. In either case, the wombling measure $\bar\Gamma[C]$ was computed throughout the $(p_{in},p_{out})$ parameter space on an $80\times 80$ grid and the best wombling line $C_w$ of minimum length $\sqrt{2}/2$ was identified following the procedure of Sec.~\ref{sec:womblingdiscrete}.}
 \end{figure}

\begin{figure}[t]
 \centering
   \begin{minipage}{.48\textwidth}
 \vspace{-0.5cm}
 \includegraphics[width=.72\textwidth]{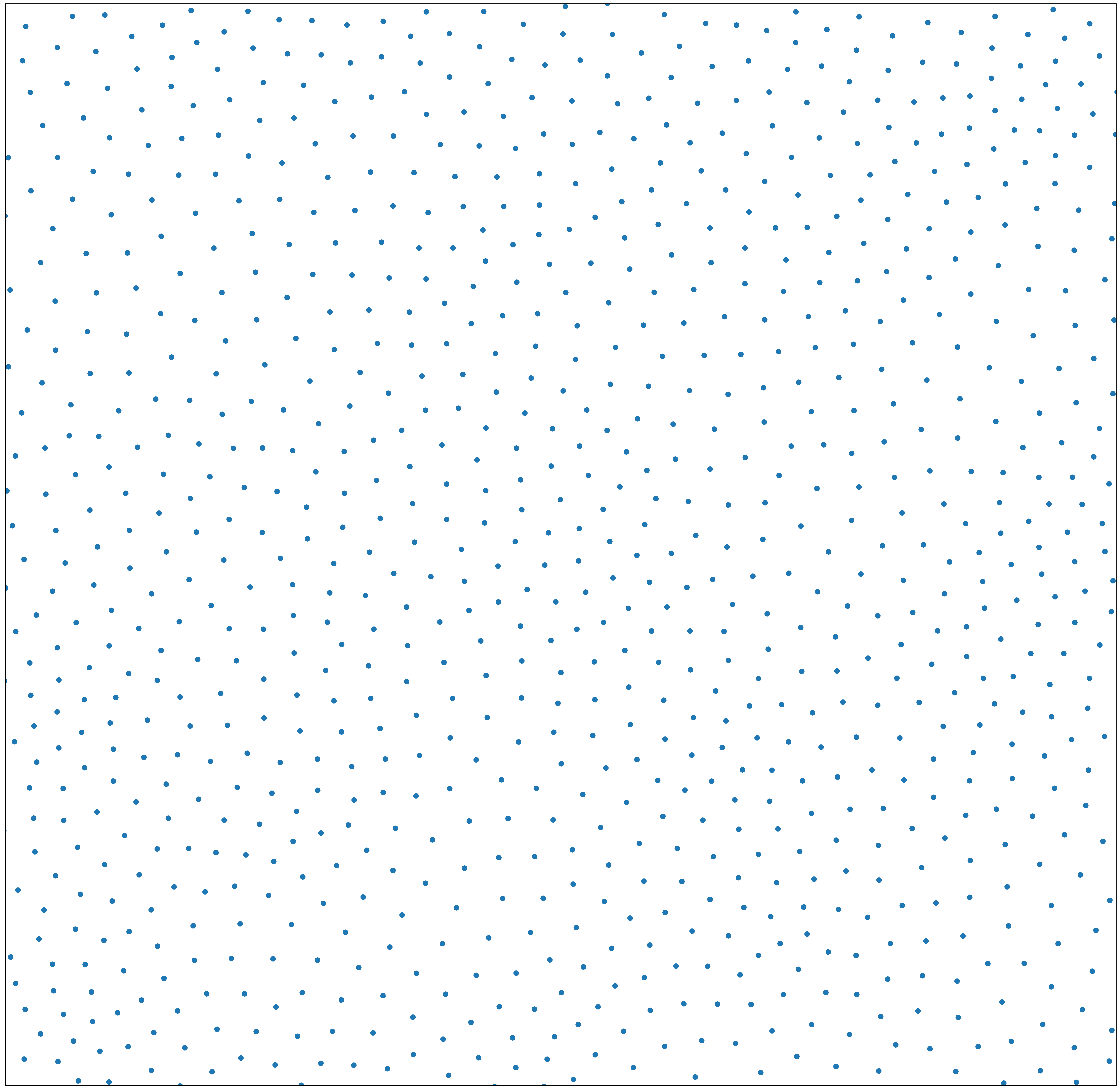} 
   \end{minipage}
 \hspace{-2cm}
   \begin{minipage}{.48\textwidth}
 \includegraphics[width=.95\textwidth]{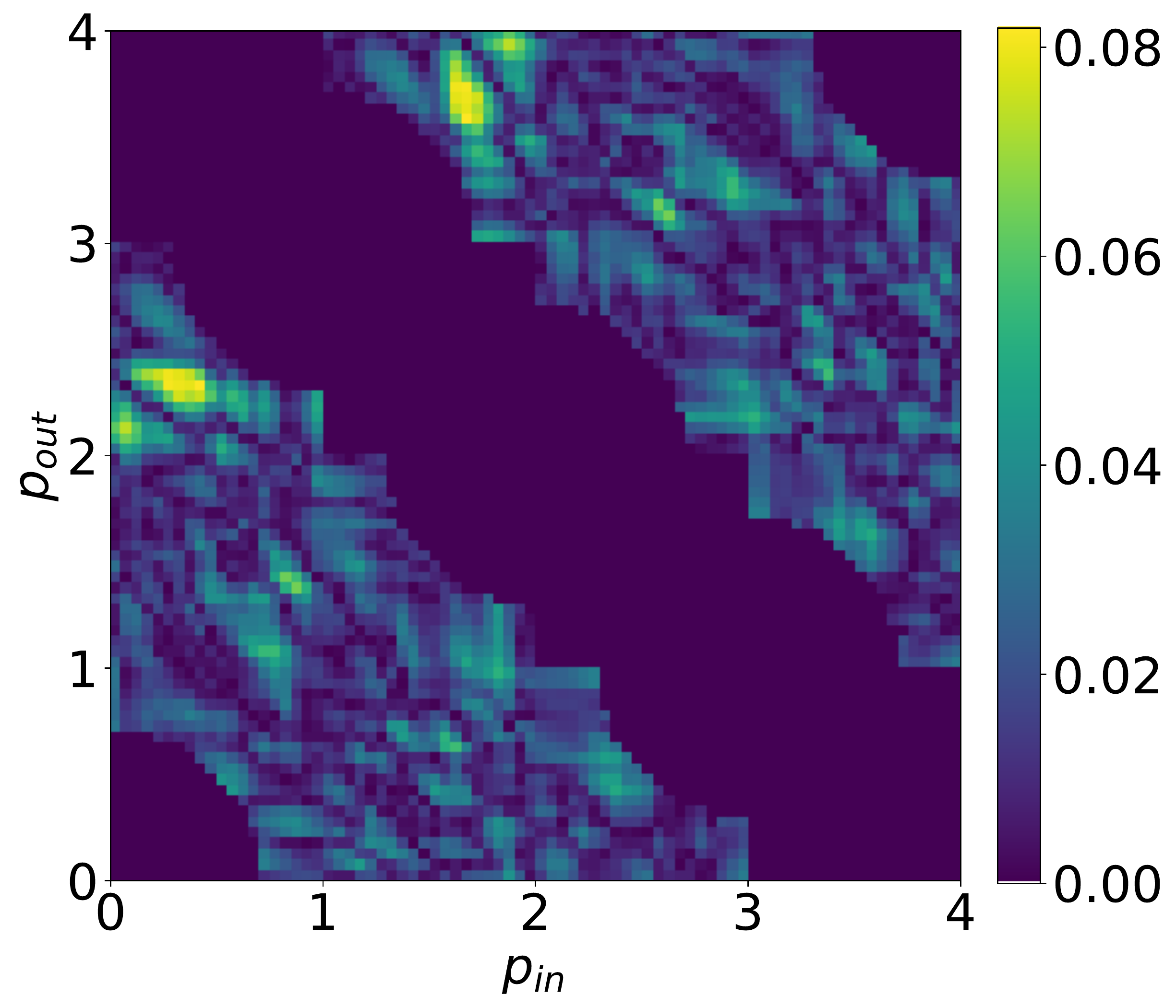}
   \end{minipage}
 \caption{\label{fig:bkndgeneric} The point data (left panel) and the $|\bar\Gamma[C]|$ heatmap (right panel) for a representative pure-background pseudo-experiment entering the blue histogram in Fig.~\ref{fig:distributions}.   }
\end{figure}

With those preliminaries, we are ready to compare the distributions of our test statistic $|\bar\Gamma_w|$ for signal and background. The blue histogram in Fig.~\ref{fig:distributions} shows the $|\bar\Gamma_w|$ distribution for 100 pure-background pseudo-experiments with $N=1000$ points each, where the data was generated from the pure-background distribution (\ref{fSMeqC}) as in the left panel of Fig.~\ref{fig:datasetsFlat}. At the preprocessing stage, we increased the number  of Lloyd iterations to 10, since we shall be interested in the case of relatively weak signals (see the related discussion in Sec.~\ref{sec:rho1p5} and Ref.~\cite{Debnath:2015wra}). Following the procedure of Sec.~\ref{sec:womblingdiscrete}, we then computed the wombling measure $\bar\Gamma[C]$ on an $80\times 80$ grid in the $(p_{in},p_{out})$ parameter space and the best wombling line $C_w$ (of minimum length $\sqrt{2}/2$) was identified and its wombling measure (\ref{eq:Gammawdef}) was entered in the histogram. The resulting distribution is relatively concentrated around a mean of 0.08, and extends up to 0.11, which sets the lower limit on the target range for signal detection. For illustration, in Fig.~\ref{fig:bkndgeneric} we show results for one typical pure-background pseudo-experiment whose value for $\bar\Gamma_w$ is equal to the mean of the distribution shown in Fig.~\ref{fig:distributions}. 

Given the background distribution from Fig.~\ref{fig:distributions}, we can now assess what types of signals might be discoverable. Obviously, the larger the signal component, the more pronounced the wombling boundary. In our conventions, the signal strength was parametrized by the $\rho$ parameter. For example, in Sec.~\ref{sec:rho5example} we saw that for $\rho=5$ we obtained $\bar\Gamma_w\sim 0.7$, while the weaker signal with $\rho=1.5$ in Sec.~\ref{sec:rho1p5} resulted in only $\bar\Gamma_w\sim 0.19$ (note that in Sec.~\ref{sec:rho1p5} the data was preprocessed with only 5 Lloyd steps; adding 5 additional steps as was done in Fig.~\ref{fig:distributions} would further reduce the value of $\bar\Gamma_w$ slightly). Given the pure-background distribution in Fig.~\ref{fig:distributions}, it is clear that in both of those examples the observed effect could not have been attributed to a background fluctuation and would represent a discovery. At the same time, a careful inspection of the left panel in Fig.~\ref{fig:linedata1p5} shows that the $\rho=1.5$ example of Sec.~\ref{sec:rho1p5} was rather ``lucky'' due to fortuitous fluctuations in the data near the theoretical boundary. In order to estimate the prospects for a more typical signal scenario, we simulate 100 pseudo-experiments with $N=1000$ data points each, generated from the distribution (\ref{fSMfLine}) with $\rho=1.5$. The corresponding distribution of the test statistic $|\bar\Gamma_w|$ for those signal pseudo-experiments is shown with the orange histogram in Fig.~\ref{fig:distributions}. We see that, on average, the values of $|\bar\Gamma_w|$ are larger in the presence of a signal --- the mean of the orange histogram is shifted to 0.10. Comparing the tails of the two distributions, we find that in 40\% of the cases, the signal is discoverable at 2 sigma and in 16\% of the cases it is discoverable at 3 sigma. These prospects can probably be further improved by optimizing the different aspects of our wombling algorithm, but such an optimization is outside the scope of this paper.

\section{Conclusions and outlook}

In this paper we reviewed and refined the existing procedures for identifying wombling boundaries in point datasets. Our interest in this topic stems from the fact that high energy physics collider data can be viewed as point data in the relevant phase space of the final state signature. For better visual illustration, we considered point data examples in two dimensions, but our technique can be readily generalized to higher dimensional data. We proposed several modifications to the standard algorithm which lead to improved detection efficiency and significance:
\begin{itemize}
\item We advocated the use of the Delaunay triangulation of the data instead of (or perhaps in addition to) the Voronoi tessellation of the data. We argued that the Delaunay tessellation is the natural framework for computing the local gradient vectors which is the first and most important step of any wombling algorithm.
\item We considered three different techniques for reducing the effect of statistical fluctuations:
\begin{enumerate}
\item Rescaling the local gradient vectors as in (\ref{Grescaling}), see Section~\ref{sec:gradrescaling}.
\item Applying Voronoi relaxation of the data via several Lloyd iterations as a preprocessing step, see Section~\ref{sec:LSU}.
\item Local averaging of the gradient vectors, which can be performed either at the level of a Voronoi cell (\ref{ave}) or at the level of a Delaunay triangle (\ref{superave}), see Section~\ref{sec:averaging}.
\end{enumerate}
\item We studied new tagging variables (dot products of neighboring averaged gradient vectors) for selecting elements of the tessellation marking the location of a wombling boundary. In Section \ref{sec:tagging} we showed that the new variables have improved selection efficiency, since they take into account the spatial correlations among neighboring gradient vectors along the wombling boundary. In Section~\ref{sec:agglomeration} we pointed out that the new variables additionally can be used to naturally connect the tagged elements into continuous boundaries.
\item In Secs.~\ref{sec:womblingcontinuous} and \ref{sec:womblingdiscrete}  we explored the idea of Refs.~\cite{Banerjee2006,Gelfand2015} to rank wombling boundary candidates $C$ by a global wombling measure, e.g., $\Gamma[C]$ from (\ref{eq:Gammadef}) or $\bar\Gamma[C]$ from (\ref{eq:Gammaavedef}). On the basis of several toy examples we showed that this approach is successful in identifying the correct boundary, and with a slight modification (\ref{eq:rescaling}) can be used even when the shape of the boundary is different from the assumed ansatz, see Sec.~\ref{sec:shape}.
\item In Sec.~\ref{sec:significance} we showed how one can estimate the statistical significance of any detected wombling boundary using a frequentist approach.
\end{itemize}

The present study complements and further expands the work of Refs.~\cite{Debnath:2015wra,Debnath:2016mwb,Debnath:2016gwz,Debnath:2018azt,DebnathPhD} in an interesting direction which, while popular in other fields of science, is still rather new to the field of high energy physics. We believe that our investigations here are only scratching the surface of what could be a very promising research thrust. In particular, the approach of treating high energy collider data as point data and studying its geometric properties is complementary to the existing binning techniques and in the long run could prove to be more suitable to the application of modern machine learning techniques \cite{Albertsson:2018maf,Bourilkov:2019yoi}. 

\acknowledgments
We thank D.~Debnath, J.~Gainer, C.~Kilic, D.~Kim and Y.-P.~Yang for useful discussions. PS is grateful to the LHC Physics Center at Fermilab for hospitality and financial support as part of the Guests and Visitors Program in the summer of 2019. The work of PS is supported by the University of Florida CLAS Dissertation Fellowship funded by the Charles Vincent and Heidi Cole McLaughlin Endowment. This work was supported in part by the United States Department of Energy under Grant No. DE-SC0010296.

\appendix

\section{Studies of non-uniform background distributions}
\label{sec:nonuniform}

The illustrative examples in the main body of the paper so far have used data generated from a uniform background distribution (\ref{fSMeqC}). In most applications outside collider physics this is a valid assumption, since the density in the bulk is at most a slowly varying function. However, in particle physics one often has to face backgrounds which are steeply falling functions parametrized by power laws or exponents. For completeness, in this appendix we shall relax the uniformity assumption about the background and shall consider two other typical situations, namely, when the background distribution is given by a (linear) power law (Section~\ref{sec:ramp}) or an exponential (Section~\ref{sec:exp}).

\subsection{Example with linearly increasing background}
\label{sec:ramp}

\begin{figure}[t]
 \centering
 \includegraphics[width=.5\textwidth]{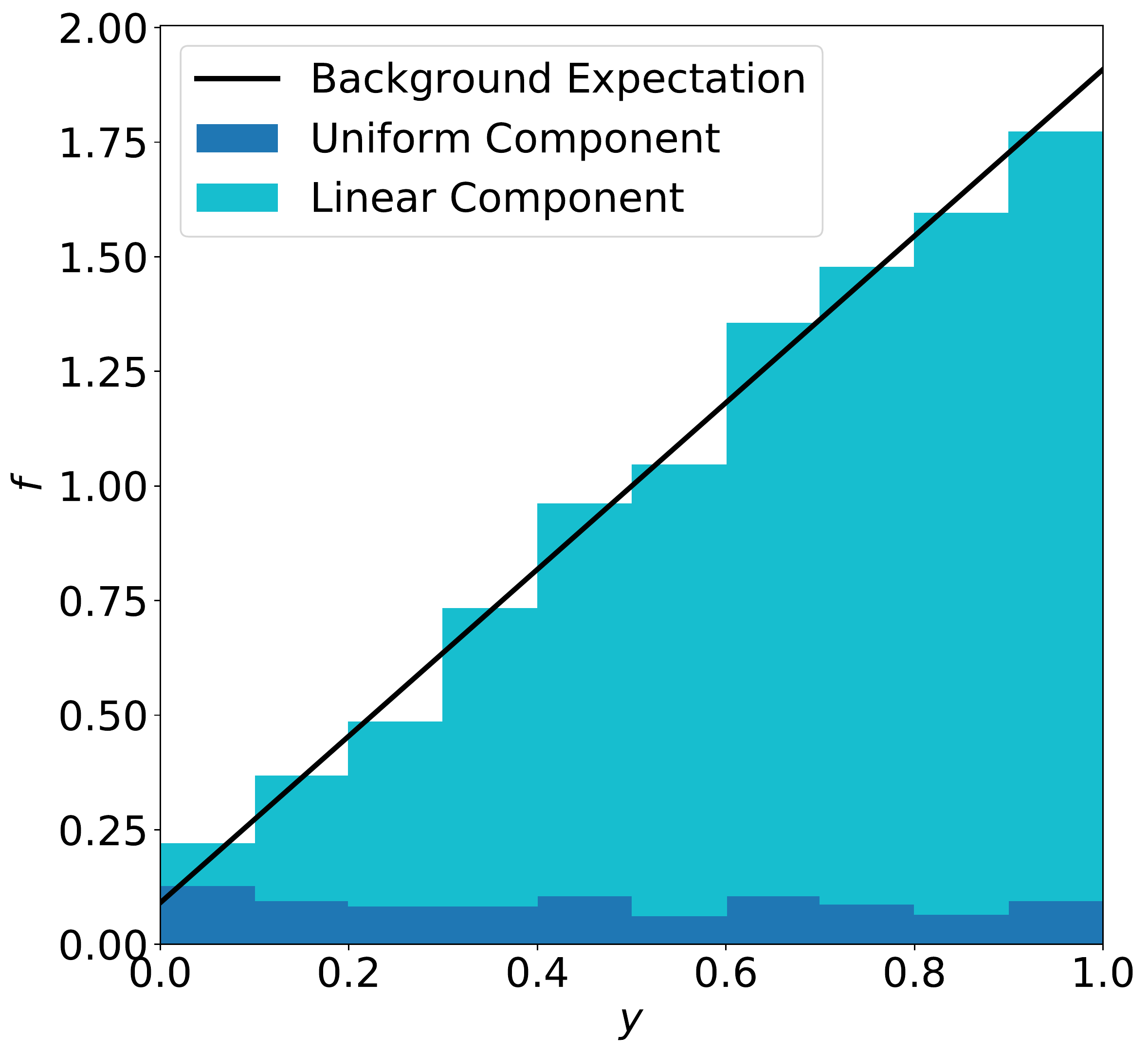}
 \caption{\label{fig:bkndDistributionRamp} 
 The background distribution (\ref{bknd:ramp}) for the example with linearly increasing background considered in Sec.~\ref{sec:ramp}, for $N_U/N_L=0.1$.  }
\end{figure}

In this section we reconsider our circular boundary example from Sec.~\ref{sec:discretecircular}, only this time 
we trade the uniform background distribution (\ref{fSMeqC}) for a linearly increasing function
\begin{equation}
f(x,y) = \frac{N_U/N_L+2y}{N_U/N_L+1}.
\label{bknd:ramp}
\end{equation}
Due to the rotational symmetry, without any loss of generality we can take the function $f$ to increase in the positive $y$ direction, as shown in eq.~(\ref{bknd:ramp}). The black solid line in Fig.~\ref{fig:bkndDistributionRamp} illustrates the $y$-dependence of this background distribution.
Note that since $f(x,y=0)>0$, we can think of (\ref{bknd:ramp}) as being made of two components: $f=f_U+f_L$, where the uniform component
$$
f_U(x,y) = \frac{1}{1+N_L/N_U},
$$
illustrated by the dark-shaded blue histogram in Fig.~\ref{fig:bkndDistributionRamp}, contains a total of $N_U$ events, while the linear component
$$
f_L(x,y) = \frac{2y}{1+N_U/N_L},
$$
depicted by the light-shaded cyan histogram in Fig.~\ref{fig:bkndDistributionRamp}, contains a total of $N_L$ events. In what follows (as well as in Fig.~\ref{fig:bkndDistributionRamp}) we fix $N_U/N_L=0.1$.

\begin{figure}[t]
 \centering
 \includegraphics[height=.42\textwidth]{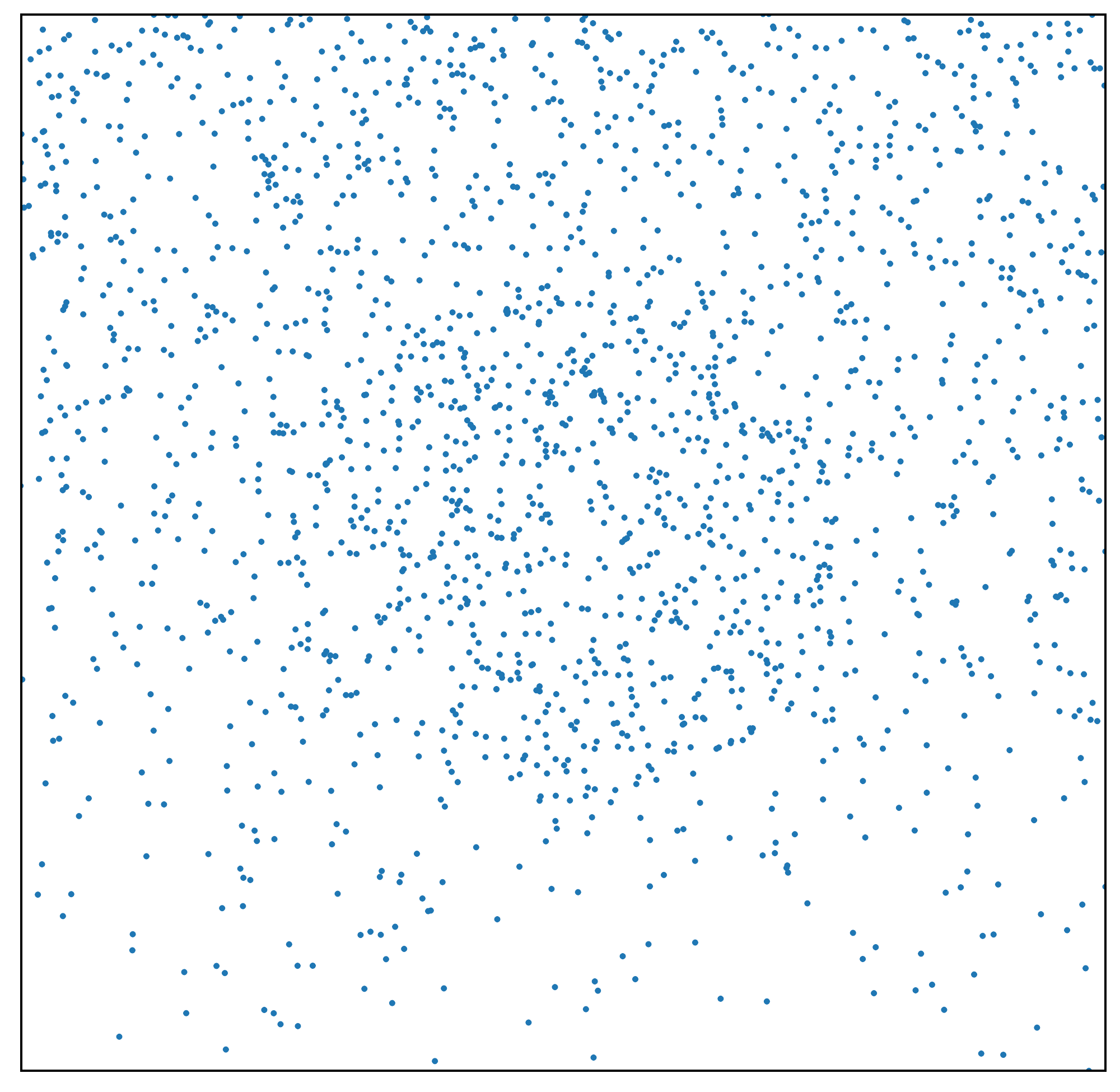}
 ~
 \includegraphics[height=.42\textwidth]{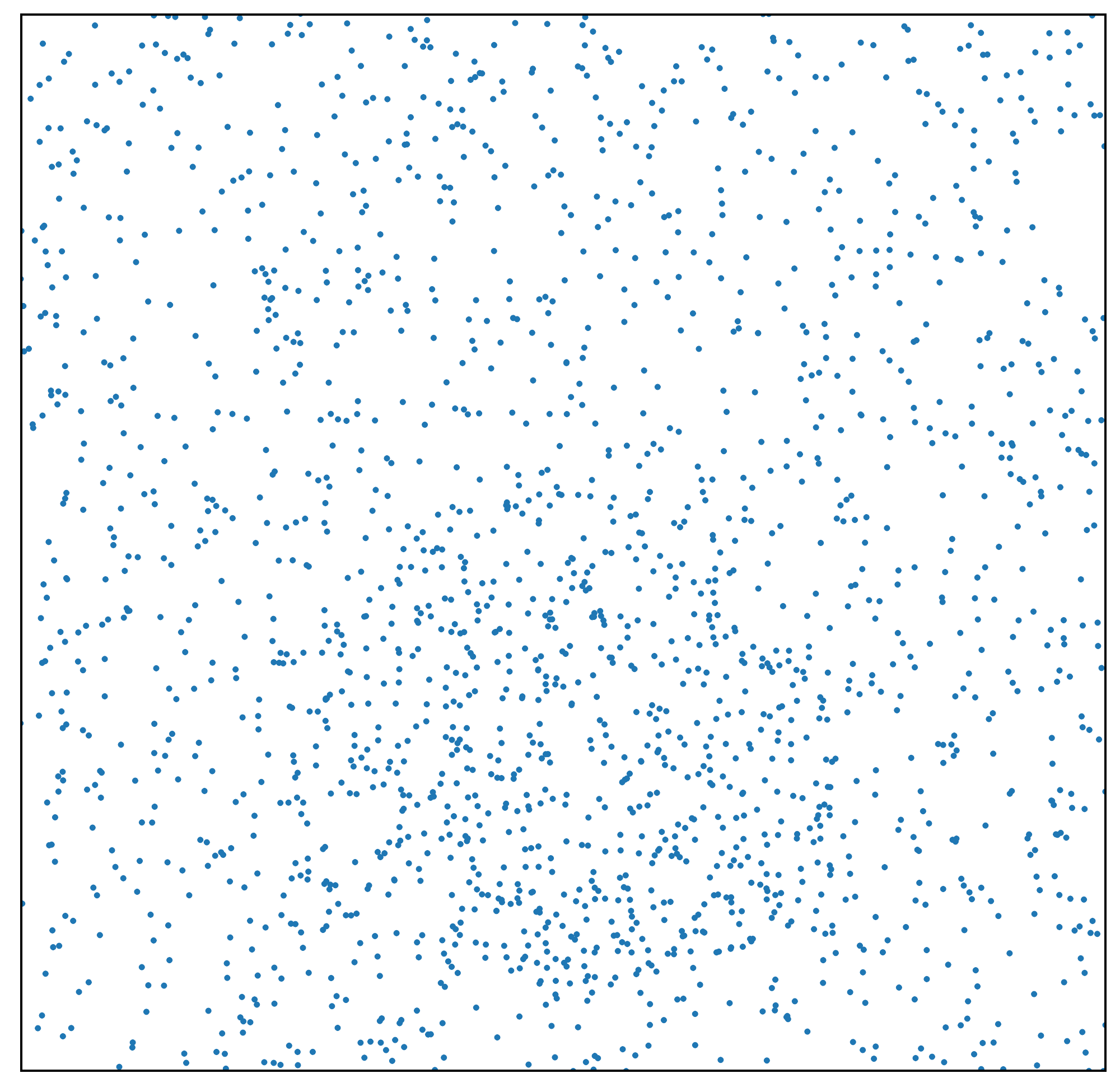}
 \caption{\label{fig:ramp_points} 
 The point data for the example with linearly increasing background considered in Sec.~\ref{sec:ramp}
 before (left panel) and after (right panel) the probability integral transform (\ref{Rtransformation}). }
\end{figure}

To this background distribution (sampled with $N_U+N_L=1500$ points) we add a uniform circular signal as before, with total number of signal points $N_S=450$. The resulting point dataset is shown in the left panel of Fig.~\ref{fig:ramp_points}.
\begin{figure}[t]
 \centering
 \includegraphics[height=.42\textwidth]{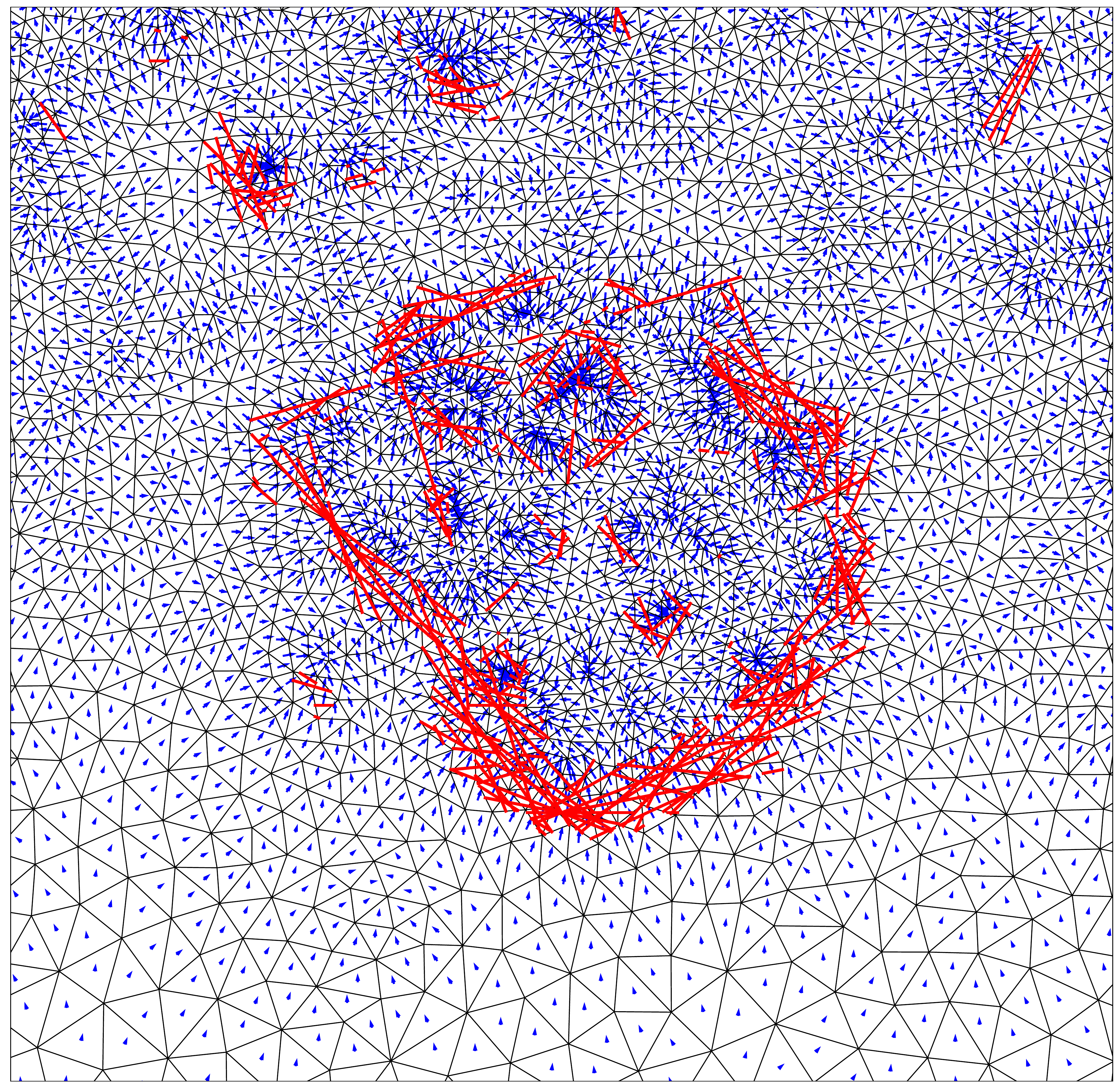}
 ~
 \includegraphics[height=.42\textwidth]{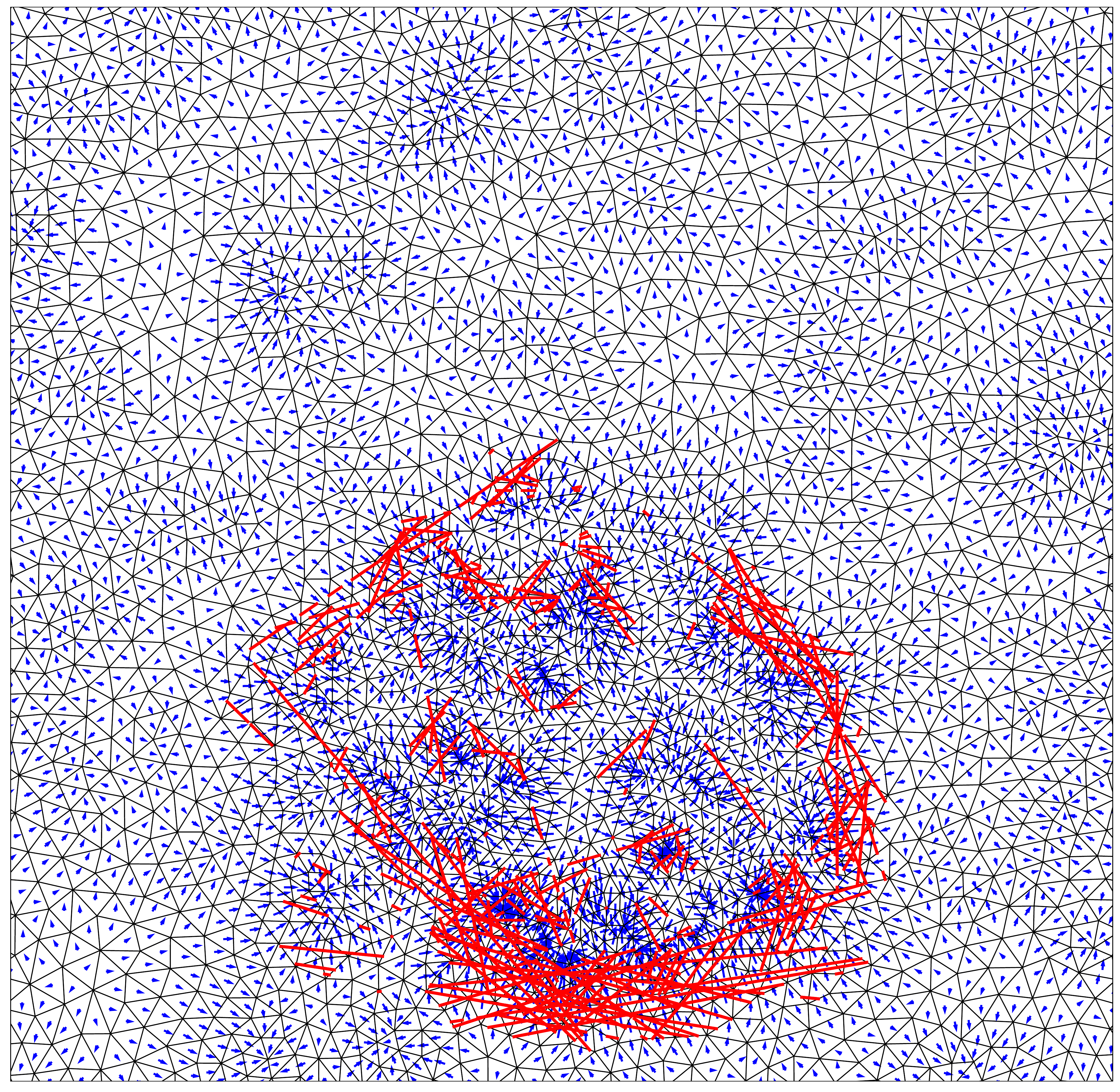}
 \caption{\label{fig:ramp_strawman} 
 The same as Fig.~\ref{fig:circleresult6}, but for the point data in Fig.~\ref{fig:ramp_points}. 
 Here we scanned on a coarser $40\times 40$ grid after applying 7 Lloyd steps. The red lines have rescaled flux values 
 (\ref{eq:rescaling}) in the top 4 percentile (with $\gamma=1$). }
\end{figure}
In principle, we could now run our previous analysis from Sec.~\ref{sec:discretecircular} directly on this dataset, and the result is depicted in the left panel of Fig.~\ref{fig:ramp_strawman}, which is the analogue of Fig.~\ref{fig:circleresult6} for this case. In producing Fig.~\ref{fig:ramp_strawman}, we applied 7 Lloyd steps and scanned on a coarser $40\times 40$ grid, which results in 4 times fewer lines compared to Fig.~\ref{fig:circleresult6} (which was made on an $80\times 80$ grid). Correspondingly, to make a fair comparison with Fig.~\ref{fig:circleresult6}, we plotted the lines within the top 4 percentile of rescaled flux values (as opposed to the top 1 percentile shown in Fig.~\ref{fig:circleresult6}). 

The result in the left panel of Fig.~\ref{fig:ramp_strawman} demonstrates that our method is able to successfully identify the boundary in the presence of non-uniform background as well. However, we also notice that there are a few isolated line segments being picked up which are located in the background region near the top portion of the plot, where the background is large and the statistical fluctuations are creating relatively large local gradients. We therefore try an alternative strategy, where we first perform a change of variables designed to flatten the background distribution \cite{Debnath:2014eaa}, after which we perform the wombling analysis on the resulting dataset. In cases where the background distribution is known analytically, e.g., as in eq.~(\ref{bknd:ramp}), the required transformation is simply the probability integral transform, which in our case reads 
\begin{equation}
y \to \frac{\left( y + b \right)^2 - b^2}{2b+1},
\label{Rtransformation}
\end{equation}
where 
$$
b=\frac{N_U}{2N_L}.
$$
Note that the transformation (\ref{Rtransformation}) preserves the location of the points at $y=0$ and $y=1$, i.e., it is a map of $[0,1]\to [0,1]$. After rescaling the $y$ values of our original dataset, we obtain the data shown in the right panel of Fig.~\ref{fig:ramp_points} --- note how the density within the background region appears much more uniform, since by construction it is sampled from a uniform distribution. At the same time, the signal region has shifted down, but its boundary is still clearly defined. Now, performing our wombling analysis on the dataset in the right panel of  Fig.~\ref{fig:ramp_points}, we obtain the result in the right panel of Fig.~\ref{fig:ramp_strawman}.  We see that the boundary is again identified, this time somewhat more cleanly, since there are fewer spurious line segments in the bulk of the background region. In summary, the two panels in Fig.~\ref{fig:ramp_strawman} show that in the case of known non-uniform backgrounds, both approaches are viable, and the choice of which one to use can be left to the individual user.

\subsection{Example with exponentially increasing background}
\label{sec:exp}

\begin{figure}[t]
 \centering
 \includegraphics[width=.5\textwidth]{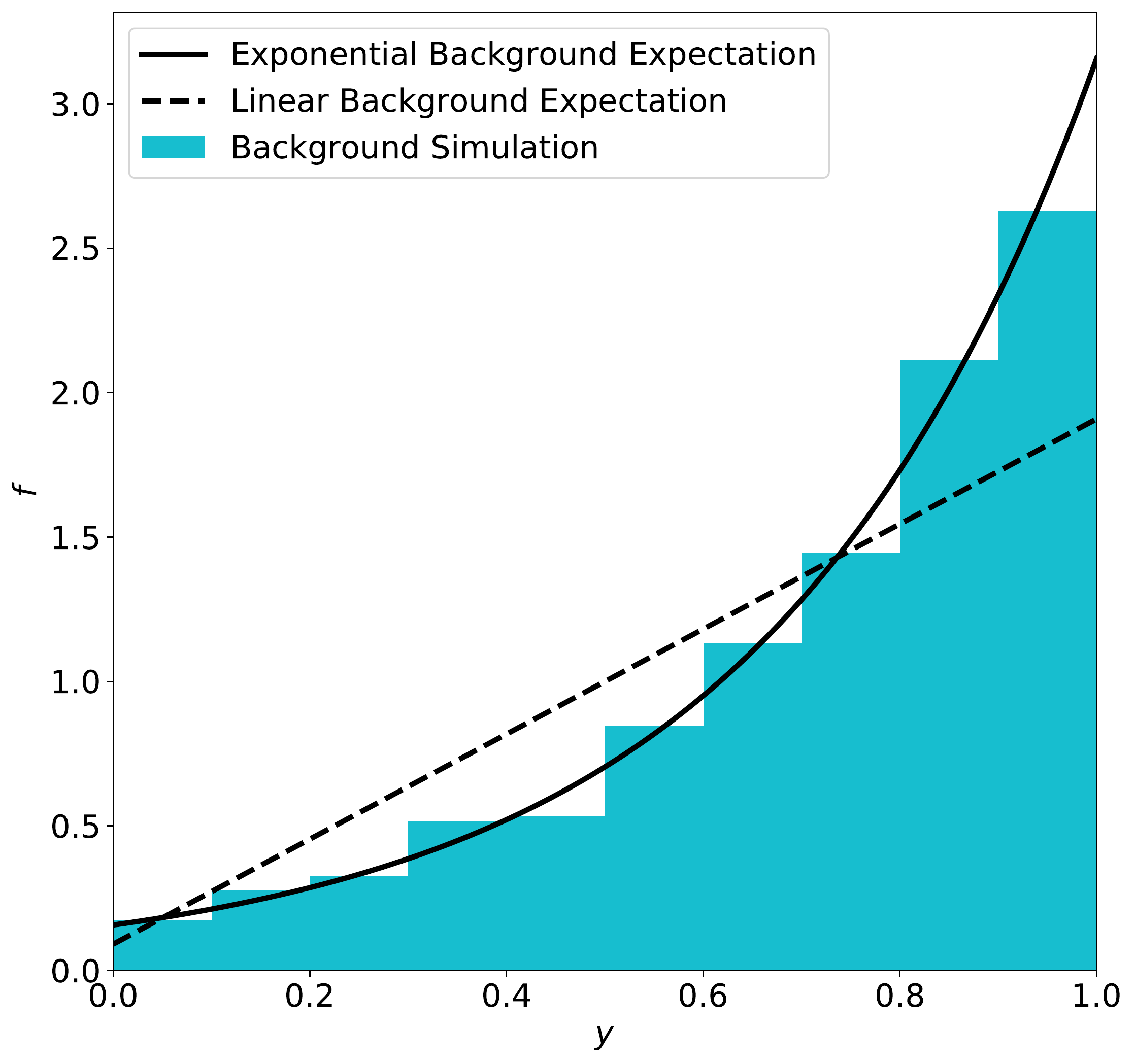}
 \caption{\label{fig:bkndDistributionExp} 
 The unit-normalized background distribution (\ref{bknd:exp}) (solid line) and the corresponding simulated data (histogram) used for the example considered in Sec.~\ref{sec:exp}. For comparison, the dashed line shows the linear background distribution used for the example in Sec.~\ref{sec:ramp}. }
\end{figure}

In this section we shall consider another example with non-uniform background, namely a background distribution given by
\begin{equation}
f(x,y)=\frac{3\, e^{3x}}{e^3-1}.
\label{bknd:exp}
\end{equation}
This function is illustrated in Fig.~\ref{fig:bkndDistributionExp} with the black solid line, together with a histogram of the simulated background data points. For comparison, we also show the linearly increasing background distribution from the previous section (the black dashed line).

We shall now repeat our wombling analysis exercise, but with an added twist, to make it more challenging. First, we shall keep the same number of background points (1500) but we shall weaken our signal down to 300 points only. Second, we shall {\em not} assume that we know the exact analytical form (\ref{bknd:exp}) of the background distribution, so that we cannot do the background flattening exactly. Instead, we shall attempt to flatten the background using ``the wrong'' linear relation (\ref{bknd:ramp}), since it resembles the actual (exponential) background --- compare the solid and dashed lines in Fig.~\ref{fig:bkndDistributionExp}.

\begin{figure}[t]
 \centering
 \includegraphics[height=.42\textwidth]{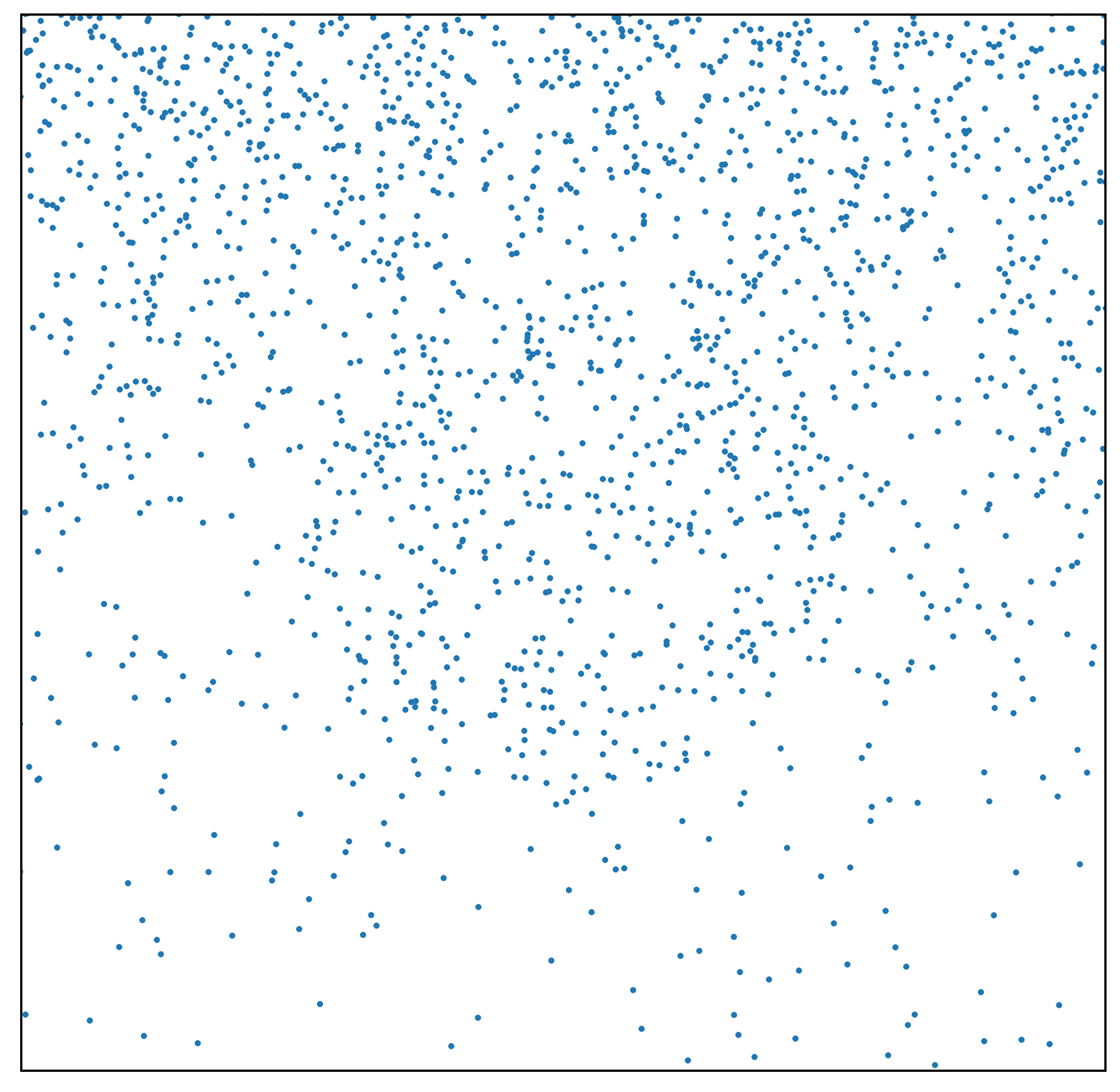}
 ~
 \includegraphics[height=.42\textwidth]{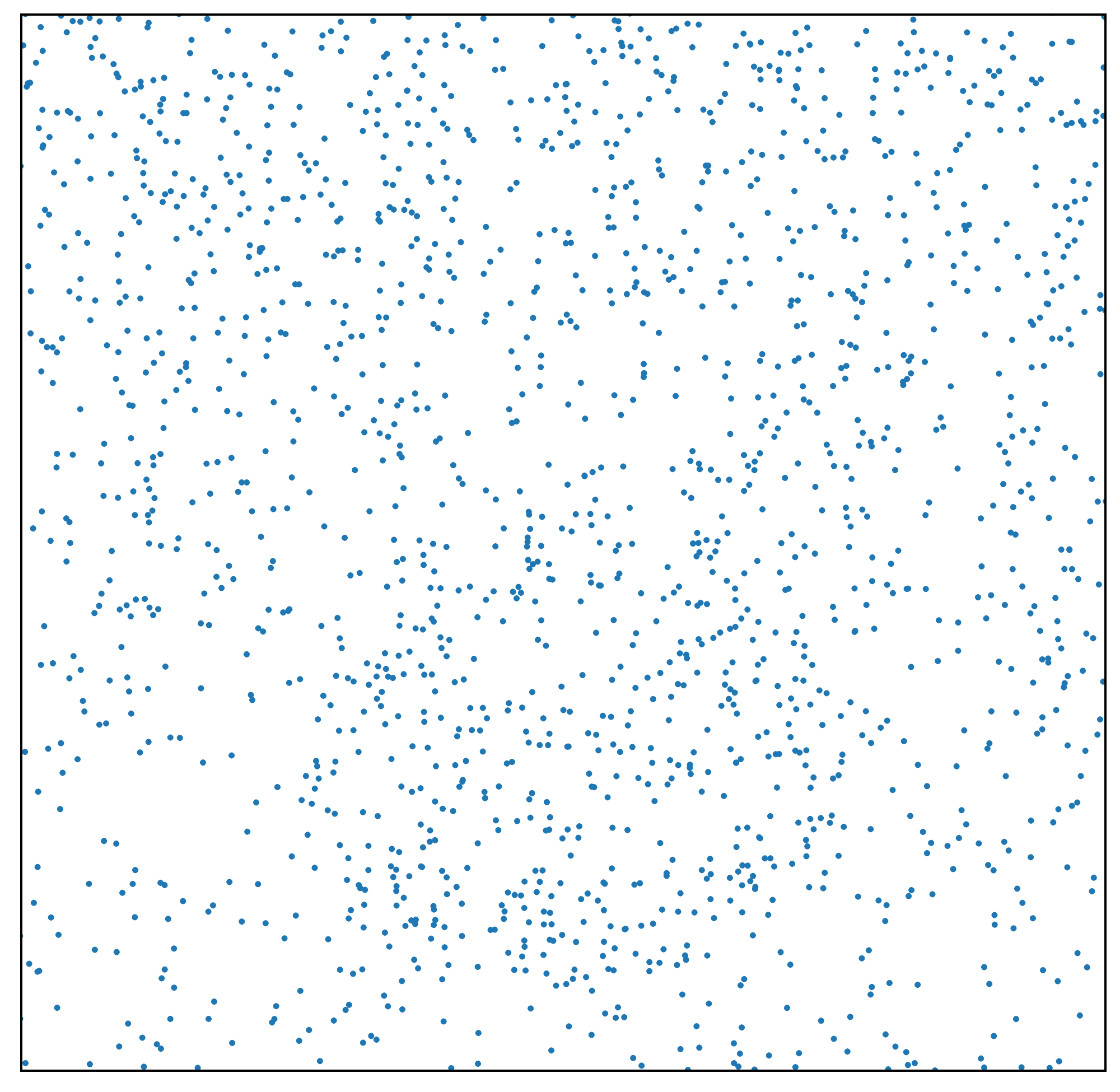}
 \caption{\label{fig:exp_points} 
 The same as Fig.~\ref{fig:ramp_points}, but for the example with exponentially increasing background considered in Sec.~\ref{sec:exp}. Here we use a weaker signal, with $N_S=300$ points, while keeping the same number of background points (1500). Note that the $y$ values in the right panel have been rescaled with the ``wrong'' transformation (\ref{Rtransformation}) corresponding to a linearly increasing background. }
\end{figure}

\begin{figure}[t]
 \centering
 \includegraphics[height=.42\textwidth]{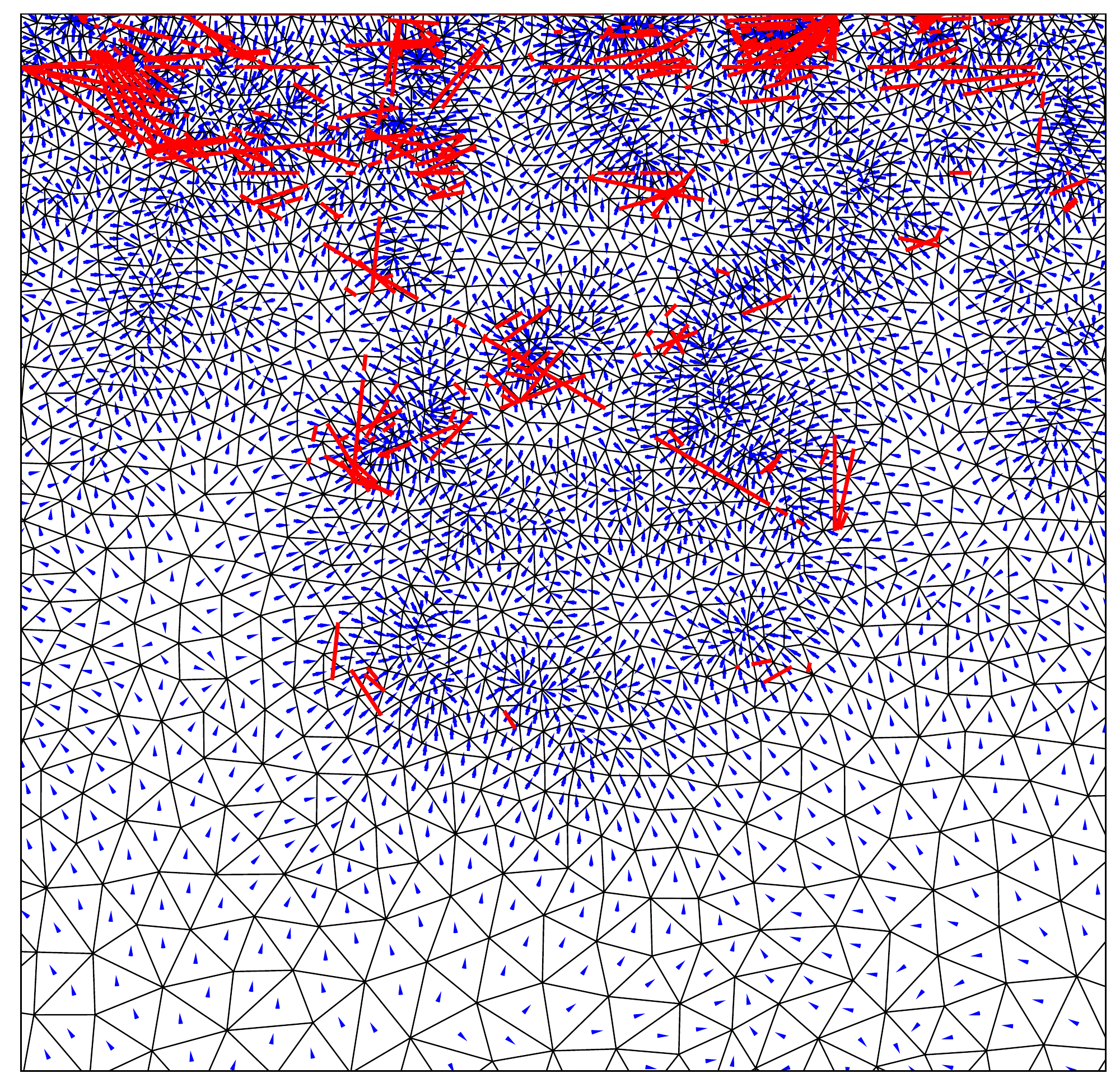}
 ~
 \includegraphics[height=.42\textwidth]{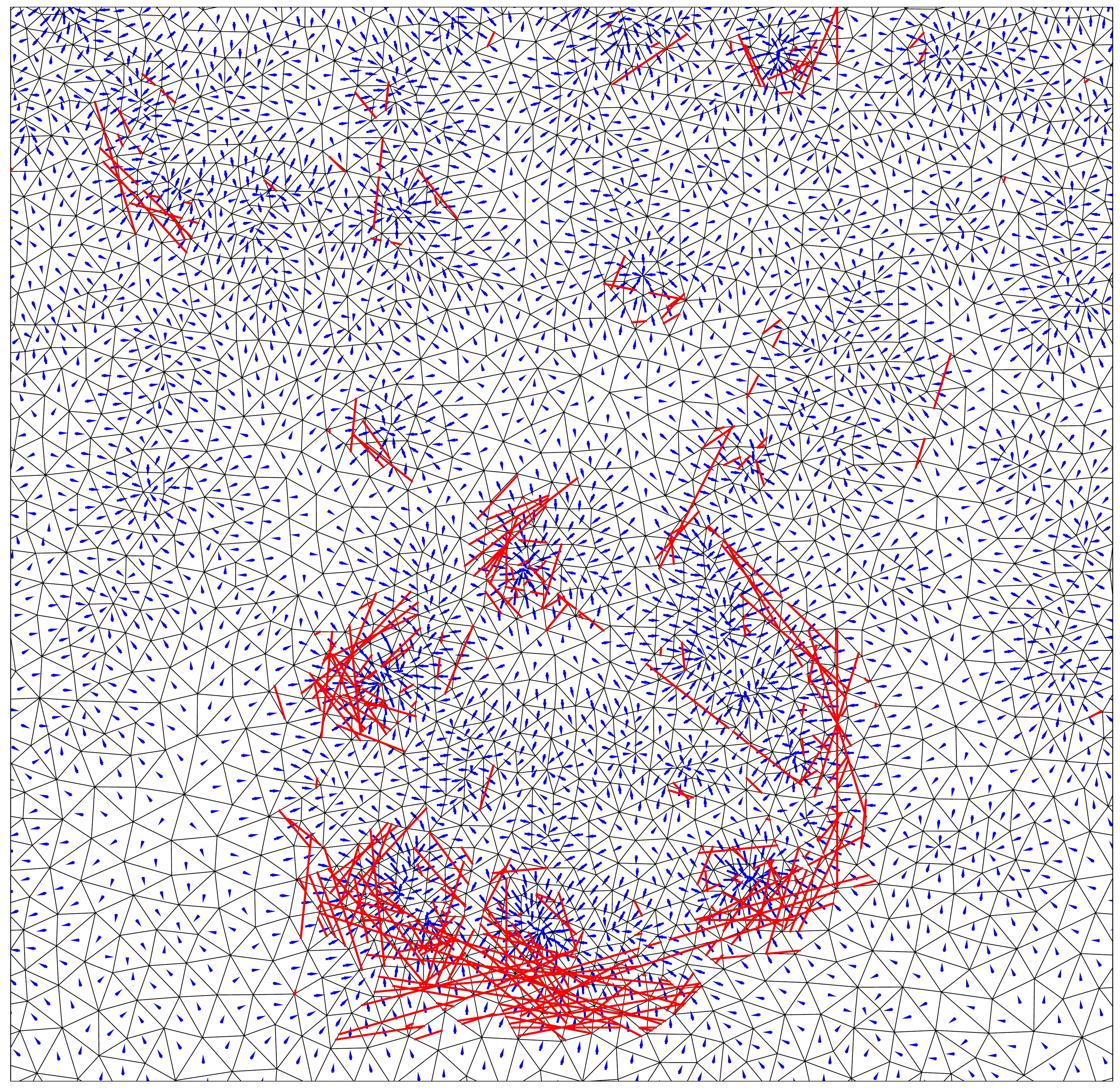}
 \caption{\label{fig:exp_strawman} 
 The same as Fig.~\ref{fig:ramp_strawman}, but for the example with exponentially increasing background considered in Sec.~\ref{sec:exp}. }
\end{figure}

The left panel in Fig.~\ref{fig:exp_points} shows the original point dataset, while the right panel in Fig.~\ref{fig:exp_points} shows the same data after rescaling the $y$ values with the ``wrong'' transformation (\ref{Rtransformation}) corresponding to a linearly increasing background. After performing our wombling analysis on these two datasets, we obtain the results shown in the respective panels of Fig.~\ref{fig:exp_strawman}. We see that even in this more challenging exercise, the boundary is still being identified properly in the right panel, where we have applied an approximate flattening transformation. In the left panel, on the other hand, the method is still doing its job - it is finding the regions with largest gradients, which in this case are in the background region, due to the exponential behavior.


\end{document}